\pdfoutput=1

\documentclass[9pt,twocolumn,twoside,dvipsnames]{pnas-new}

\templatetype{pnasresearcharticle} 

\newbox{\orcidlogo}
\sbox{\orcidlogo}{\large\includegraphics[height=1.8ex]{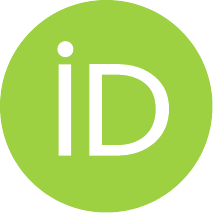}}
\newcommand{\orcid}[1]{\href{https://orcid.org/#1}{\usebox{\orcidlogo}\,}}

\title{grenedalf: population genetic statistics for the next generation of pool sequencing}

\author[1,2,*]{\orcid{0000-0002-1340-9644} Lucas Czech}
\author[3]{\orcid{0000-0002-3199-1447} Jeffrey P. Spence}
\author[1,4,5,6,7*]{\orcid{0000-0001-5711-0700} Moisés Expósito-Alonso}

\affil[1]{Department of Plant Biology, Carnegie Institution for Science, Stanford, USA}
\affil[2]{Section for GeoGenetics, Globe Institute, University of Copenhagen, Denmark}
\affil[3]{Department of Genetics, Stanford University, Stanford, USA}
\affil[4]{Department of Biology, Stanford University, Stanford, USA}
\affil[5]{Department of Global Ecology, Carnegie Institution for Science, Stanford, USA}
\affil[6]{Department of Integrative Biology, University of California Berkeley, Berkeley, USA}
\affil[7]{Howard Hughes Medical Institute, University of California Berkeley, Berkeley, USA}
\affil[*]{To whom correspondence should be addressed. E-mail: \href{mailto:lczech@carnegiescience.edu}{\color{black}{lczech@carnegiescience.edu}} and \href{mailto:moisesexpositoalonso@gmail.com}{\color{black}{moisesexpositoalonso@gmail.com}}}

\newcommand{\mainauthor}{Czech \textit{et al.}}
\leadauthor{\mainauthor}

\setboolean{displaywatermark}{false}

\usepackage{listings}
\usepackage{lastpage}
\usepackage{verbatimbox}

\usepackage{caption}
\usepackage[labelformat=simple]{subcaption}

\usepackage{siunitx}

\usepackage{enumitem}

\renewcommand{\thesubsection}{\arabic{section}.\arabic{subsection}}

\newcommand\todo[1]{}

\newcommand{\figref}[1]{Figure~\ref{#1}}

\newcommand{\fst}{F\textsubscript{ST}}

\newcommand\toolname{\textsc}

\newcommand{\codeline}[1]{\texttt{\lstinline|#1|}}

\newcommand\fileformat{\texttt}

\keywords{
}

\begin{abstract}

    \textbf{Summary.} 
    Pool sequencing is an efficient method for capturing genome-wide allele frequencies from multiple individuals, with broad applications such as studying adaptation in Evolve-and-Resequence experiments, monitoring of genetic diversity in wild populations, and genotype-to-phenotype mapping. 
    Here, we present grenedalf, a command line tool written in C++ that implements common population genetic statistics such as $\theta$, Tajima's D, and \fst{} for Pool sequencing. It is orders of magnitude faster than current tools, and is focused on providing usability and scalability, while also offering a plethora of input file formats and convenience options.
    
    \vspace{0.6em}
    \textbf{Availability and implementation.}
    grenedalf is published under the GPL-3, and freely available at \href{https://github.com/lczech/grenedalf}{github.com/lczech/grenedalf}.

    \vspace{0.6em}
    \textbf{Contact.}
    \href{mailto:lczech@carnegiescience.edu}{lczech@carnegiescience.edu} and \\
    \href{mailto:moisesexpositoalonso@gmail.com}{moisesexpositoalonso@gmail.com}.

    \vspace{0.6em}
    \textbf{Supplementary information.} Supplementary data are available online.

\end{abstract}

\begin{document}

\verticaladjustment{-2pt}

\maketitle
\thispagestyle{fancy}
\ifthenelse{\boolean{shortarticle}}{\ifthenelse{\boolean{singlecolumn}}{\abscontentformatted}{\abscontent}}{}


\section{Introduction and Motivation}
\label{sec:IntroMotivation}

Pool sequencing, or Pool-seq, is a cost-effective high-throughput sequencing method for obtaining genome-wide allele frequencies across multiple individuals simultaneously \citep{Schlotterer2014-mx}.
The approach is commonly used in large-scale genomic studies to estimate genetic diversity and variation within a population across space and time, or to identify genetic changes that are associated with trait evolution or environmental adaptation across populations. 
This makes it suitable for applications such as studying adaptation in Evolve-and-Resequence (E\&R) studies, genotype-to-phenotype mapping, or pooled genome scans and mutant screens. \todo{cite E\&R papers here?}

The pooling of a finite number of individuals from the population, as well as the finite number of reads being sequenced from each individual, introduce two levels of sampling noise in allele counts \citep{Ferretti2013}. Typical population genetic statistics, such as measures of diversity ($\theta$, Tajima's D) and differentiation (\fst{}), hence need to be adapted to correct for the induced biases. Existing software tools that implement these corrections are \toolname{PoPoolation} \citep{Kofler2011a,Kofler2011b}, \toolname{poolfstat} \cite{Hivert2018,Gautier2022-zd}, and \toolname{npstat} \citep{Ferretti2013}. These tools however lack usability, do not scale to contemporary large datasets, and do not support haplotype-corrected frequencies in low-coverage E\&R experiments such as those from HAF-pipe or other HARP-based pipelines \citep{Tilk2019-jr,Kessner2013-ph}.

We present \toolname{grenedalf}, a command line tool to compute widely-used population genetic statistics for Pool-seq data. It aims to solve the shortcomings of previous implementations, and is several orders of magnitude faster, scaling to thousands of samples \citep{Czech2022}. Further, it improves usability, accepts many standard file formats, and offers many convenience options. 


\section{Estimators of Population Genetic Statistics}
\label{sec:StatisticsEstimators}

We re-implemented consistent estimators of population diversity and differentiation, namely nucleotide diversity $\theta_\pi$, Watterson's $\theta$, Tajima’s D, and Nei's and Hudson's \fst{}, which account for the noises introduced by the two finite sampling processes of individuals and reads in Pool-seq.
Several of these estimators were previously available in multiple software packages implemented in Perl \citep{Kofler2011a, Kofler2011b}, R \cite{Hivert2018,Gautier2022-zd}, or C \cite{Ferretti2013}.  Because of implementation differences of estimates available in these packages, we re-derived population genetic estimates and examined their differences (see Supplement). 

Most commonly, our input are sequence reads or read-derived allele counts, as those fully capture the effects of both sources of noise, which can then be corrected for.  Our implementation however can also be used with inferred or adjusted allele frequencies as input, for instance using information from the haplotype frequencies of the founder generation in E\&R experiments \citep{Tilk2019-jr,Kessner2013-ph}.  These can elevate the effective coverage, and thus improve the calling of low-frequency alleles, which can otherwise be difficult to distinguish from sequencing errors \citep{Ferretti2013}.  With these reconstructed allele frequencies, the correction for read depth is less relevant, but the correction for pool size remains important. It is hence convenient to be able to use the same framework for these data, which existing implementations do not offer. 



\subsection*{Genetic Diversity ($\pi$)}

Our implementation of the Pool-seq estimators for $\theta_\pi$ and Watterson's $\theta$ largely follows the approach by PoPoolation. We have however updated some of the equations with computationally more efficient but otherwise equivalent alternatives, and have improved the numerical stability.

However, in our attempt to re-derive a Pool-seq estimator for Tajima's D, we noticed several long-standing issues in the existing estimator.  In short, it seems that Pool-seq data might not allow meaningful estimates of this statistic; see the Supplement for details.  In addition, we noticed several implementation bugs in 
\toolname{PoPoolation} \citep{Kofler2011a}, up until and including v1.2.2 of the tool.  We discussed these with the authors, and the bugs have since been fixed (pers.~comm. with R.~Kofler). 
If conclusions of studies depend on numerical values of Tajima's D computed with \toolname{PoPoolation}, we recommend reanalyzing the data.  
Due to these statistical issues, we generally advise to be cautious when applying and interpreting Tajima's D with Pool-seq data.  





\subsection*{Population Differentiation (\fst{})}

We also show that the estimators for \fst{} as implemented in \toolname{PoPoolation2} (which we call the ``Kofler'' and ``Karlsson'' estimators) are biased upward for low read depths and for small sample pool sizes (see Supplement).
We hence developed unbiased estimators for the average pairwise diversity within populations, $\pi_\text{within}$, the pairwise diversity between populations, $\pi_\text{between}$, and the pairwise diversity across the combined populations, $\pi_\text{total}$, that take the particular biases and assumptions of Poo-seq data into account.
With these, we compute two variants of \fst{}, following Nei \citep{Nei1973} and Hudson \citep{Hudson1992}:

\begin{align*}
    \text{F}_\text{ST}^\text{Nei}    = 1 - \frac{\pi_\text{within}}{\pi_\text{total}} 
    \hspace{2em}\text{and}\hspace{2em}
    \text{F}_\text{ST}^\text{Hudson} = 1 - \frac{\pi_\text{within}}{\pi_\text{between}}
\end{align*}

We provide thorough derivations and analyses of these estimators in the Supplements, and showcase their application in \citep{Czech2022}.



\section{Features and Data Processing Flow}
\label{sec:FeaturesDataProcessing}

Beyond implementing a variety of population genetic estimates, our C++ software library and command line tool were designed to address several bioinformatics challenges which limit the next generation of pool-sequencing applications: (1) Flexible and modular architecture. Different Pool-seq softwares use different file formats. We separated file format reading and transformations from computations and algorithms. File format transformations are seamless, and new formats can be included independent of downstream analyses. (2) Usability. Software for large-scale individual genotypes (i.\,e., VCFtools, PLINK, etc.) provide convenience tools for merging, filtering, and manipulating datasets. We provide these for Pool-seq formats. \todo{not fully implemented yet - focussed on the stats so far. but coming soon!} (3) Speed. Current tools are too inefficient to allow modern Pool-seq datasets and experiments to grow into hundreds or thousands of population samples. Our highly optimized routines provide orders of magnitude gains in speed. 

\begin{figure}[!htb]
    \centering
    \includegraphics[width=.8\linewidth]{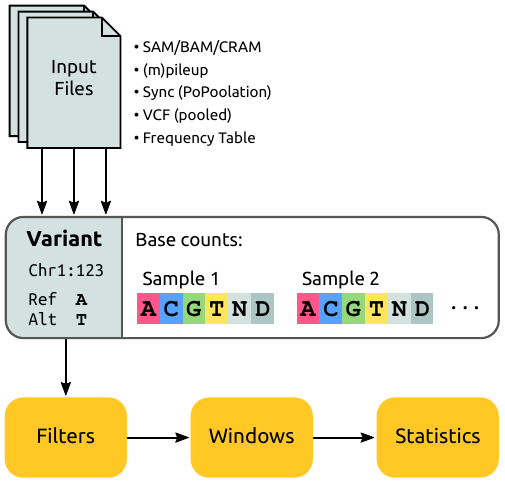}
    \caption{
        \textbf{Summary of the data flow from input files to stream through a genome and compute statistics.}
        Different input file types are supported with their idiosyncratic options, which all are represented by a uniform data type that we call a (potential) \codeline{Variant}. A \codeline{Variant} describes a single position on a chromosome, here, position \codeline{123} on chromosome \codeline{Chr1}, and stores the reference and alternative base for file formats that support them (and otherwise infers them from the two most common bases at the position, or from a provided reference genome file). This is similar to the data of the \fileformat{sync} format.
        For each sample of the input (e.\,g., read groups in SAM files, columns in mpileup files, or sample frequencies from tabular formats), the nucleotide base counts (\texttt{ACGT}) of the pooled reads are stored, including counts for "any" (\texttt{N}) and "deletion" (\texttt{D}), which are however ignored in most statistics.
        The stream of \codeline{Variants} along the genome is then filtered using a cascade of filters, such as sub-setting to regions of interest and numerical quality filters. Next, the data stream can be assembled into different types of windows, such as sliding windows, single positions, or entire chromosomes. Finally, the desired statistics are computed per window.
    }
\label{fig:Dataflow}
\end{figure}

In the following, we provide an overview of the data processing flow, which is summarized in \figref{fig:Dataflow}. 


\subsection*{File Formats}

Two commonly used file formats for Pool-seq data are the \fileformat{(m)pileup} \citep{Li2009b} and \fileformat{sync} format. The latter is a simple allele count format introduced in \toolname{PoPoolation2} \cite{Kofler2011b}, which is usually obtained by converting from \fileformat{bam} via \fileformat{(m)pileup} to \fileformat{sync}, requiring an additional data transformation step to analyze the data.

In contrast, and in addition to these formats, \toolname{grenedalf} can directly work with other standard file formats such as \fileformat{sam}/\fileformat{bam} \citep{Li2009b}, \fileformat{cram} \citep{Fritz2011},  \fileformat{vcf} (using the \texttt{"AD"} allelic depth field) \citep{Danecek2011-qk}, and a variety of simple table formats, for reading allele counts or allele frequencies from pool sequencing data. All formats can also optionally be gzipped (decompression is done asynchronously for speed), and their idiosyncratic options (such as filtering by read flags or splitting by read groups for \fileformat{sam}/\fileformat{bam}) are supported.  This eliminates the need for intermediate file conversions, reduces overhead for file bookkeeping, disk space, and processing time (see Supplement), and increases user convenience.

Note that not all data types are well suited for the Pool-seq approach. For instance, a widespread practice is to use a variant calling tool on the data before downstream analyses. However, many standard variant callers were developed for individual instead of pooled data, meaning that their statistical assumptions might be violated in Pool-seq \citep{Czech2022}.  Furthermore, formats such as \fileformat{vcf} only store variant sites in the first place, so one cannot distinguish if a missing site is invariant or did not meet the minimum data quality threshold. A mask file as explained below can be use to remedy this. We still support \fileformat{vcf} as a convenience, but recommend to ensure that the variant calling was conducted appropriately for the Pool-seq approach.  For this reason, it is often beneficial to directly work off the ``raw'' data, such as \fileformat{sam}/\fileformat{bam} files or \fileformat{sync} files that were not already filtered for SNPs, when running \toolname{grenedalf}.

If a reference genome is provided, it is used to fill in the reference bases when using file formats that do not store these. When multiple input files are provided (even of different formats, and with missing data), they are traversed in parallel, using either the intersection or the union of the genomic positions present in the files, and internally combined as if they were one file with multiple samples. Samples can furthermore be grouped by merging their counts, for instance to combine different sequencing runs into an (artificial) pool.



\subsection*{Filters}

After parsing the input, a variety of filters can be applied to the data stream, either per-sample or across samples. 
First, we can apply sample sub-setting, and sub-setting to chromosomes or genomic regions within or across chromosomes, using a variety of formats (\fileformat{bed}, \fileformat{GFF2}/\fileformat{GFF3}/\fileformat{GTF}, \fileformat{map}/\fileformat{bim} (PLINK), \fileformat{vcf}, or simple text formats). The region filters completely remove genomic positions from the data stream, to speed up the downstream steps.

Next, a mask can be specified, in order to pre-select loci of interest. This is for instance useful when an external filtering was applied to the data beforehand; the mask then specifies which positions were considered ``valid'' by that filter. This is important in order to correctly compute the per-window averages of the statistical estimators, where we need to know the number of high quality loci (independently of whether they are SNPs or invariant positions).

Lastly, users can specify a variety of numerical quality filters, such as minimum allele count and minimum or maximum read depth.  Some of the numerical filter settings are highly relevant for the computed estimators; see the Supplement for details.  

After all specified filters have been applied, we execute a simple SNP detection, based on the base counts (\texttt{ACGT}). Any locus that has (after filtering) exactly one base with non-zero count is considered invariant, while a locus with two or more non-zero counts is considered a SNP. Additional filters can be specified for this step, such as a minimum allele frequency, or sub-setting to only biallelic SNPs (exactly two counts are non-zero).  This SNP detection mechanism is hence independent of the input file format, and simply uses the available data. For instance, with \fileformat{sam}/\fileformat{bam} files, we typically have data at (almost) all loci, and can hence use this to distinguish invariant sites from low quality or missing sites.



\subsection*{Windowing}

The data is then assembled into windows along the genome. We implemented different types of windows, depending on the analysis needs, namely, representing (a) intervals of a fixed number of bases, (b) a fixed number of variants (SNPs) per window, (c) user-defined regions that can be potentially nested or overlapping, such as genes or LD blocks, (d) single SNPs, (e) whole chromosomes, and (f) whole genome.
Existing tools only offered one or two of these types of windows.
The first three of these types keep data in memory proportional to the window sizes, which is necessary for overlapping windows and for sliding windows to allow a stride between windows smaller than the window size.
The remaining window types (single SNPs, whole chromosomes, and whole genome) instead directly stream through the input data, thereby keeping the memory footprint to a minimum. This is a distinguishing feature compared to, e.\,g., \toolname{poolfstat}, which reads whole files into memory, and hence does not scale to large datasets with many samples (see Supplement).


\subsection*{Statistics Computation}

Finally, with the data stream processed as described above, the desired statistical estimators are computed. Typically, the statistics are then averaged over the window, in order to obtain per-base-pair estimates. To account for the characteristics of the input data (with missing data; only containing variant loci; etc), we offer different policies for the window averaging: (a) the window size (likely an underestimation), (b) the number of all available loci in the input, (c) the number of ``valid'' high-quality positions (i.e., the number of loci that passed all quality filters; invariants and SNPs), (d) the number of SNPs only (likely an overestimation), and (e) no averaging (i.e., simply report the sum of all per-site values; this allows the user to apply custom averaging later on).
These policies are covering the most common use cases of data types. When the data has sufficient coverage, we recommend to use (c), the number of high-quality positions. If specified, this also takes the mask into account, so that SNP-only input data can be properly normalized per window.


\section{Performance Comparison and Implementation}
\label{sec:PerformanceImplementation}

\begin{figure}[!htb]
    \centering
    \includegraphics[width=1.0\linewidth]{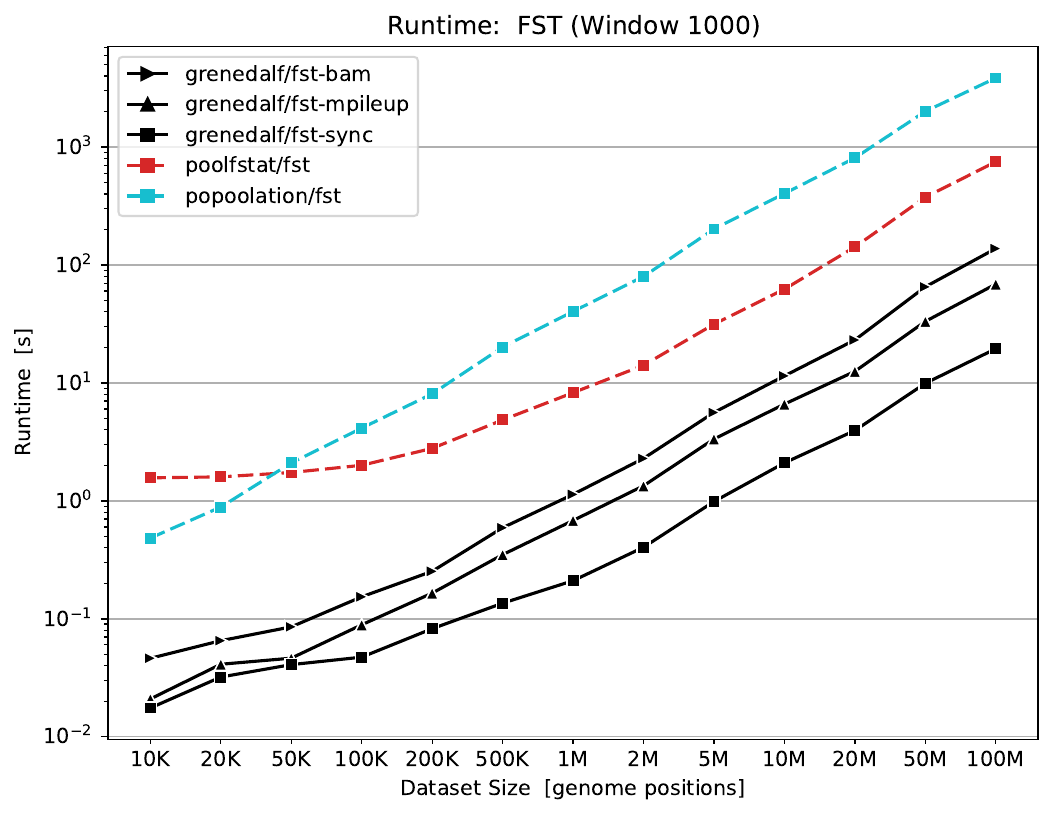}
    \caption{
        \textbf{Runtime benchmark (log-log scaled) for computing \fst{} for different input data sizes on real-world data.}
        Here we compare the single-threaded runtime of \toolname{grenedalf} for computing \fst{} on subsets of two samples from \textit{A. thaliana} to existing implementations. 
        With the \fileformat{sync} format, \toolname{grenedalf} is about 40x faster than \toolname{poolfstat}, and about 200x faster than \toolname{PoPoolation}. Even with the computationally more demanding \fileformat{pileup} and \fileformat{bam} formats (which the other two tools do not support as input), \toolname{grenedalf} is significantly faster. The gain over existing tools is even greater for larger datasets, and when using multiple threads. See Supplement for all benchmarks. 
    }
\label{fig:Benchmark}
\end{figure}

In \figref{fig:Benchmark}, we compare the runtime of existing tools to \toolname{grenedalf}, which is more than two orders of magnitude faster than previous implementations on real-world data.  More detailed benchmarks are available in the Supplement. 
Overall, these improvements enable the analysis of datasets much larger, as for instance required in our GrENE-net.org experiment \cite{Czech2022}.  Furthermore, this will allow for novel types of applications that were previously not feasible, such as running bootstrapping (either over reads, or genomic positions, or both) to obtain confidence intervals for the statistics of interest.

Performance in runtime and memory was one of the major design goals. For instance, the file parsing is highly optimized and executed with asynchronous buffers. All data is read in streams, so that the number of input files, and their sizes, do not significantly affect the amount of required memory. Processor-intensive steps, such as file parsing and the statistics computations, are multi-threaded with a shared thread pool to leverage modern multi-core systems, and we paid close attention to selecting appropriate data structures for efficiency.
Particular care was given to the implementation of the statistics; we optimized computations towards CPU-level parallelism, increased overall numerical stability and range, extended the range of valid inputs for aspects such as the involved binomial computations, 
and replaced some expensive subroutines by fast closed-form expressions or lookup-tables.

The core implementation of the command line tool \toolname{grenedalf} is part of \toolname{genesis}, our high-performance software library for working with phyogenetic and population genetic data \citep{Czech2020}. 
Written in modern C++, \toolname{genesis} is the best-scoring code across 48 scientific code bases in comprehensive software quality benchmarks \citep{Zapletal2021-cv}. 
A key feature of the underlying software design is its flexibility and modularity.  The design allows for further additions of file formats and statistics algorithms without the need to alter any other software component. Therefore, any new addition benefits from the overall architecture and efficiency, and components of the software can be combined as needed.
This structure permits users to  use the command line tool \toolname{grenedalf} directly on their datasets, but they can also use the functionalities of \toolname{genesis} for their own method development. 



\section{Conclusion and Outlook}
\label{sec:Conclusion}

We presented \toolname{grenedalf}, a command line tool for computing population genetic statistics, which scales to modern pool sequencing datasets, and which provides a plethora of input file formats and convenience options.

In the future, given the ease with which statistics computations can be incorporated into our modular software design, we aim to re-implement more of the existing Pool-seq statistics, such as $f$ statistics \citep{Hivert2018,Gautier2022-zd}, and implement a Pool-seq-based GWA tool \citep{Schlotterer2014-mx}.
An under-explored area is the incorporation of short indels, which can potentially be treated as another type of count-based variation.
Furthermore, we want to integrate \toolname{grenedalf} with our short-read processing and variant calling pipeline \toolname{grenepipe} \citep{Czech2022-vf}, which already supports estimating allele frequencies from Pool-seq data via the HAF-pipe tool \citep{Tilk2019-jr,Kessner2013-ph}. 
To this end, it will also be beneficial to develop a proper file format for allele frequencies from Pool-seq, akin to the \fileformat{vcf} for individual sequencing. 



\vspace{-0.5em}
\section*{Appendices}

\small 

grenedalf is published under the GPL-3, and freely available at \href{https://github.com/lczech/grenedalf}{github.com/lczech/grenedalf}.

\subsection*{Supplementary Documents}

\begin{enumerate}[label=(\Alph*)]
    \item Pool-Sequencing corrections for population genetic statistics; includes assessment of biases of the statistics.
    \item Software Comparison. Benchmarking the computational requirements of \toolname{grenedalf} versus existing tools.
\end{enumerate}

\subsection*{Competing Interests}

The authors declare that they have no competing interests.

\subsection*{Author’s Contributions}

L.C. developed and tested the software, helped to develop the novel statistics, analyzed the data, and wrote the manuscript. J.P.S. developed and re-derived statistics, analyzed the data, and helped to write the manuscript. M.E.-A. provided funding, helped to develop the novel statistics, and helped to write the manuscript.


\subsection*{Acknowledgments}

We thank Robert Kofler and Nicola De Maio for discussing their equations and implementation and for fixing bugs in \toolname{PoPoolation}.
The soundtrack for this work was provided by Howard Shore.
This project was initiated as part of a collaborative network, Genomics of rapid Evolution to Novel Environments (GrENE-net.org), from where it inherits its name.
This work was funded by the Carnegie Institution for Science,  the Office of the Director of the National Institutes of Health’s Early Investigator Award with award \#1DP5OD029506-01 (M.E.-A.).

\normalsize

\section*{References}
\bibliography{01-manuscript}

\end{document}




\begingroup
\let\center\flushleft
\let\endcenter\endflushleft
\maketitle
\endgroup
\vspace*{-3.5em}
\textsuperscript{a}{Department of Plant Biology, Carnegie Institution for Science, Stanford, USA.\\}
\textsuperscript{b}{Section for GeoGenetics, Globe Institute, University of Copenhagen, Denmark.\\}
\textsuperscript{c}{Department of Genetics, Stanford University, Stanford, USA.\\}
\textsuperscript{d}{Department of Biology, Stanford University, Stanford, USA.\\}
\textsuperscript{e}{Department of Global Ecology, Carnegie Institution for Science, Stanford, USA.\\}
\textsuperscript{f}{Department of Integrative Biology, University of California Berkeley, Berkeley, USA.\\}
\textsuperscript{g}{Howard Hughes Medical Institute, University of California Berkeley, Berkeley, USA.\\}
\textsuperscript{*}{To whom correspondence should be addressed: \href{mailto:lczech@carnegiescience.edu}{\color{black}{lczech@carnegiescience.edu}},
and \href{mailto:moisesexpositoalonso@gmail.com}{\color{black}{moisesexpositoalonso@gmail.com}}}




\vspace*{2em}
\setlength{\fboxsep}{2em}
\fbox{%
  \parbox{48em}{
        \large
        \textbf{Version: 2024-06-07}
        \\ \\
        The latest version of this document can be found at 
        \url{https://github.com/lczech/pool-seq-pop-gen-stats}
        \\
        When citing this document or equations in it, 
        we recommend to also mention the version,
        as section and equation numbering might change.
  }%
}
\vspace*{2em}

\section*{Abstract}
\label{supp:sec:Abstract}

This document describes our assessment of the specific pool-sequencing (Pool-seq) equations for population genetic measures of diversity (such as $\theta_\pi$, $\theta_\text{Watterson}$, Tajima's $D$), and differentiation (such as $\mathfst$).
The aim of these equations is to correct for the noises of pool sequencing, specifically limited sample size (number of individuals pooled, or pool size~$\samplesize$) and limited read depth (number $\readdepth$ of reads obtained from those individuals that cover a given locus).

We re-render some approaches originally presented and implemented in \toolname{PoPoolation} \cite{Kofler2011a} and \toolname{PoPoolation2} \cite{Kofler2011b}, as well as other publications.  We re-derive estimators of genetic diversity, and suggest some improvements that fix standing issues when working with Pool-seq data. Furthermore, we (re-)derive two consistent (asymptotically unbiased) estimators for \fst{} for pool sequencing data, using two distinct definitions of \fst{}, that correct for both sources of noise mentioned above.  We also evaluate the different estimators for \fst{} here using simulated data, showing their particular biases.  Lastly, we derive novel estimators for \fst{} that additionally take sequencing error into account.  

Our main findings are as follows. (1) Correcting for Pool-seq noise in Tajima's D in order to obtain results that are numerically comparable to values obtained from individual sequencing is difficult, and the approach of \toolname{PoPoolation} has statistical shortcomings. We hence recommend not interpreting their values of Tajima's D quantitatively, and instead compare $\theta_{\pi}$ and $\theta_{w}$ directly without Pool-seq corrections. (2) Some of the established Pool-seq estimators for \fst{} are biased upward for small pool sizes and low read depth; we hence recommend using the estimators presented here instead.

The estimators presented here are implemented in our tool 
grenedalf: \url{https://github.com/lczech/grenedalf}


\pagebreak

\tableofcontents

\vspace*{5em}
\textbf{Note on sources:}
\begin{itemize}
    \item This document is partially based on the PoPoolation equations document \texttt{correction\_equations.pdf} as found in their code repository; we provide a copy at \url{https://github.com/lczech/popoolation/blob/master/files/correction_equations.pdf}. This document derives some of the equations implemented, but also contains equations that are not implemented in \toolname{PoPoolation} but may be of interest for a deeper understanding of the topic.
    \item Furthermore, we incorporated several implementation details from the reverse-engineered code of \toolname{PoPoolation} and \toolname{PoPoolation2}. We want this document to represent the equations that are actually computed when running these programs (and their re-implementations in \toolname{grenedalf}), as we feel that they need a more thorough assessment than what is available in the current literature.
    \item Lastly, for the avid reader, apart from these documents and the original publications of \toolname{PoPoolation} \cite{Kofler2011a} and \toolname{PoPoolation2} \cite{Kofler2011b}, we also recommend the \toolname{npstat} \cite{Ferretti2013} and \toolname{poolfstat} \cite{Hivert2018,Gautier2022-zd} publications, which contain further renderings of some of the equations.
\end{itemize}

\pagebreak


\section{Preliminaries}
\label{supp:sec:Definitions}



\subsection{Pool Sequencing Data and Notation}
\label{supp:sec:Definitions:sub:PoolSequencing}

We first define the input that we assume to be given for all subsequent equations.
In the software implementation of the equations, these are based on the input data, or set by the user as parameters.

$\samplesize$ :
Pool size, as given in the dataset (and in the software, provided by the user). This is the number of individuals (haplotypes) that were pooled together for the sequencing of a population. Note on the pool size for different levels of ploidy:  The pool sequencing approach assumes the pool size to be the number of distinct haplotypes in the population that have been sampled. Hence, for instance for 100 diploid individuals, the number of haplotypes (and hence the pool size) is 200.

$\readdepth$ :
Observed read depth: The number of reads sequenced from the pool at a focal position in the genome. This is the main data that we are using. NB: In PoPoolation, npstat, as well as previous versions of this document (up until, and including, version 2023-09-12), this is called ``coverage'' instead of read depth. This term however is ambiguous and can both mean ``coverage depth'' (number of reads at each base of the genome) and ``coverage breadth'' (proportion of the genome that is covered by reads).



$b$ :
Minimum allele count, provided by the user.
We do not want to consider SNPs with fewer than $b$ alternative reads in the data, as they might be sequencing errors.
Note that we assume $b$ to be a user-provided constant,
and hence leave it out of (most) function arguments for simplicity.

For most equations we will consider a single pool sequencing sample (one population, or one experiment), but in cases where we consider multiple samples (multiple pool-sequenced populations), we will denote the different samples with subscripts in parentheses, such as $\samplesize_{(1)}, \samplesize_{(2)},\ldots$ or $\readdepth_{(1)}, \readdepth_{(2)}, \ldots$ to denote different populations or experiments.

Beyond these data-specific and user-specified parameters, we will use the following notation.  In general, we will use capital letters for random variables in our model of pool sequencing, which will be described shortly in \secref{supp:sec:model}.  We will also differentiate sample quantities (\ie those that are computed from data) from their population counterparts (\ie parameters that describe the population from which the data were sampled) with hats.  For example, we differentiate the frequency of the $A$ allele in the whole population, $f_A$, from the empirical frequency of $A$ nucleotides in our pool sequencing sample, $\empfreq_A$.

$\tau$ :
Index over nucleotides, with $\tau \in \left\{ \text{A}, \text{C}, \text{G}, \text{T} \right\}$ being implicit in any summations.

$C_\tau$ :
Nucleotide counts, \ie how many reads have a certain nucleotide $\tau$ at a given position in the genome. Note that we generally do not make assumptions about the focal position in the genome being biallelic, unless stated otherwise.

$\readdepth = \sum_\tau C_\tau$ : The (total) read depth at a given position.

$f_\tau$ :
Nucleotide frequencies in the population, \ie the proportion of haplotypes that have nucleotide $\tau$ at the focal position.

$\empfreq_\tau$ :
Empirical nucleotide frequencies, \ie $\empfreq_\tau := C_\tau / \readdepth$. Naturally, $\sum_\tau \empfreq_\tau = 1$.


$m$ :
Index of summation over potential levels of read depth $\readdepth$.

$k$ :
Index of summation over potential pool sizes $\samplesize$.




We will also make use of the generalized harmonic numbers, which frequently arise in coalescent theory.  Define $a_1$ and $a_2$ as the sum of (squared) reciprocals of the first $n$ positive integers:
\begin{align}
    \label{eq:an}
    a_1(n) &:= \sum_{k=1}^{m} \frac{1}{k}
    \\
    \label{eq:bn}
    a_2(n) &:= \sum_{k=1}^{m} \frac{1}{k^2}
\end{align}

We use this notation as a compromise between Equation (3.6) of \citeay{Hahn2018} and the notation of $a_n$ and $b_n$ used in \citeay{Achaz2008} for these quantities.

\subsection{Statistical Model of Pool Sequencing}
\label{supp:sec:model}

We consider a simple model of pool sequencing as a series of multinomial samplings from a population.  In particular we assume the following model, where $N_\tau$ is the (unobserved) number of individuals in the pool with the $\tau$ allele:

\begin{align*}
(N_A, N_C, N_G, N_T) &\sim \text{Multinomial}\left(\samplesize, (f_A, f_C, f_G, f_T)\right)\\
(C_A, C_C, C_G, C_T) | (N_A, N_C, N_G, N_T) &\sim \text{Multinomial}\left(\readdepth, \left(\frac{N_A}{\samplesize}, \frac{N_C}{\samplesize},\frac{N_G}{\samplesize},\frac{N_T}{\samplesize}\right)\right)
\end{align*}

Throughout, we will repeatedly make use of the following results:
\begin{align}
\mathbb{E}\left[\empfreq_\tau\right] = f_\tau, \label{eq:frequnbiased}
\end{align}
and
\begin{align}
\mathbb{E}\left[\left(\frac{\readdepth}{\readdepth-1}\right)\left(\frac{\samplesize}{\samplesize-1}\right)\left(1-\sum_\tau\empfreq_\tau^2\right)\right] = 1-\sum_\tau f_\tau^2. \label{eq:hetunbiased}
\end{align}

To see Equation~\ref{eq:frequnbiased} we can use the tower property:
\[
\mathbb{E}\left[\empfreq_\tau\right] = \mathbb{E}\left[\mathbb{E}\left[\empfreq_\tau \mid N_\tau \right]\right] = \mathbb{E}\left[\frac{N_\tau}{n}\right] = f_\tau.
\]
To obtain Equation~\ref{eq:hetunbiased}, we make use of the following fact about binomial distributions.  Suppose $X$ is binomially distributed with parameters $m$ and $p$, then $X$ can be represented as the sum of $m$ independent, identically distributed Bernoulli random variables, $Y_1,\ldots,Y_m$ that take the value $1$ with probability $p$ and $0$ with probability $1-p$.  We can then see:
\begin{align*}
\mathbb{E}\left[X(m-X)\right] &= \mathbb{E}\left[\sum_{i=1}^m\sum_{j=1}^mY_i(1-Y_j)\right]\\
&= m\mathbb{E}\left[Y_1(1-Y_1)\right] + m(m-1)\mathbb{E}\left[Y_1(1-Y_2)\right]
\end{align*}
where the second equality follows from the linearity of expectation and the fact that all of the $Y_i$s are identically distributed.  Then, since $Y_i$ is either $0$ or $1$, $Y_i(1-Y_i)$ must be zero.  Finally, by independence,
\[
\mathbb{E}\left[Y_1(1-Y_2)\right] = \mathbb{E}\left[Y_1\right]\left(1-\mathbb{E}\left[Y_2\right]\right) = p(1-p).
\]
Therefore,
\[
\mathbb{E}\left[X(m-X)\right]  = m(m-1)p(1-p)
\]

We can now make use of this result twice along with the tower property to obtain Equation~\ref{eq:hetunbiased}:
\begin{align*}
\mathbb{E}\left[\left(\frac{\readdepth}{\readdepth-1}\right)\left(\frac{\samplesize}{\samplesize-1}\right)\left(1-\sum_\tau\empfreq_\tau^2\right)\right] &= \sum_\tau\left(\frac{\readdepth}{\readdepth-1}\right)\left(\frac{\samplesize}{\samplesize-1}\right) \mathbb{E}\left[\empfreq_\tau(1-\empfreq_\tau)\right]\\
&=  \sum_\tau\left(\frac{\readdepth}{\readdepth-1}\right)\left(\frac{\samplesize}{\samplesize-1}\right) \left(\frac{1}{\readdepth^2}\right) \mathbb{E}\left[C_\tau\left(c-C_\tau\right)\right]\\
&= \sum_\tau\left(\frac{1}{\readdepth(\readdepth-1)}\right)\left(\frac{\samplesize}{\samplesize-1}\right) \mathbb{E}\left[\mathbb{E}\left[C_\tau\left(c-C_\tau\right) \mid N_\tau\right]\right]\\
&= \sum_\tau\left(\frac{\samplesize}{\samplesize-1}\right) \mathbb{E}\left[\frac{N_\tau}{\samplesize}\left(1-\frac{N_\tau}{\samplesize}\right)\right]\\
&= \sum_\tau\frac{1}{\samplesize(\samplesize-1)}\mathbb{E}\left[N_\tau\left(\samplesize - N_\tau\right)\right]\\
&= \sum_\tau f_\tau(1-f_\tau) = 1 - \sum_\tau f_\tau^2.
\end{align*}

The last equality holds as $\sum_\tau f_\tau = 1$.


\section{Pairwise Heterozygosity \texorpdfstring{$\theta_\pi$}{Theta Pi}}
\label{supp:sec:ThetaPi}

First, we derive estimators for the pairwise heterozygosity, $\theta_\pi$, also called Tajima's $\pi$.  Formally, pairwise heterozygosity is the probability that two haploids drawn \emph{from the population} have different alleles at a position chosen uniformly at random.  We always assume that the population is large so that for a particular site, $\theta_\pi = 1 -\sum_\tau f_\tau^2$, which can be seen by noting that the probability that we pick two individuals with the $\tau$ allele is $f_\tau^2$.

If we were to sequence individuals, it would be straightforward to derive unbiased estimators for $\theta_\pi$, but instead we obtain a pool of reads, which introduces two major issues.  The first is that we no longer have access to individuals.  If we had sequenced individuals, we could estimate $\theta_\pi$ by looking at two individuals and asking if they have different alleles at a position, but in pool sequencing, we can only look at two different reads and ask if they have different alleles.  The issue is that we cannot know whether those reads were from the same individual or from different individuals, so a naive estimator of $\theta_\pi$ that just looks at pairwise differences between reads will be biased because in some fraction of times those reads will have come from the same individual, and hence must have the same allele (ignoring sequencing error).  The second major issue is that positions where only a small number of reads have a particular allele could be an artifact of sequencing error.  As such, PoPoolation restricts to sites where there are either exactly $0$ or at least $b$ reads supporting each allele.

Below we derive PoPoolation's estimator of pairwise heterozygosity, $\thetapi$, which addresses both of the issues listed above.  Dealing with the bias from pool-sequencing is straightforward.  Restricting to positions where there are at least $b$ reads with the minor allele is more complex, however, and requires additional assumptions that will almost certainly not be met in practice.  In particular, the $\thetapi$ estimator crucially relies on the assumption that the population is evolving neutrally and is at equilibrium -- in technical terms, the derivation relies on the sample frequency spectrum being proportional to $1/k$, which is to say that the probability that $k$ out of $\samplesize$ individuals have the derived allele at a segregating site has probability proportional to $1/k$.  In practice, natural selection, gene flow, or fluctuations in effective population size can all cause deviations from this assumption, and these deviations will result in $\thetapi$ being a biased estimator of $\theta_\pi$.



\subsubsection*{PoPoolation Pool-seq estimator, \texorpdfstring{$\thetapi$}{Theta Pi hat}}
\label{supp:sec:ThetaPi:sub:PoolSequencing}

%

PoPoolation begins by defining %
\begin{align}
    \label{eq:ThetaPiFreq2}
    \thetapi^{\samplesize\to\infty} &:= \frac{\readdepth}{\readdepth-1} \left(1 - \sum_\tau \empfreq_\tau^2 \right),
\end{align}
which is equal to the left hand side of Equation~\ref{eq:hetunbiased} up to a factor of $\frac{\samplesize}{\samplesize-1}$.  This indicates that this is a biased estimator of the heterozygosity at this site, but the bias is on the order of $O\left(\frac{1}{\samplesize}\right)$, which means that PoPoolation assumes an extremely large pool size, which is why we use the $\samplesize\to\infty$ superscript.

At this point, the PoPoolation equations document begins to simplify the above equation, and then breaks it down for biallelic SNPs. However, their (and our) implementation differs from this, and uses the above equation that works with any (not just biallelic) SNPs. We hence do not introduce these simplifications here.
Note however that the computation is still only conducted on biallelic sites, as the correction term introduced below assumes this. This means, we are assuming that multiallelic sites are rare, which is true for small mutation rates.

\popoolissue{We think that it is makes sense to compute only on biallelic SNPs, as this is a theoretical derivation of $4N_e\mu$, which assumes the infinitesimal model (which does not allow for multi-allelic loci). Still, we briefly wanted to check in with you here. Our guess is that the computation in code is just done for all four nucleotides to keep the code simple -- but the function only gets called for biallelic sites, so it behaves as if it was only computed for biallelic SNPs. Is that right?}



\subsubsection*{Bias of $\thetapi^{\samplesize\to\infty}$ when restricting to sites with at least $b$ minor allele reads}
\label{supp:sec:ThetaPi:sub:PoolSequencing:sub:ExpectedValue}

Other than the bias from the finite pool size, $\thetapi^{\samplesize\to\infty}$ is a reasonable estimator of the pairwise heterozygosity at a given site.  To obtain an overall estimate of $\theta_\pi$, we could average $\thetapi^{\samplesize\to\infty}$ across all sites.  PoPoolation seeks to avoid including sites that appear to be segregating solely because of sequencing error, however, which adds a complication.  We expect that sequencing errors should be rare, and so it would be extremely unlikely to see $b$ or more reads supporting a particular allele solely due to sequencing error.  As such, we can remove all sites with fewer than $b$ reads supporting the minor allele.  This is problematic, however, and biases the estimator --- while removing such sites certainly removes most or all sites that are segregating solely due to sequencing error, it also removes a large number of sites where the mutation just happens to be at low frequency.

The approach that PoPoolation takes is to assume neutrality and an equilibrium demography (all variation is neutral, and the population is of constant size and in mutation-drift equilibrium), calculate the bias of $\thetapi^{\samplesize\to\infty}$, and introduce a term to correct for that bias.  Despite using sequencing error as a motivation, PoPoolation ignores sequencing error in the following derivation.  The derivation will assume biallelic sites, and all expectations in this section will be with respect to both the randomness in the Pool-Seq experiment as well as the randomness in the evolutionary process.  

To begin, we compute the expectation of $\thetapi^{\samplesize\to\infty}$ when we treat sites with fewer than $b$ reads supporting the minor allele as not contributing to $\thetapi^{\samplesize\to\infty}$.  Since we are assuming that we are only looking at biallelic sites, we will use $M$ to denote the number of reads supporting the derived allele.  Letting $\mathcal{E}_b$ be the event that at least $b$ reads support the minor allele, we see
\begin{align}
\mathbb{E}\left[\thetapi^{\samplesize\to\infty} \mathbb{I}\left\{\mathcal{E}_b\right\}\right] &= \sum_{m=b}^{\readdepth-b} \mathbb{E}\left[\thetapi^{\samplesize\to\infty} \mid M=m \right] \mathbb{P}(M = m)\\
&= \mathbb{P}\left(\text{Site is a SNP}\right)\sum_{m=b}^{\readdepth-b} \mathbb{E}\left[\thetapi^{\samplesize\to\infty} \mid M=m \right] \mathbb{P}\left(M = m \mid \text{Site is a SNP}\right)\\
&=  2\mathbb{P}\left(\text{Site is a SNP}\right)\sum_{m=b}^{\readdepth-b} \frac{m(\readdepth-m)}{\readdepth(\readdepth-1)} \mathbb{P}\left(M = m \mid \text{Site is a SNP}\right)
\label{eq:ExpectationThetaPi}
\end{align}
In words, the expected value is computed by summing all possible SNP counts (that exceed the minimum count $b$) that can occur in a pool with read depth $\readdepth$, weighted by the probability to have each of those counts.  As we are using the derived allele count $m$ in the equation above, and either the derived or ancestral allele could be the minor allele, we ``sandwich'' our potential values for the read depth between $b$ and $\readdepth-b$.

The two probabilities used above are computed as follows.

$\mathbb{P}\left(\text{Site is a SNP}\right)$ is the probability of observing a SNP in our pool of $\samplesize$ individuals.  Here we invoke our neutrality and equilibrium assumptions, where a classical result from coalescent theory gives:
\begin{align}
\label{eq:PSNP}
\mathbb{P}\left(\text{Site is a SNP}\right) = 1- \prod_{k=1}^{\samplesize-1} \frac{1}{1+\frac{\theta_\pi}{k}} \approx \theta_\pi a_1(n-1)
\end{align}

where the final approximation ignores terms on the order of  $\theta_\pi^2$, which implicitly assumes that the mutation rate is small.

$\mathbb{P}\left(M = m \mid \text{ Site is a SNP }\right)$ is the probability of observing $m$ as the derived allele count in a SNP with $\readdepth$ reads from a pool of $n$ individuals.  We can further break this down by conditioning on the number of individuals in the pool that have the derived allele, which we will call $\widetilde{M}$:
\begin{align}
    \label{eq:Pmcn}
    \mathbb{P}\left(M = m \mid \text{site is a SNP}\right) &= \sum_{k=1}^{n-1} \mathbb{P}\left(M=m \mid \widetilde{M} = k\right)\mathbb{P} \left( \widetilde{M} = k \mid \text{site is a SNP}\right)\\
    &= \frac{1}{a_1(n-1)}  \sum_{k=1}^{n-1} \frac{1}{k}  \mathbb{P}\left(M=m \mid\widetilde{M} = k\right)
\end{align}
where the second line follows from another classic result in coalescent theory that (assuming neutrality, equilibrium, and that $\theta_\pi$ is small) the probability that $k$ individuals in a pool of size $\samplesize$ have the derived allele given that a site is s SNP is $\frac{1}{ka_1(n-1)}$.

Finally, $ \mathbb{P}\left(M=m \mid \widetilde{M} = k\right)$ is the probability of having $m$ reads support the derived allele when $k$ individuals in the pool have the derived allele.  From our binomial sampling model of pool sequencing, this is:

\begin{align}
    \label{eq:Pmcnk}
     \mathbb{P}\left(M=m \mid \widetilde{M} = k\right) &= \binom{\readdepth}{m} \left(\frac{k}{\samplesize}\right)^m \left(\frac{\samplesize-k}{\samplesize}\right)^{\readdepth-m}
\end{align}
In words, this follows a binomial distribution, with $m$ successes out of $c$ trials
with a success probability of $k/n$ for each trial.
That is, we compute how likely it is to observe $m$ counts of the allele in $c$ reads,
given a frequency $k/n$ of that allele in the pool.
The count of the other allele is implicitly $c-m$.

Starting from \eqnref{eq:ExpectationThetaPi}, we can now put this together:
\begin{align}
\nonumber
\mathbb{E}\left[\thetapi^{\samplesize\to\infty} \mathbb{I}\left\{\mathcal{E}_b\right\}\right]  &=2\theta_\pi \sum_{m=b}^{\readdepth-b} \frac{m(\readdepth-m)}{\readdepth(\readdepth-1)} \sum_{k=1}^{n-1} \frac{1}{k} \binom{\readdepth}{m} \left(\frac{k}{\samplesize}\right)^m \left(\frac{\samplesize-k}{\samplesize}\right)^{\readdepth-m}
\end{align}

This shows that if we use our estimator $\thetapi^{\samplesize\to\infty}$ but treat as zero sites where the minor allele is not supported by at least $b$ reads, then the estimate will be biased by a factor of
\[
2\sum_{m=b}^{\readdepth-b} \frac{m(\readdepth-m)}{\readdepth(\readdepth-1)} \sum_{k=1}^{n-1} \frac{1}{k} \binom{\readdepth}{m} \left(\frac{k}{\samplesize}\right)^m \left(\frac{\samplesize-k}{\samplesize}\right)^{\readdepth-m}
\]

which we will now use to eliminate said bias.



\subsubsection*{Unbiased estimate of $\theta_\pi$ when restricting to sites with at least $b$ minor allele reads}
\label{supp:sec:ThetaPi:sub:PoolSequencing:sub:FinalApprox}

Based on the previous section, we can construct an unbiased estimator for $\theta_\pi$ when restricting to sites with at least $b$ reads supporting the minor allele, which we denote by $\widehat{\theta}_{\pi,\text{pool}}$.  Let $\mathcal{S}_{bc}$ be the set of sites where the read depth is $c$, and at least $b$ reads support the minor allele, and let $\thetapi^{\samplesize\to\infty}(\ell)$ be the above estimator evaluated at the $\ell^\text{th}$ site (where above we suppressed the dependence on site because we only considered one site at a time).  Finally, let $\mathcal{C}$ denote the set of distinct observed read depth levels across the genome.  We then define  $\widehat{\theta}_{\pi,\text{pool}}$ as
\begin{align}
    \label{eq:ThetaPiPoolEstimate}
     \widehat{\theta}_{\pi,\text{pool}} &:= \frac{1}{\# \text{sites}} \cdot \sum_{c\in \mathcal{C}}
    \frac{
       \sum_{\ell \in \mathcal{S}_{bc}} \thetapi^{\samplesize\to\infty}(\ell)
    }{
        2\sum_{m=b}^{\readdepth-b} \frac{m(\readdepth-m)}{\readdepth(\readdepth-1)} \sum_{k=1}^{n-1} \frac{1}{k} \binom{\readdepth}{m} \left(\frac{k}{\samplesize}\right)^m \left(\frac{\samplesize-k}{\samplesize}\right)^{\readdepth-m}
    }
\end{align}
%

The approach to compute this for a window is to sum up all values across the SNPs in the window, and potentially compute an average to obtain a per-base-pair estimate of nucleotide diversity.
This is the equation as implemented in \toolname{PoPoolation} as the measure called \texttt{pi},
and implemented in our \toolname{grenedalf} as well. Note that the denominator only depends on the total read depth $\readdepth$ and the pool size $\samplesize$, and hence only needs to be computed once per read depth level, yielding a significant computational speedup.


\subsubsection*{Window averaging}
\label{supp:sec:ThetaPi:sub:PoolSequencing:sub:WindowAvg}

In order to properly compute the window average to obtain an estimate on the per-base-pair-scale, we divide the sum of the above estimator for all SNPs in the window, and divide by an appropriate denominator. We ideally want to use the number of high-quality positions in the window for this, i.e., all positions that are not missing, passed all potential quality filters, etc. In other words, we want to normalize by the number of positions where we did in fact call a variant, or are certain enough that it is an invariant position. In good data conditions, this is the denominator that is recommended to use.

However, not all input data types might contain the full information that is necessary to calculate this. For instance, formats such as VCF are by their very nature often used to only list the variant positions (SNPs), and so we lack information on the other positions. We would need to know which of these have not been called as variants simply because they are indeed invariant, versus which ones were of insufficient quality to make a proper call. This information can for instance be provided via some external source, such as a mask file. However, in the absence of this information, common strategies in practice are: (1) Simply use the window length itself as the denominator for averaging, which likely produces an underestimation, in particular in low coverage or low quality conditions; (2) Use the number of SNPs instead, which is definitely an overestimate of the per-base-pair diversity (and instead gives a per-SNP estimate), but when interpreted with this in mind, can still be a relevant metric to compare regions to each other.

\section{Watterson's Theta, \texorpdfstring{$\widehat{\theta}_w$}{}}
\label{supp:sec:ThetaWatterson}





Under neutrality and equilibrium demography, $\theta_\pi$ is equal to $4N_e\mu =: \theta$, where $\theta$ is four times the effective population size $N_e$ times the per-generation mutation rate $\mu$, and hence, the estimator in the previous section can be used as an estimate of the population-scaled mutation rate.  An alternative estimate of the mutation rate is Watterson's theta, $\widehat{\theta}_w$, which is based on the number of segregating sites instead of the pairwise heterozygosity.  If we had sequenced $n$ individuals, the usual definition of $\widehat{\theta}_w$ is:
\[
\widehat{\theta}_w := \frac{S}{\text{\# sites} \cdot a_1(n-1) }
\]
where $S$ is the number of segregating sites; see e.\,g., Equation (3.5) of \citeay{Hahn2018}. In the pool sequencing case we might hope to just count up the number of sites that have more than one allele to use in place of $S$, but sequencing error is deeply problematic here.  Any place where any read has a sequencing error will look like a segregating site, vastly inflating our estimate of the true number of sites that are segregating in our sample.  PoPoolation therefore follows a similar approach as in the case of the pairwise heterozygosity, by only considering a site to be segregating if at least $b$ reads support each of two alleles.  Again, this restriction introduces bias, because it will miss many sites where one of the alleles really is at a low frequency --- but still segregating --- in the pool.  As before, we will compute this bias and then correct for it, again assuming neutrality, and equilibrium demography, and that the mutation rate is small enough to only focus on biallelic sites.

Formally, let $M_\ell$ be the number of reads supporting the derived allele at site $\ell$, and define
\begin{align}
    S_b(\ell) :=
    \begin{cases}
        1 & \text{if } b \le M_\ell \le \readdepth-b
        \\
        0 & \text{otherwise}
    \end{cases}
\end{align}

We then define our naive Pool-Seq analog of Watterson's $\theta$ as
\[
\widehat{\theta}_w :=\frac{1}{\text{\# sites} \cdot a_1(\samplesize-1)}\sum_{\ell} S_b(\ell)
\]

Reasoning the same as above, we get the expected value of $\widehat{\theta}_w$ as
\begin{align}
    \nonumber
    \mathbb{E}\left[\widehat{\theta}_w\right]
    &= \frac{1}{a_1(\samplesize-1)} \cdot \mathbb{P}\left(\text{Site is a SNP}\right)
        \cdot \sum_{m=b}^{\readdepth-b} \mathbb{P}\left(M = m \mid \text{Site is a SNP}\right)
    \\
    \intertext{
        with the two probability terms again as in \eqnref{eq:PSNP} and \eqnref{eq:Pmcn}, thus leading to
    }
    \label{eq:ExpectationThetaW}
    &= \frac{1}{a_1(n-1)} \cdot \theta\sum_{m=b}^{\readdepth-b}\sum_{k=1}^{n-1} \frac{1}{k} \binom{\readdepth}{m} \left(\frac{k}{\samplesize}\right)^m \left(\frac{\samplesize-k}{\samplesize}\right)^{\readdepth-m}
\end{align}
To accommodate different read depth levels throughout the genome, denote the set of different read depth levels as $\mathcal{C}$, and the set of sites with read depth $c$ as $\mathcal{S}_c$. Equation~\ref{eq:ExpectationThetaW} immediately suggests the following unbiased estimator, which we denote $\widehat{\theta}_{w, \text{pool}}$:
\begin{align}
    \label{eq:CorrectedThetaEstimate}
    \widehat{\theta}_{w, \text{pool}} &:= \frac{1}{\text{\# sites}} \cdot \sum_{c \in \mathcal{C}}
    \frac{
        \sum_{\ell \in \mathcal{S}_c} S_b(\ell)
    }{
         \sum_{m=b}^{\readdepth-b}\sum_{k=1}^{n-1} \frac{1}{k} \binom{\readdepth}{m} \left(\frac{k}{\samplesize}\right)^m \left(\frac{\samplesize-k}{\samplesize}\right)^{\readdepth-m}
    }
\end{align}

Again, the approach to compute this for a window is to sum up all values across the SNPs in the window, and normalize as explained above.  As before, the denominator for the per-site estimate in \eqnref{eq:CorrectedThetaEstimate} only depends on the read depth $\readdepth$, and hence only needs to be computed once per read depth level that is present in the data.  This is the equation as implemented in \toolname{PoPoolation} as the measure called \texttt{theta}, and implemented in our \toolname{grenedalf} as well.






\section{Tajima's D}
\label{supp:sec:TajimaD}

{
\color{Maroon}
\setlength{\fboxsep}{2em}
\fbox{%
  \parbox{48em}{
        In the following, we describe our assessment of the pool-sequencing corrected equations for Tajima's D as presented in \toolname{PoPoolation}. We however find that their statistical approach has some shortcomings, as described below, and suspect that these significantly change the results. We hence recommend to \emph{not} use the correction terms, but instead use an uncorrected direct comparison of $\theta_\pi$ and $\theta_w$. This will not allow for a quantitative interpretation of the values in terms of the classical (non-pool) Tajima's D, but avoids biases of the statistics. See \secref{supp:sec:TajimaD:sub:RealDeal} for details. We still describe the equations here, for completeness, and elaborate on their issues below.
  }%
}
}

Above, we have defined pool-sequencing corrected estimators $\widehat{\theta}_{\pi,\text{pool}}$ and $ \widehat{\theta}_{w, \text{pool}}$.
Now, we want to use them to define a test akin to Tajima's D for pool sequencing.
We are here again following the PoPoolation approach, and re-derive their equations.


\subsection{Pool-Sequencing Correction}
\label{supp:sec:TajimaD:sub:PoolSequencingCorrection}

The PoPoolation equations document derives the following estimator.
To the best of our knowledge, this is however not implemented in \toolname{PoPoolation};
instead, they compute Tajima's D as presented in the following \secref{supp:sec:TajimaD:sub:Classic}.
We still introduce the approach here, for reference, and in the hope that it might be helpful.

First, we define:
\begin{align}
    \widehat{d}_\text{pool} ~&:=~ \widehat{\theta}_{\pi,\text{pool}} ~-~ \widehat{\theta}_{w, \text{pool}}
\end{align}
and use this to define our statistic:
\begin{align}
    \label{eq:BasicTajimaD}
    D_\text{pool} ~&:=~ \frac{ \widehat{d}_\text{pool} }{ \sqrt{ \widehat{\text{Var}}( \widehat{d}_\text{pool}) }},
\end{align}
with a plug-in estimate of the variance of $ \widehat{d}_\text{pool}$.
To derive this plug-in estimate of the variance, we start with the standard expansion of the variance:
\begin{align}
    \nonumber
    \text{Var}( \widehat{d}_\text{pool} ) &= \mathbb{E}\left[ \widehat{d}_\text{pool}^2 \right] - \mathbb{E}\left[\widehat{d}_\text{pool}\right]^2
\end{align}
At this point, we use that $\mathbb{E}[\widehat{d}_\text{pool}] = 0$ (i.e., unbiased) for populations at equilibrium, since both $\widehat{\theta}_{\pi,\text{pool}}$ and $\widehat{\theta}_{w, \text{pool}}$ are unbiased estimates of $\theta$.  Therefore, $\mathbb{E}[\widehat{d}_\text{pool}]^2 = 0$.

Then, following PoPoolation, we compute the variance as:

\begin{align}
    \nonumber
    \text{Var}( \widehat{d}_\text{pool} ) &= \mathbb{E}( \widehat{d}_\text{pool}^2 )
    \\
    \nonumber
    &= \mathbb{P}\left(\text{Site is a SNP}\right) \sum_{m=b}^{c-b} \mathbb{E}\left[\widehat{d}_{\text{pool}}^2 \mid M=m \right] \cdot \mathbb{P}\left(M = m \mid \text{Site is a SNP}\right)
    \intertext{which can be resolved using equations \eqnref{eq:PSNP} and \eqnref{eq:Pmcn} from previous sections:}
    \label{eq:VarDPool}
    &= \theta \sum_{m=b}^{c-b}  \mathbb{E}\left[\left(\widehat{\theta}_{\pi,\text{pool}} -  \widehat{\theta}_{w,\text{pool}} \right)^2 \Big\vert\ M=m\right]    \cdot \sum_{k=1}^{n-1} \frac{1}{k}  \binom{c}{m} \left(\frac{k}{n}\right)^m \left(\frac{n-k}{n}\right)^{c-m}
\end{align}
Both $\widehat{\theta}_{\pi,\text{pool}}$ and $\widehat{\theta}_{w, \text{pool}}$ are completely determined once we fix $M$ to be $m$, so the only unknown quantity is $\theta$.
PoPoolation suggests using $\widehat{\theta}_{\pi,\text{pool}}$
on the same window on which we are computing $D_\text{pool}$, as our estimate of $\theta$ to obtain a plug-in estimate of the variance:
\[
\widehat{\text{Var}}( \widehat{d}_\text{pool} )  := \widehat{\theta}_{\pi, \text{pool}} \sum_{m=b}^{c-b} \mathbb{E}\left[\left(\widehat{\theta}_{\pi,\text{pool}} -  \widehat{\theta}_{w,\text{pool}} \right)^2 \Big\vert\ M=m\right]
    \cdot \sum_{k=1}^{n-1} \frac{1}{k}  \binom{c}{m} \left(\frac{k}{n}\right)^m \left(\frac{n-k}{n}\right)^{c-m}
\]

This assumes that all individuals contribute the same number of reads to the pool.
As stated above, both $\widehat{\theta}_{\pi,\text{pool}}$ and $\widehat{\theta}_{w, \text{pool}}$ are completely determined once we fix $M$ to be $m$, so these values can be pre-computed and stored to achieve a computational speedup.

The first summation in \eqnref{eq:VarDPool} involves computing $\widehat{\theta}_{\pi,\text{pool}}$ and 
$\widehat{\theta}_{w,\text{pool}}$ repeatedly $c-2b$ many times, 
with each of these computations involving the compututation of their respective denominators.
However, as $c$ remains constant throughout this computation, these denominators (the correction terms)
are shared across positions along the genome with the same read depth and number of derived alleles, so that we can re-use them across positions with identical values of those, to gain a $\approx C$-fold speedup.

At this point, the PoPoolation equation document also introduces an approach to compute Tajima's D based on the above in windows.
We here skip this part for brevity.







\subsection{Integration with Classic Tajima's D}
\label{supp:sec:TajimaD:sub:Classic}

On large windows, the classic Tajima's D is not a measure of significance (in number of standard deviations away from the null hypothesis), but instead is a measure of the magnitude of the divergence from neutrality.  This is because all loci are considered completely linked, even if they are not in reality.  However, the above pool-sequencing Tajima's D instead consideres all loci as completely unlinked. This ignores positive correlations between linked sites, so it gives a different numerical result that has a much higher absolute value compared to classic Tajima's D.

Now, we intend to derive a correction term for the pool-sequence Tajima's D to obtain values that are comparable to classic Tajima's D in non-small windows, that is, we want a measure of the magnitude of the divergence from neutrality.
We again follow the PoPoolation approach, and here derive the equations that are actually implemented. 

Note that this is the part that we believe to be difficult, and where the presented equations do not achieve the desired goal of producing values that are numerically comparable to classical Tajima's D for individual-based sequencing.  The equations are presented here to the best of our understanding, for completeness. In \secref{supp:sec:TajimaD:sub:RealDeal}, we then outline why these are not solving the issue, and present some ideas for moving forward.


\subsubsection*{Approach by Achaz}

\toolname{PoPoolation} uses a modified version of the $Y^*$ test of \citeay{Achaz2008},
which was originally developed as a test for neutrality despite the presence of sequencing errors.
This test only works when excluding singletons, that is, we set $b:=2$ for this part.

Following \toolname{PoPoolation} and \citeay{Achaz2008}, we first define:
\begin{align}
f^*(n) &:= \frac{n-3}{a_1(n-1) \cdot (n-1)-n}
\intertext{which is then used to define:}
\alpha^*(n) &:= f^{*2} \cdot\left( a_1(n-1) - \frac{n}{n-1} \right) + f^* \cdot\left( a_1(n-1) \cdot \frac{4(n+1)}{(n-1)^2} - 2 \cdot\frac{n+3}{n-1} \right) - a_1(n-1) \cdot\frac{8(n+1)}{n(n-1)^2} + \frac{n^2+n+60}{3n(n-1)}
\intertext{and:}
\nonumber
\beta^*(n) &:= f^{*2} \cdot\left( a_2(n-1) - \frac{2n-1}{(n-1)^2} \right) + f^* \cdot\left( a_1(n-1) \cdot \frac{8}{n-1} - a_1(n-1) \cdot\frac{4}{n(n-1)} - \frac{n^3 + 12n^2 -35n +18}{n(n-1)^2} \right) \\
\label{eq:betastar}
&- a_1(n-1) \cdot\frac{16}{n(n-1)} + a_1(n-1) \cdot\frac{8}{n^2(n-1)} + \frac{2(n^4 + 110n^2 -255n + 126)}{9n^2(n-1)^2}
\end{align}
Note that these equations were originally developed for data from individuals,
and hence here, $n$ denotes the number of individuals \emph{as if} we were doing individual sequencing.

NB: The \toolname{PoPoolation} document recommends to counter-check the correctness of their equation with the original of \citeay{Achaz2008}.
In fact, \toolname{PoPoolation}  introduced a mistake in the last term of $\beta^*$,
which we have fixed here. Above is the (hopefully) correct one, following \citeay{Achaz2008}.
Note that the mistake only concerns the \toolname{PoPoolation}  equations document, but not their implementation.
However, in their implementation, \toolname{PoPoolation} instead has a slight mistake in the first term of $\beta^*$, which should not have a significant effect on the computed value though.

\popoolissue{The above is indeed not a big one, but we thought it's good to mention it.}

\popoolissue{A more serious issue occurred in the computation of alpha*. The code actually computes this as beta*, and never calls the actual alpha* function, see \href{https://github.com/lczech/popoolation/blob/092e7a6f7ee4910c1bec4377e0adccc353175bc8/Modules/VarMath.pm\#L104}{here}. This is a bug that Robert and I have discussed before, and I think it is fixed now.}

\popoolissue{Another small issue: The computation of alpha* requires ``effective coverage'' (read depth) to be larger than 1, see \href{https://github.com/lczech/popoolation/blob/092e7a6f7ee4910c1bec4377e0adccc353175bc8/Modules/VarMath.pm\#L226}{here}, but tests this against $n$, which is the number of individuals. Are we missing something here?}

\popoolissue{Furthermore, we tested the computation of alpha*, and it also gives NaN values for input 2 and 3, so the test should in fact be $>= 4$ instead.}


\subsubsection*{The number of individuals sequenced}

The only unresolved parameter is $n$, which corresponds to the number of individuals sequenced ---
if we were to do individual sequencing.
In our case of pool sequencing, according to \toolname{PoPoolation},
we can simply substitute this with the expected number of distinct individuals sequenced.

To this end, we use the read depth $\readdepth$, as well as the pool size $n$, which we here use as our substitute for the number of individuals sequenced.
Then, we define $\tilde{n}$ as the expected number of individuals from our pool that have been sequenced:
\begin{align}
    \label{eq:IndivSeqBruteForce}
    \tilde{n}
    &= \sum_{k=1}^{t} \sum_{j=1}^{k} (-1)^{k-j} \cdot k \binom{n}{k} \binom{k}{j} \left(\frac{j}{n}\right)^\readdepth
\end{align}
where $t=\text{max}(\readdepth, n)$; if $n$ is much larger than $\readdepth$, we can assume $\tilde{n} \approx \readdepth$.

\popoolissue{We think that there is either an error here, or a bug in the code. The code in PoPoolation actually uses the minimum of coverage (aka read depth) and pool size in the above, instead of the maximum as stated above, see \href{https://github.com/lczech/popoolation/blob/092e7a6f7ee4910c1bec4377e0adccc353175bc8/Modules/VarMath.pm\#L156}{here}. We are not sure which one is correct though, or what the implications of this are.}

\popoolissue{Furthermore, the original equations document states that ``$\tilde{n}$ is obtained averaging $\tilde{n}$ over the window W'', which we also stated above. However, looking at the code \href{https://github.com/lczech/popoolation/blob/092e7a6f7ee4910c1bec4377e0adccc353175bc8/Modules/VarMath.pm\#L127}{here}, there is no averaging over the window, and instead it's just computed once. The function for averaging exists in the same PoPoolation function as well, see \href{https://github.com/lczech/popoolation/blob/092e7a6f7ee4910c1bec4377e0adccc353175bc8/Modules/VarMath.pm\#L117-L125}{here}, but is commented out, and also contains a bug, as \texttt{\$cov} is not used. We are unsure how much averaging over the window vs just one value actually changes the results, and hence wanted to bring it up here.} 

Computing the expected number of distinct individuals sequenced corresponds to the following statistical question:
Given a set of integers $A = \{1, \ldots, n\}$ (corresponding to individuals),
pick a set $B$ of $C$ elements from set $A$ with replacement (corresponding to reads);
what is the expected number of distinct values (individuals) that have been picked in $B$ (that we have reads from)?

\toolname{PoPoolation} computes this value by brute force using \eqnref{eq:IndivSeqBruteForce},
that is, by trying all possible ways to pick numbers from the set.
However, there exists a closed form solution to this question, which yields massive speedups for larger read depths, which we have implemented.

One way to arrive at the closed form expression is as follows:
Define an indicator random variable $I_i$ for $1 \leq i \leq n$ as $1$
if individual $i$ is present in the set $B$ (that is, if individual $i$ has been sequenced), and as $0$ if not.
Then, the size of set $B$ is simply $\sum_{i=1}^{n} I_i$.

The probability that $I_i$ equals $1$ (that is, that individual $i$ has been sequenced)
for any $i$ is given by:
\begin{align}
    P(I_i = 1) &= 1-\left( \frac{n-1}{n} \right)^\readdepth
\end{align}
In words, this is the complement of \emph{not} picking $i$ in all of the $\readdepth$ picks from set $A$.

The expected size of the set $B$ can then be computed by linearity of expectation for all $i$,
yielding our closed form expression:
%
%
\begin{align}
    \label{eq:NTildeClosedForm}
    \tilde{n} = n \left( 1 - \left( \frac{n-1}{n} \right)^\readdepth \right)
\end{align}
This is the equation that we compute in our implementation to arrive at $\tilde{n}$ for a given read depth $\readdepth$ and pool size $n$.

Note that $\tilde{n}$ is bounded by both $n$ and $c$, meaning that $\forall n \geq 1$ and $\forall c \geq 1$ we have
\begin{align}
    \label{eq:NTildeBounds}
     \tilde{n} \leq n \qquad\mbox{and}\qquad \tilde{n} \leq c
\end{align}
The first inequality follows readily from the fact that $0 \le (\frac{n-1}{n})^c \le 1$. \\
The second inequality can be shown using Bernoulli's inequality:
\begin{align}
     \left( \frac{n-1}{n} \right)^c = \left(1 - \frac{1}{n} \right)^c \geq 1 - \frac{c}{n}
\end{align}
With this, we get
\begin{align}
    \tilde{n} = n \left( 1 - \left( \frac{n-1}{n} \right)^\readdepth \right) \leq n \left( 1 - \left( 1 - \frac{c}{n} \right) \right) = c
\end{align}
This also follows from our definition of $\tilde{n}$, the expected number of sequenced individuals. That number cannot be greater than the pool size $n$ itself, and it can also not exceed the number $c$ of reads that we sequenced from that pool.

\todo{UNRESOLVED ISSUE: This is computed using the coverage $\readdepth$ (aka read depth) in the above equations (and in the PoPoolation equations), but seems to be called with $b$ (min coverage) in their code (and in mine). Jeff, is that yet another bug (number three then, on top of the two already mentioned two sections below) that they introduced in their code? it seems weird to compute this based on the min coverage, than the actual coverage at the site.  JPS: I'm actually not totally sure what the PoPoolation people are doing here.  I guess they're plugging an estimate of $n$ into the f alpha and beta, which is fine, but could be biased.  My guess is that you would want to use $\readdepth$ and not $b$, but as I said I'm not totally sure what they're doing here. Update 1/4/23 -- I think that they should be using $c$ here and not $b$. But, I should say that just plugging in an estimate for $n$ in the Achaz equations will result in some bias as equations 20, 21, and 22 are nonlinear in n.}


\popoolissue{Here seem to be more issues in the PoPoolation code. Looking again at the line where the call to compute $n_\text{base}$ is located, \href{https://github.com/lczech/popoolation/blob/092e7a6f7ee4910c1bec4377e0adccc353175bc8/Modules/VarMath.pm\#L127}{here}, the function is called with the min coverage (read depth), instead of the actual coverage as stated in the equations here. We are not sure what effect this has, but think that this is an issue.}

\popoolissue{Furthermore, that call to \texttt{nbase\_buffer} takes two arguments, \texttt{\$n} and \texttt{\$mincoverage}, but the inner function of \texttt{get\_nbase\_buffer} only does one \texttt{shift}, hence ignoring the second argument. Robert changed that line after our conversation, but I am not sure that everything is fixed yet, see below.}
 
\popoolissue{At this point also, see \href{https://github.com/lczech/popoolation/blob/092e7a6f7ee4910c1bec4377e0adccc353175bc8/Modules/VarMath.pm\#L145}{here}, the variable is again called \texttt{\$cov}, indicating that we want to take the coverage (and not the min coverage). But because of the missing shift, this value is in fact the pool size again, from the call to the inner function mentioned above. Unless we are misunderstanding perl function call rules (which might well be the case), there seems to be something really off here. In summary: The function is computed with coverage=poolsize, but with a detour that first also makes it seem that coverage=min coverage by accident.}

\popoolissue{Lastly, we noted that the computation of $\tilde{n}$ generally results in some non-integer value. However, if we understand the equations for alpha* and beta* correctly, they expect integer values. We did some quick test of this, and it seems that some low non-integer values of pool sizes and coverages can even lead to negative results then. This should not be an issue in practice, but it still seems that something is not quite right here.}

\subsubsection*{Final estimator for D}

Now that we have a way of computing a reasonable value for the number of individuals sequenced,
we can finally define the estimator:
\begin{align}
    \tilde{D}_\text{pool} &:=
    \frac{
        \widehat{\theta}_{\pi, \text{pool}} - \widehat{\theta}_{w, \text{pool}}
            }{
        \sqrt{ {|W|}^{-1} \cdot \alpha^*(\tilde{n}) \widehat{\theta}_{w, \text{pool}} ~+~ \beta^*(\tilde{n}) \widehat{\theta}_{w, \text{pool}}^2 }
    }
\end{align}
%
following PoPoolation and \citeay{Achaz2008}.
This requires $b=2$; furthermore, PoPoolation suggests using ``not too small'' windows.
We are using the size $|W|$ of the window here, that is, the total length along the window in bases, which is typically much larger than the number of SNPs in the window. As we use a difference in two estimators of $\Theta$ in the numerator here, and incorporate the window size in the denominator, we do not need to apply an additional normalization here.

\todo{UNRESOLVED ISSUE: That is what i get from their code. Is that the correct term to use here? should it be the number of SNPs in the window instead?  JPS: I have no idea.  I would need to go through the Achaz paper in some detail to figure it out.  My guess is number of SNPs is right.}

The above is the estimator as implemented in \toolname{PoPoolation} and also offered in our implementation. However, note that the original implementation has some unresolved issues, as described below in \secref{supp:sec:TajimaD:sub:Bugs}. We furthermore believe that this estimator does not fully achieve its goal of being quantitatively comparable to classical individual-based Tajima's D, as described in \secref{supp:sec:TajimaD:sub:RealDeal}.
Before describing these more fundamental problems though, we first have a look at the assumptions and biases of the approach as described above.




\subsection{Assumptions and Biases}
\label{supp:sec:TajimaD:sub:AssumptionsBiases}

In the above computation of the correction term for Tajima's D for pool sequencing,
several assumptions were made that lead to the resulting estimator being conservative,
\ie yielding smaller values than what would be expected from individual sequencing of samples.
Based on the explanation in the \toolname{PoPoolation} equations document (most of the text in this section is adapted from there), we explore the underlying assumptions and biases.

The locally fluctuating read depth is replaced by the minimum read depth.
This makes the variance estimator larger, and therefore leads to conservative estimates of Tajima's D.

The random number of different individuals sequenced under a given read depth $C$
is replaced by its expected value $\tilde{n}$.
This assumption should not affect the results much:
If the pool size is large compared to the read depth, sequencing the same individual more than once is uncommon.

Furthermore the number of different individuals sequenced will have a low variance.
As we are working with the minimum read depth, $\tilde{n}$ will be biased downwards,
tending to give a conservative estimate of the variance.

At different positions, the subsets from the pool that are sequenced might be different.
Their coalescent histories will be correlated but not identical.
As the classical equations for Tajima's D are for single samples sharing a common coalescent history,
there is less correlation in the data than assumed with the classical formula.
This again should make the variance approximation more conservative.

Summing up, the approximate variance in the above equations provides a conservative approximation,
and the values for Tajima's D will tend to be smaller than those that would be expected
for an experiment based on individual sequencing of single samples.

Lastly, the \toolname{PoPoolation} code repository contains a plot showing the correlation between the classical Tajima's D
and the corrected Tajima's D using the equations described above;
please see \href{https://github.com/lczech/popoolation/raw/master/files/correlation_classic_correctedTajimasD.png}{here},
where the x-axis corresponds to the classical value, and the y-axis the the corrected one.
This plot has been made with real-world data from Drosophila with a read depth (in their naming: coverage) of 12,
a window size of 500 and a minimum count of 1, and corroborates our sense that the \toolname{PoPoolation} estimator is conservative.





\subsection{PoPoolation Bugs}
\label{supp:sec:TajimaD:sub:Bugs}

From our assessment of the \toolname{PoPoolation} code, and from personal communication with Robert Kofler, we found that the implementation of the above $\tilde{D}_\text{pool}$ in \toolname{PoPoolation} $\leq$ v1.2.2 contains several bugs, which significantly alter the numerical results of the computation of Tajima's D.
We want to thank Robert for his support regarding our questions, and for quickly fixing the bugs that we have pointed out.  

As mentioned above, based on further observed results, we however suspect that there might be other statistical issues in their approach, as described below in \secref{supp:sec:TajimaD:sub:RealDeal}. These would need a more careful examination to be verified. At the time being, we have decided to not look further into this, for the reasons described below.

\todo{important unmentioned and unresolved bugs:  (1) using min b for the n tilde computation seems super wrong as well.  (2) Harmonic numbers appear in the Achaz formulae, but we plug in the “effective” number of individuals which is generally not an integer. But the usual harmonic numbers are defined only for integers.  (3) the last position of a window is accidentally used as the first of the following instead.}


Note: In version 2023-04-12 of this document, we also included a Figure where we examined the effect of the bugs that we had identified until then, by comparing values of Tajima's D with and without these bugs, for different pool sizes, minimum read depths (back then, unfortunately called ``coverage''), and window sizes.  We have now decided to remove that figure again, because with the potentially remaining open statistical issues of their approach, the comparison is rendered futile.

As a consequence of these unresolved issues, we recommend not using this Pool-seq corrected estimator of Tajima's D, see also below.
We hence also recommend rerunning any analyses of Tajima's D that have been conducted with past versions of \toolname{PoPoolation}, particularly if any results rely on the exact numerical values.  If necessary, such analyses should be re-evaluated with the insights described here.

\subsection{Conclusion and Comprehensive Approach}
\label{supp:sec:TajimaD:sub:RealDeal}

Above, the goal was to derive an estimator of Tajima's D for Pool-seq data which is quantitatively comparable to classical Tajima's D on individual data.  The overall difficulty is to find an estimator for the variance that we can use in \eqnref{eq:BasicTajimaD}.  
In a sense, there is a philosophical question about what Tajima's D even should be in the pool sequencing case, because the denominator should be the standard deviation (square root of the variance) of the numerator, but in the Pool-seq context, it is unclear what variance we mean.  Using the approach by Achaz, and plugging in the ``effective number of individuals sequenced'', as suggested by Kofler et al, is a somewhat reasonable idea, albeit not a perfect one. 


Firstly, the equations above leave it open what read depth to use for the computation of $\tilde{n}$ in \eqnref{eq:NTildeClosedForm} for computing the Achaz approach.  In \toolname{PoPoolation}, they use the minimum allele count $b$ as provided by the user, instead of empirical read depth in the window.  We suspect that this leads to a major distortion of the results:  As shown in \eqnref{eq:NTildeBounds}, the value of $\tilde{n}$ is bounded by both the read depth and the pool size; however, a requirement of the approach is that $b := 2$ (see \secref{supp:sec:TajimaD:sub:Classic}), which hence means that $\tilde{n} \leq 2$, which will rarely reflect the expected number of individuals sequenced.

Hence, empirical alternatives for the read depth likely yield more reasonable results. For instance, in order to use Tajima's D for outlier detection, one typically wants to be conservative. In that case, we should assume loci to be as correlated as possible, and hence use the smallest read depth in the given window instead, and plug that into the Achaz approach.

Furthermore, if we ignore all of the noise from pool sequencing, then the denominator should just be based on the equations by Achaz as explained above, but using the pool size $n$ instead of the effective pool size $\tilde{n}$.  This probably works well enough if read depth is high, as the noise induced by pool sequencing is relatively small compared to the noise introduced by demography (see below).  If however we want to include the noise from pool sequencing, which in particular is relevant at low read depths, then it becomes quite complicated, and the approach as described above starts to break down.

For a comprehensive estimator of the variance, we would have to know about the covariance of frequencies across sites. In the Pool-seq context, this has two components: (1) The first is a demographic component --- how has the randomness of the populations affected these two sites? The approach by Achaz answers that question (in isolation).  (2) The second is a Pool-seq component --- how does the randomness in the allele frequencies at these two sites covary?

These components are tricky to reconcile: By sequencing some subset of individuals at one site, and a different subset of individuals at another site, the estimated frequencies will covary in a different way than if the ``same'' subset of individuals were sequenced at both sites.  We would hence need to somehow integrate over the number of overlapping individuals for every pair of sites, which is not feasible with Pool-seq data.

Hence, finding a comprehensive solution that \emph{actually} yields the intended quantitatively comparable values of Tajima's D might involve knowledge of population history and simulations under a neutral model. We hence do not further pursue this approach at the time being.

In conclusion, we suggest the following: 

\begin{itemize}
    \item The approach of \toolname{PoPoolation} to use the minimum read depth $b$ to compute $\tilde{n}$ is likely not reasonable.
    \item Instead, using the pool size to replace the whole term, i.\,e., $\tilde{n} = n$, is better under high read depths. This basically assumes that the number of individuals sequenced is equal to the pool size.
    \item Alternatively, for a more conservative approach, we can compute $\tilde{n}$ as in \eqnref{eq:NTildeClosedForm}, by using pool size $n$, but set $c$ to the minimum empirical read depth found in the given window.
    \item By far the easiest and safest option in our opinion is to (for now) abandon the goal of quantitative comparability with the classic Tajima's D, and instead directly use the difference of $\theta_{\pi,\text{pool}}$ and $\theta_{w, \text{pool}}$.  As long as read depth is relatively equal along the regions being compared, this should enable the detection of outliers and other features of interest.
\end{itemize}

Lastly, we want to invite the research community to further scrutinize these approaches, and provide suggestions for how to move forward, in order to get a Pool-seq equivalent of Tajima's D that is quantitatively comparable to the classic estimator.


\section{Fixation Index \texorpdfstring{\fst}{FST} for Pool-Seq}
\label{supp:sec:FST}

In this section, we will derive unbiased estimators of various measures of heterozygosity in two populations for Pool-sequencing data.
These will then be combined to obtain ``sample-size'' and ``pool-size'' corrected estimators of two definitions of \fst.
On top of these two estimators for \fst{} in the pool-sequencing context, we also walk through the two existing estimators as suggested by \citeay{Kofler2011b} and \citeay{Karlsson2007}.
Both are implemented in \toolname{PoPoolation2}, and are called the ``classical'' or ``conventional pool sequencing'' approach, and the ``Karlsson approach adapted to digital data'', respectively, in \citeay{Kofler2011b}.
We compare all four approaches to each other, and show that the ``classical'' approach is biased for lower read depths or small pool sizes, and the Karlsson approach is biased for small pool sizes (bias on the order of 1/pool size).
See also \citeay{Hivert2018} for an assessment of \fst{} in the pool-sequencing context.


\label{supp:sec:FST:sub:Definition}

There are several non-equivalent \emph{definitions} of $\text{F}_\text{ST}$.
The overall goal is to measure some degree of differentiation between two populations, which can be represented as a proportion of variation that cannot be explained by variation within populations.
What is unclear is a proportion of \emph{what} variation?
There are two natural candidates leading to two related, but distinct definitions of $\text{F}_\text{ST}$.
The first definition, which we will call $\text{F}_\text{ST}^\text{Nei}$ following \citeay{Nei1973}, considers the proportion of the total variation in the two populations.
This statistic is also called $\text{G}_\text{ST}$, see for example Equation (5.5) of \citeay{Hahn2018}.
The second definition, which we will call $\text{F}_\text{ST}^\text{Hudson}$ following \citeay{Hudson1992}, considers the proportion of the variation between populations, see also \citeay{Cockerham1969} and \citeay{Weir2002}.
This second definition is also considered in \citeay{Karlsson2007}, which we examine below in \secref{supp:sec:FST:sub:Karlsson}.

To make this more formal, we can consider the probability that two haploids carry different alleles.
We could consider drawing the two haploids from the same population (with the population chosen at random), which we call $\pi_\text{within}$; or we could consider drawing the two haploids from \emph{different} populations, which we call $\pi_\text{between}$; or finally we could consider drawing the two haploids totally at random from either population (potentially the same populations, potentially different populations) which we call $\pi_\text{total}$.
See \citeay{Bhatia2013} for more background information on this.

Our two definitions of $\text{F}_\text{ST}$ are then

\begin{align}
    \text{F}_\text{ST}^\text{Nei}    &:= 1 - \frac{\pi_\text{within}}{\pi_\text{total}} \label{eq:popnei}\\
    \text{F}_\text{ST}^\text{Hudson} &:= 1 - \frac{\pi_\text{within}}{\pi_\text{between}} \label{eq:pophudson}
\end{align}

If we consider a single locus with up $4$ alleles, with frequencies $f_{\tau(p)}$ (possibly zero) with  $\tau$ denoting the allele with $\tau \in \left\{A, C, G, T\right\}$ and $p$ denoting the population with subscripts $1$ and $2$, we can calculate the various $\pi$s as follows

\begin{align}
    \label{eq:PiDefs:PiWithin}
    \pi_\text{within} &= \frac{1}{2}\left[\left(1 - \sum_\tau f_{\tau(1)}^2\right) + \left(1 - \sum_\tau f_{\tau(2)}^2\right)\right] \\
    \label{eq:PiDefs:PiBetween}
    \pi_\text{between} &= 1 - \sum_\tau f_{\tau(1)}f_{\tau(2)} \\
    \label{eq:PiDefs:PiTotal}
    \pi_\text{total} &= \frac{1}{2}\pi_\text{within} + \frac{1}{2}\pi_\text{between}
\end{align}

which are then used in our above definitions of \fst.


\subsection{Unbiased estimators of the \texorpdfstring{$\pi$s}{pi's}}
\label{supp:sec:FST:sub:EstimatorsPi}

Since both definitions of $\text{F}_\text{ST}$ rely on these $\pi$s, we will need to derive unbiased estimates for them.
We will show below that the following are unbiased estimators of the corresponding quantities without hats:

\begin{align}
    \nonumber
    \widehat{\pi}_\text{within} &:=
        \frac{1}{2}
        \Bigg[                                 \left( \frac{n_{(1)}}{n_{(1)}-1} \right) \left( \frac{\readdepth_{(1)}}{\readdepth_{(1)}-1} \right) \left( 1 - \sum_{\tau} \empfreq_{\tau(1)}^2 \right) \Bigg. \\
    \label{eq:Fst:UnbiasedPiWithin}
        \Bigg. &+ \phantom{\frac{1}{2} \Bigg[} \left( \frac{n_{(2)}}{n_{(2)}-1} \right) \left( \frac{\readdepth_{(2)}}{\readdepth_{(2)}-1} \right) \left( 1 - \sum_{\tau} \empfreq_{\tau(2)}^2 \right) \Bigg] \\
    \label{eq:Fst:UnbiasedPiBetween}
    \widehat{\pi}_\text{between} &:=
        1 - \sum_{\tau} \empfreq_{\tau(1)}\empfreq_{\tau(2)} \\
    \label{eq:Fst:UnbiasedPiTotal}
    \widehat{\pi}_\text{total} &:=
        \frac{1}{2}\widehat{\pi}_\text{within} + \frac{1}{2}\widehat{\pi}_\text{between}
\end{align}

In the following, we derive these estimators.


\subsubsection*{Unbiased estimator of \texorpdfstring{$\widehat{\pi}_\text{within}$}{Pi Within}}
\label{supp:sec:FST:sub:EstimatorsPi:sub:PiWithin}

We have derived previously that
\[
\mathbb{E}\left[\left(\frac{n_{(1)}}{n_{(1)}-1}\right)\left(\frac{\readdepth_{(1)}}{\readdepth_{(1)}-1}\right)\left(1 - \sum_{\tau}\empfreq_{\tau(1)}^2\right) \right] = \left(1 - \sum_\tau f_{\tau(1)}^2\right)
\]
within a single population.  It follows immediately that averaging these estimators across the two populations is unbiased for $\pi_\text{within}$.


\subsubsection*{Unbiased estimator of \texorpdfstring{$\widehat{\pi}_\text{between}$}{Pi Between}}
\label{supp:sec:FST:sub:EstimatorsPi:sub:PiBetween}

Since the two pools are independent, we have that
\[
\mathbb{E}\left[\widehat{\pi}_\text{between}\right] = 1- \sum_\tau \mathbb{E}\left[\empfreq_{\tau(1)}\right]\mathbb{E}\left[\empfreq_{\tau(2)}\right]
\]
The frequency of alleles within a pool is an unbiased estimate for the frequency in the population, so
\[
\mathbb{E}\left[\empfreq_{\tau(p)}\right] = f_{\tau(p)}
\]
showing that $\widehat{\pi}_\text{between}$ is unbiased for $\pi_\text{between}$.


\subsubsection*{Unbiased estimator of \texorpdfstring{$\widehat{\pi}_\text{total}$}{Pi Total}}
\label{supp:sec:FST:sub:EstimatorsPi:sub:PiTotal}

That $\widehat{\pi}_\text{total}$ is unbiased for $\pi_\text{total}$ follows immediately from the definition of $\pi_\text{total}$ in \eqnref{eq:PiDefs:PiTotal} and the unbiasedness of $\widehat{\pi}_\text{within}$ and $\widehat{\pi}_\text{between}$.


\subsubsection*{Intuition and Reasoning}
\label{supp:sec:FST:sub:EstimatorsPi:sub:Intuition}

In the above equations, for each of the two populations, we apply a double Bessel's correction for our estimate of $\widehat{\pi}_\text{within}$ in \eqnref{eq:Fst:UnbiasedPiWithin}, to correct for the elevated variance that comes from the double sampling process of picking individuals from a population, and of picking reads from those individuals (during sequencing). Because in both steps, we pick finite numbers, resulting in higher variance in the estimates, which hence needs correcting. 

However, the estimate of $\widehat{\pi}_\text{between}$ in \eqnref{eq:Fst:UnbiasedPiBetween} does not have those corrections. One way to gain some intuition about why the Bessle's corrections are not needed in that case is as follows.

One way to think about $\pi$ in a single population is as the probability to get different alleles when picking two random haploids from the population.  In a large population, it does not matter whether we pick those two haploids with replacement or without replacement: in that case, even if we pick the two haploids with replacement, it is unlikely that we would happen to pick the same haploid twice.  In that sense, $\pi$ is measuring the probability that two different haploids have different alleles.  
 
Now, to estimate the population-level $\pi_\text{within}$, we could look at a smaller sample of haploids, and calculate the probability that two randomly chosen haploids have different alleles. However, if our sample size is small, then the difference between sampling with or without replacement becomes relevant, because there is a $\sfrac{1}{n}$ chance that we pick the same haploid twice if we are sampling with replacement. If we pick the same haploid twice, they obviously cannot have different alleles.  This means that if we compute our probabilities by sampling with replacement, then we underestimate by a factor of $1 - \sfrac{1}{n} = \frac{n-1}{n}$ because in $\sfrac{1}{n}$ of our comparisons we are actually comparing a haplotype to itself.  As a result, we would need to multiply our estimate by $\frac{n}{n-1}$ to undo this underestimation.  Something similar happens again when we move to pool sequencing, where our equations are (implicitly) in terms of sampling with replacement, but now we are sampling reads, and so we end up (by the same argument) with needing to multiply by a factor of $\frac{c}{c-1}$.  

This however is different in the case of $\pi_\text{between}$.  In that case, we are comparing a haploid from one population to a haploid from a different population.  Hence, we are comparing a single haploid (or single read) from one population to a single haploid (or read) from another population.  If we only draw a single sample, then there is no difference between sampling with or without replacement, and so our equations (which again are implicitly based on sampling with replacement) do not need to be corrected to match the idea of sampling without replacement.  

In summary, when considering two entities at a time from the same pool/population, we need to include Bessel's corrections such as for pool size and read depth. However, if we are only considering one entity at a time from each pool/population, then we do not need those corrections.


\subsection{Final asymptotically unbiased estimators of \texorpdfstring{$\text{F}_\text{ST}$}{FST} per SNP and per window}
\label{supp:sec:FST:sub:EstimatorFST}

These estimators then immediately suggest the following ratio estimators for the different definitions of $\text{F}_\text{ST}$:

\begin{align}
    \widehat{\text{F}}_\text{ST}^\text{Nei} &:= 1 - \frac{\widehat{\pi}_\text{within}}{\widehat{\pi}_\text{total}}\\
    \widehat{\text{F}}_\text{ST}^\text{Hudson} &:= 1 - \frac{\widehat{\pi}_\text{within}}{\widehat{\pi}_\text{between}}
\end{align}

All of this has been for a single site, but we are often interested in combining information across SNPs within a window $W$ (or possibly genome wide).
In such a case, define $\widehat{\pi}^\ell_\text{within}$ to be $\widehat{\pi}_\text{within}$ as above but for SNP $\ell \in W$.
Define $\widehat{\pi}^\ell_\text{between}$ and $\widehat{\pi}^\ell_\text{total}$ analogously.
We then combine information across the SNPs in the window $W$ as

\begin{align}
    \widehat{\text{F}}_\text{ST}^\text{Nei}    &= 1 - \frac{\sum_{\ell \in W} \widehat{\pi}^\ell_\text{within}} {\sum_{\ell \in W} \widehat{\pi}^\ell_\text{total}} \\
    \widehat{\text{F}}_\text{ST}^\text{Hudson} &= 1 - \frac{\sum_{\ell \in W} \widehat{\pi}^\ell_\text{within}} {\sum_{\ell \in W} \widehat{\pi}^\ell_\text{between}}
\end{align}

See \citeay{Bhatia2013} for a practical and theoretical justification for using this ``ratio of averages'' instead of using an ``average of ratios''.
These are our asymptotically unbiased estimators for \fst{} for Pool-seq data, which take the finite sampling of individuals from the population, and the finite sampling of reads from each individual in the pool, into account.

Furthermore, as explained above in \secref{supp:sec:ThetaPi:sub:PoolSequencing:sub:WindowAvg}, when computing values per window, we here too need to take missing and low quality data into account when working with real-world data. To this end, we again want to normalize the involved values of $\pi$ using only the high-quality positions, that is, the variant and invariant positions that passed all quality filters. However, we are dealing with an estimator here that operators on two pools simultaneously. Hence, the window normalization needs to be applied individually per pool in \eqnref{eq:Fst:UnbiasedPiWithin}, by using the number of positions in each population to normalize the two components of the equation separately from each other. Then, in \eqnref{eq:Fst:UnbiasedPiBetween}, we instead normalize using the number of positions that passed the quality filters in \emph{both} populations.

In the following, we describe our assessment of the estimators previously implemented in \toolname{PoPoolation2}. Note that these two estimators do not use the window normalization as explained here, as they do not correct for the read depths per pool individually either. They hence implicitly assume that the window normalization would cancel out anyway, and do not apply it at all. This makes our estimators as described above better suited for data with low coverage or low quality.


\subsection{Estimator of \texorpdfstring{$\text{F}_\text{ST}$}{FST} as implemented in PoPoolation2}
\label{supp:sec:FST:sub:PoPoolation2Estimator}

The implementation in \toolname{PoPoolation2} \cite{Kofler2011b} offers two ways to estimate \fst{}:
What they call the ``classical'' or ``conventional'' approach by \citeay{Hartl2007}, and an approach adapted to digital data following \citeay{Karlsson2007}.
In this and the next section, we discuss these estimators.
We later show that they are biased, and hence recommend using the estimates as introduced above instead.
For comparability and historical backwards compatibility, we however still offer both these estimators in our implementation in \toolname{grenedalf}.


First, we present the ``classical'' approach as implemented in \toolname{PoPoolation2}, labelled with superscript ``Kofler'' here, following \citeay{Kofler2011b}.
We compute \fst{} for two subpopulations, which we here again denote with subscripts $(1)$ and $(2)$, and the total population with $(T)$. We expect pool sizes $n >= 2$.

For each SNP in a given window, \toolname{PoPoolation2} computes:

\begin{align}
    \label{eq:PoPoolation2FstPi:1}
    \widehat{\pi}_{(1)}^\text{Kofler}  &:= \frac{\readdepth_{(1)}}{\readdepth_{(1)}-1} \cdot  \left( 1 - \sum_\tau \empfreq_{\tau(1)}^2 \right) \\
    \label{eq:PoPoolation2FstPi:2}
    \widehat{\pi}_{(2)}^\text{Kofler}  &:= \frac{\readdepth_{(2)}}{\readdepth_{(2)}-1} \cdot  \left( 1 - \sum_\tau \empfreq_{\tau(2)}^2 \right) \\
    \label{eq:PoPoolation2FstPi:T}
    \widehat{\pi}_{(T)}^\text{Kofler}  &:= \frac{\readdepth_{(T)}}{\readdepth_{(T)}-1} \cdot  \left( 1 - \sum_\tau \empfreq_{\tau(T)}^2 \right) \\
    \nonumber
    \mbox{with} \\
    \nonumber
    \readdepth_{(T)} &:= \mbox{min} \left( \readdepth_{(1)}, \readdepth_{(2)} \right) \\
    \nonumber
    \empfreq_{\tau(T)} &:= \frac{1}{2} \left( \empfreq_{\tau(1)} + \empfreq_{\tau(2)} \right)
\end{align}

These quantities are accumulated over the window $W$, with $(\ell)$ denoting the above at SNP $\ell$

\begin{align}
    \label{eq:PoPoolation2FstWindow:1}
    \widehat{\pi}_{W(1)}^\text{Kofler} &= \frac{n_{(1)}}{n_{(1)}-1} \cdot \sum_{\ell\in W} \widehat{\pi}_{(1)}^\text{Kofler}(\ell)\\
    \label{eq:PoPoolation2FstWindow:2}
    \widehat{\pi}_{W(2)}^\text{Kofler} &= \frac{n_{(2)}}{n_{(2)}-1} \cdot \sum_{\ell\in W} \widehat{\pi}_{(2)}^\text{Kofler}(\ell)\\
    \label{eq:PoPoolation2FstWindow:T}
    \widehat{\pi}_{W(T)}^\text{Kofler} &= \frac{n_{(T)}}{n_{(T)}-1} \cdot \sum_{\ell\in W}  \widehat{\pi}_{(T)}^\text{Kofler}(\ell)\\
    \nonumber
    \mbox{with} \\
    \nonumber
    n_{(T)} &= \mbox{min} \left( n_{(1)}, n_{(2)} \right)
\end{align}

Finally, the estimate of \fst is computed as:

\begin{align}
    \label{eq:PoPoolation2FstEst}
    \widehat{\text{F}}_\text{ST}^\text{Kofler} &= \frac{ \widehat{\pi}_{W(T)}^\text{Kofler} - \frac{1}{2} \left(     \widehat{\pi}_{W(1)}^\text{Kofler} +     \widehat{\pi}_{W(2)}^\text{Kofler}  \right)}{ \widehat{\pi}_{W(T)}^\text{Kofler}}
\end{align}

Note that following the notation of the previous section, $\frac{1}{2}\left(\widehat{\pi}_{W(1)}^\text{Kofler} + \widehat{\pi}_{W(2)}^\text{Kofler}\right)$ is identical to the windowed version of $\widehat{\pi}_\text{within}$ and hence is unbiased.  Unfortunately, $\widehat{\pi}_{W(T)}^\text{Kofler}$ however is not an unbiased estimator of $\pi_\text{total}$.  There are two main issues.  First, recall that $\pi_\text{total} = \frac{1}{2}\left(\pi_\text{between}  + \pi_\text{within} \right)$.  The naive estimator of $\pi_\text{between}$ turns out to be unbiased as shown in the previous section, but yet in the PoPoolation estimator, there is a Bessel correction acting on the whole estimator --- this effectively biases the contribution of the part estimating $\pi_\text{between}$ upward.  Second, $\pi_\text{within}$ has two components corresponding to the two populations.  Since these have different read depths and different sample sizes, the components estimating each of these have different biases that need to be corrected separately.  Instead, PoPoolation uses the minimum read depth and the minimum sample size for both.  This again will result in an upward bias for the sample with higher read depth and/or larger pool size.  Overall, these two errors can result in substantial upward bias for $\widehat{\pi}_{W(T)}^\text{Kofler}$, which will result in substantial upward bias for $\widehat{\text{F}}_\text{ST}^\text{Kofler}$, as we show later.  This was also pointed out by \citeay{Hivert2018}. It is therefore recommended to use the estimator presented in the previous section instead.




\subsection{Asymptotically Unbiased Estimator of \texorpdfstring{\fst}{FST} by Karlsson \textit{et al.}}
\label{supp:sec:FST:sub:Karlsson}

Another estimator for \fst{} that is offered in \toolname{PoPoolation2} is based on the equations used in \citeay{Karlsson2007}, see the last page of the Supplemental Information of Karlsson \textit{et al.} for their derivation.
We here briefly also go through the derivation.

We here call this estimator using the superscript ``Karlsson'', which is again defined for two subpopulations denoted with subscripts $(1)$ and $(2)$.
We are here only looking at biallelic SNPs.
Instead of $\tau$ for the four nucleotides, we hence use $q$ for the frequency of the major (reference) allele and $p$ for the frequency of the minor (alternative) allele.  We will also use $\widehat{q}$ and $\widehat{p}$ as the corresponding empirical estimates of $q$ and $p$ (\ie the number of reads supporting the major or minor allele divided by the read depth).

We start with the definition of $\text{F}_\text{ST}^\text{Karlsson}$ from Karlsson \textit{et al.} for the SNPs in a window $W$:

\begin{align}
    \label{eq:FstK}
    \text{F}_\text{ST}^\text{Karlsson} &= \frac{\sum_W N_k}{\sum_W D_k}
\end{align}

where the the numerator $N_k$ and denominator $D_k$ for a single site $k$ in $W$ are:

\begin{align}
    \label{eq:FstNk}
    N_k &= p_{(1)} \cdot ( q_{(2)} - q_{(1)} ) ~+~ p_{(2)} \cdot ( q_{(1)} - q_{(2)} ) \\
    \label{eq:FstDk}
    \nonumber
    D_k &= p_{(1)} q_{(2)} + q_{(1)} p_{(2)} \\
        &= N_k + p_{(1)} q_{(1)} + p_{(2)} q_{(2)}.
\end{align}

It is not obvious, but follows from simple algebra that this definition is equivalent to $ \text{F}_\text{ST}^\text{Hudson} $ defined above.

These are estimated as follows, using the numerator $\hat{N}_k$ and denominator $\hat{D}_k$ at a single site:

\begin{align}
    \label{eq:FstKnh}
    \widehat{N}_k &= \left( \widehat{q}_{(1)} - \widehat{q}_{(2)} \right)^2 - \left( \frac{h_{(1)}}{\readdepth_{(1)}} + \frac{h_{(2)}}{\readdepth_{(2)}} \right) \\
    \label{eq:FstKdh}
    \widehat{D}_k &= \widehat{N}_k + h_{(1)} + h_{(2)}
    \intertext{with two additional helpers:}
    \nonumber
    h_{(1)} &= \frac{\readdepth_{(1)}}{(\readdepth_{(1)}-1)}\widehat{q}_{(1)}\widehat{p}_{(1)} \\
    \nonumber
    h_{(2)} &= \frac{\readdepth_{(2)}}{(\readdepth_{(2)}-1)}\widehat{q}_{(2)} \widehat{p}_{(2)}
    \end{align}

And finally, these are used to define estimator $\widehat{\text{F}}_\text{ST}^\text{Karlsson}$ for a window $W$:

\begin{align}
    \label{eq:FstEstK}
    \widehat{\text{F}}_\text{ST}^\text{Karlsson} &= \frac{\sum_W \widehat{N}_k}{\sum_W \widehat{D}_k}
\end{align}

According to Karlsson \textit{et al.}, when the read depths $\readdepth_{(1)}$ and $\readdepth_{(2)}$ (called ``sample sizes'' there) are equal, the estimator reduces to the estimator of \fst{} given by \citeay{Weir2002}.
Karlsson \textit{et al.} further state that by the Lehmann-Scheff\'{e} theorem \cite[Theorem 4.2.2]{Bickel1977}, it follows that $\widehat{N}_k$ and $\widehat{D}_k$ are uniformly minimum variance unbiased estimators of $N_k$ and $D_k$, respectively, and hence conclude that their estimator \^{F}\textsubscript{ST,K} is also asymptotically unbiased.

This estimator is very similar (after some algebra) to our $\widehat{\text{F}}_\text{ST}^\text{Hudson}$.
It however assumes the pool size to be infinite, that is, it is missing the corrections for pool size.


\subsection{Comparison of the Estimators and their Biases}
\label{supp:sec:FST:sub:Comparison}

As mentioned above, both the Kofler estimator in \toolname{PoPoolation2} and the Karlsson estimator have biases.
We here explore their effects via simulations, and further show that our estimators are approximately unbiased under the Pool-seq assumptions so long as sequencing error rates are low.


\subsubsection*{Simulation Setup}
\label{supp:sec:FST:sub:Comparison:sub:Simulations}

To test the accuracy of the different estimators of $\text{F}_\text{ST}$ we performed two sets of simulations.  In both sets of simulations we assumed a biallelic mutation model, and a simple model of sequencing.  Our sequencing model assumes that (1)~each position is independent, (2)~the number of reads at a given position is Poisson distributed with the mean of that distribution being the ``Read Depth'' of that simulation, (3)~each read is equally likely to come from any individual in the pool, and (4)~with probability ``Seq Error'' a read shows the opposite allele as the individual it was sampled from, but otherwise matches the individual's allele, and this error process is independent across reads.  Across both sets of simulations, we vary the sequencing error in $\left\{0, 10^{-5}, 10^{-4}, 10^{-3}\right\}$, the number of haploid individuals in the pool in $\left\{10, 100, 1000\right\}$, and the mean sequencing depth in $\left\{10, 100, 1000\right\}$.  In both sets of simulations, we obtain population-level allele frequencies at each site, and then compute the true values of either $\neifst$ or $\hudsonfst$ using \cref{eq:popnei,eq:pophudson,eq:PiDefs:PiWithin,eq:PiDefs:PiTotal,eq:PiDefs:PiBetween}.
We here use ``0'' and ``1'' to denote the major and minor allele, respectively.

In the first set of simulations, which we call the ``simple'' simulations, we simulate 1000 sites in two populations.  Let $f_j^{(p)}$ be the frequency of one of the ``1'' allele at position $j$ in population $p$.  To obtain correlated allele frequencies in the two populations we simulated the frequencies as
\begin{align*}
f_j^{(1)} &\overset{i.i.d.}{\sim} \text{Beta}(0.1, 0.5) \\
f_j^{(2)} | f_j^{(1)} &\sim \text{Beta}\left(\frac{\gamma f^{(1)}_j}{1-f^{(1)}_j+10^{-10}}, \gamma\right),
\end{align*}
where $\gamma$ controls the degree of correlation between the frequencies in the two populations.  As $\gamma$ increases, the correlation between the frequencies decreases.  This parameterization is chosen so that $\mathbb{E}\left[f_j^{(2)} \mid f_j^{(1)}\right] = f_j^{(1)}$ up to the error introduced by the $10^{-10}$ which is included for numerical stability.  Then, to obtain the frequency of the ``1'' allele in our pool, we performed independent binomial sampling at each site for each population.  Finally, we obtained the number of reads with the ``0'' and ``1'' alleles as described above.  We varied $\gamma$ in $\left\{0.1, 0.325, 0.55, 0.775, 1\right\}$, and performed 10 simulations each for each setting of $\gamma$, sequencing error, pool size, and sequencing depth, resulting in 1800 simulations.

In the second set of simulations we used \texttt{msprime} \citep{baumdicker2022efficient,kelleher2016efficient} to simulate data under a population genetic model.  In this model, we simulate two populations of constant size that diverged at a time, $t_\text{div}$, in the past, with the ancestral population having the same, constant size.  Across simulations, we varied $t_\text{div}$ in $\left\{0, 0.25, 0.5, 0.75, 1.0\right\}$.  We used a population-scaled mutation rate, $\theta$, of $0.001$, a population-scaled recombination rate, $\rho$, of $0.0001$, and simulated sequences of length $1$ Mb.  Since \texttt{msprime} is a coalescent simulator, it samples individual-level data.  As such, for a given simulation we directly obtain the alleles of the individuals at each of the $10^6$ sites for the individuals that make up our pool sequencing sample.  We then draw reads for each of those sites as described above.  To obtain population-level allele frequencies we simulate an additional 1000 haploid individuals in each population beyond those that make up the pool, and we use the empirical frequencies of the alleles in this 1000 $+$ pool size sample as the ``true'' allele frequencies in each population.


\subsubsection*{Evaluation and Results}
\label{supp:sec:FST:sub:Comparison:sub:Results}

\begin{figure*}[p]
    \centering
    \includegraphics[width=.69\linewidth]{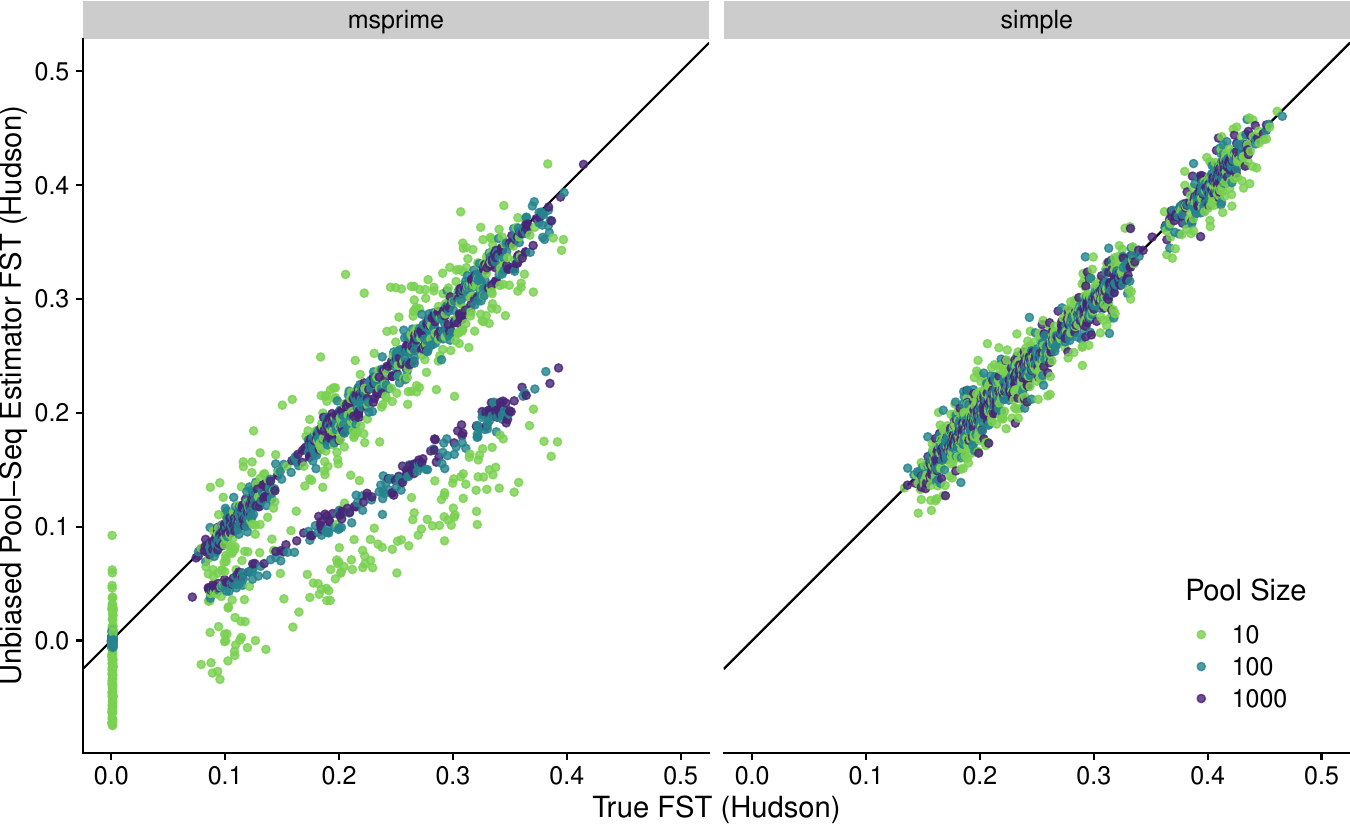}
    \includegraphics[width=.69\linewidth]{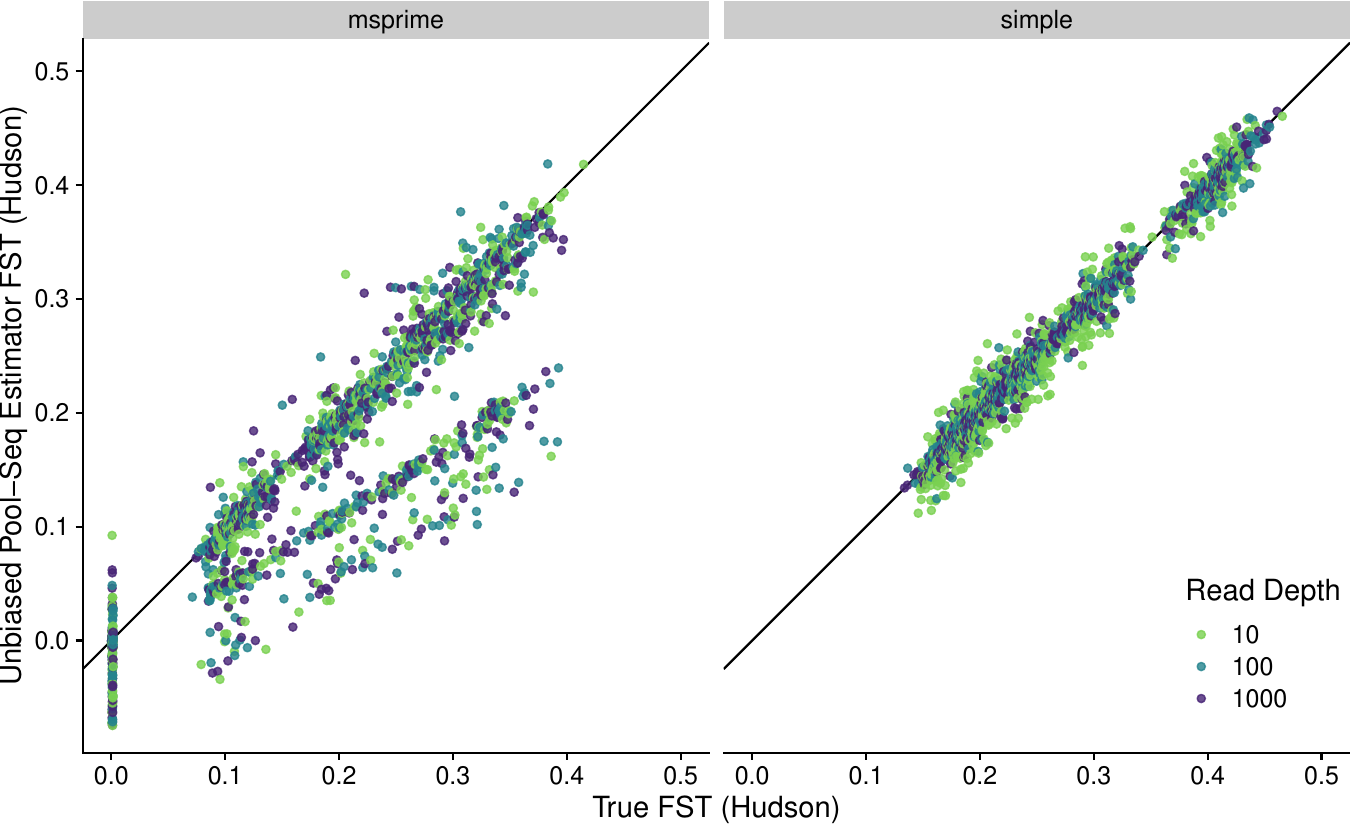}
    \includegraphics[width=.69\linewidth]{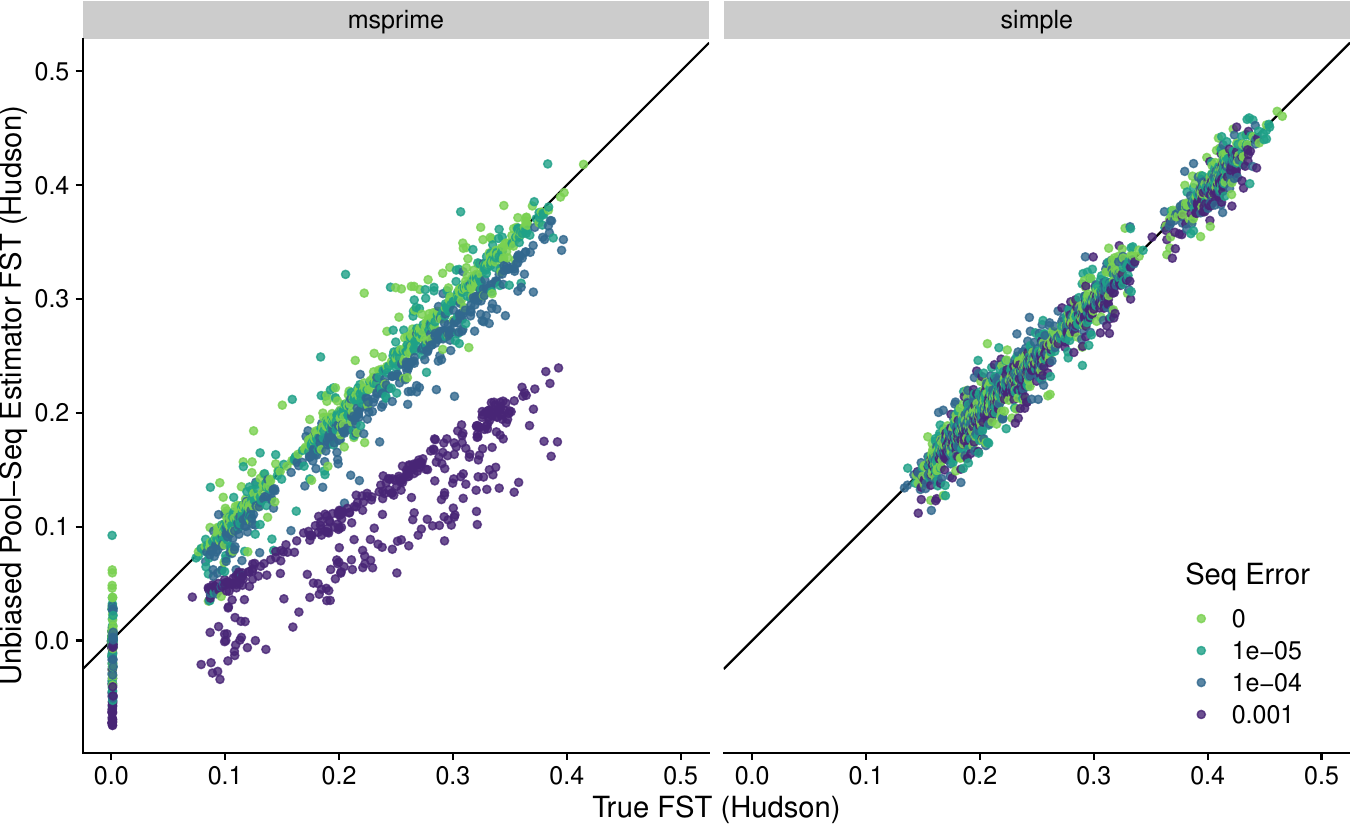}
    \vspace*{-1em}
    \caption{
        Unbiased Pool-Seq Estimator for FST (Hudson) vs true FST (Hudson).
    }
\label{fig:UnbiasedHudson}
\end{figure*}

\begin{figure*}[p]
    \centering
    \includegraphics[width=.69\linewidth]{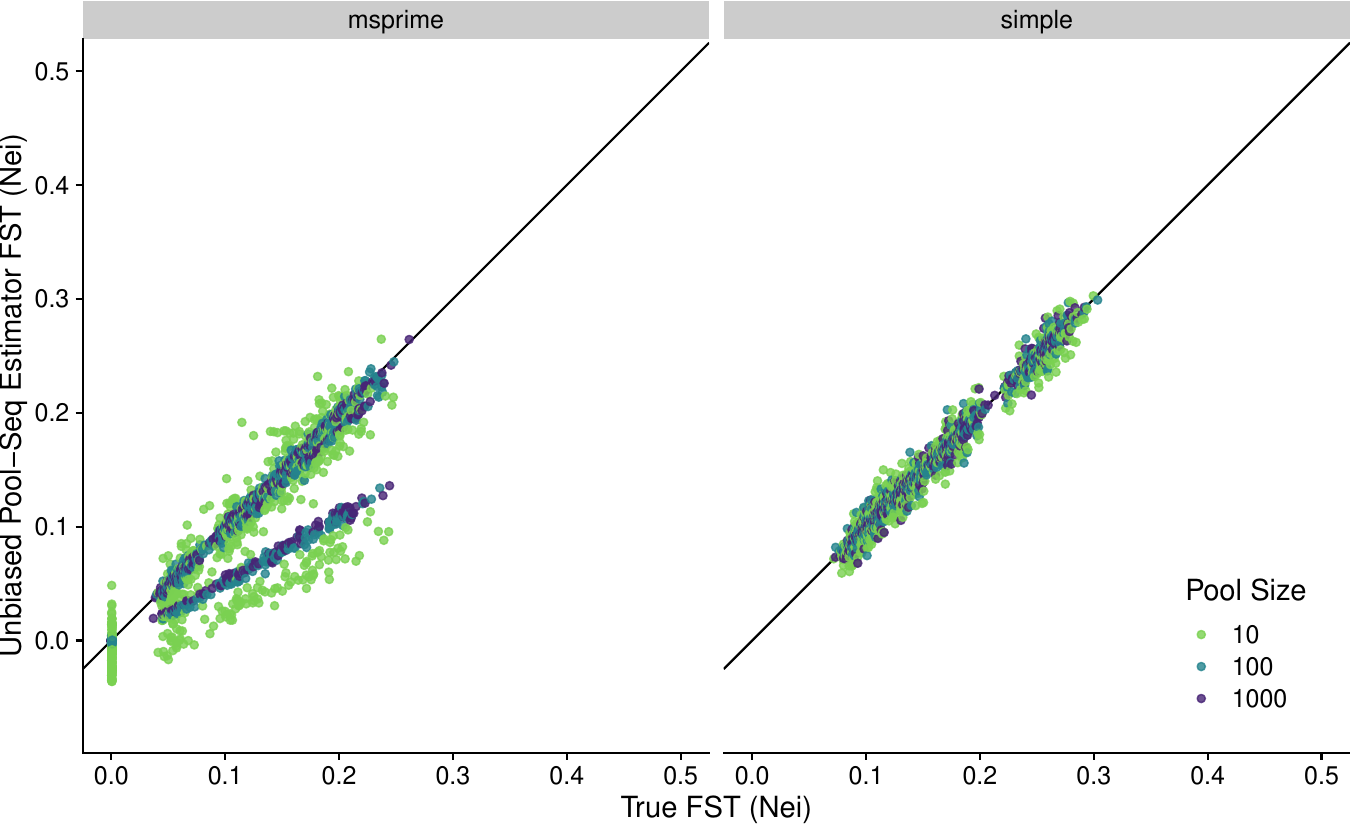}
    \includegraphics[width=.69\linewidth]{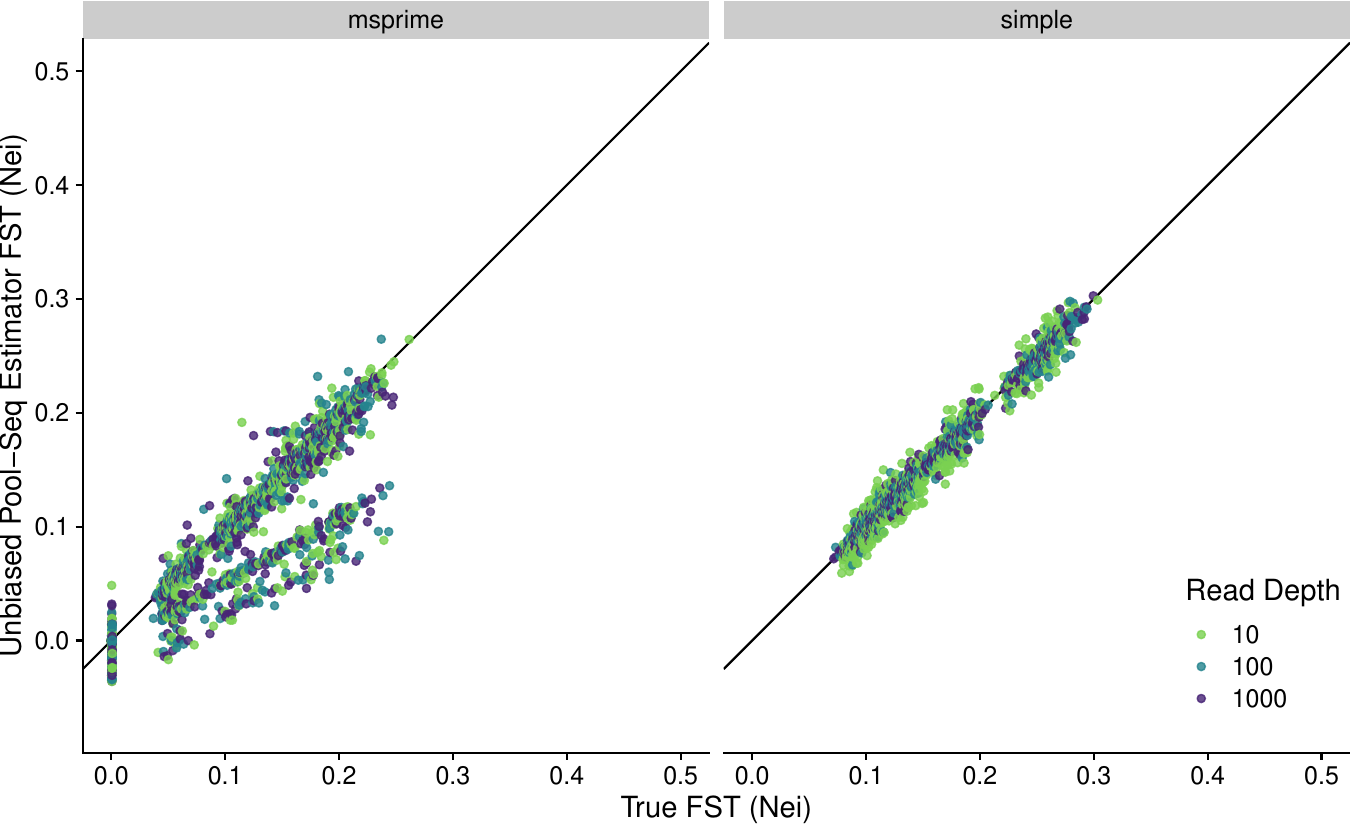}
    \includegraphics[width=.69\linewidth]{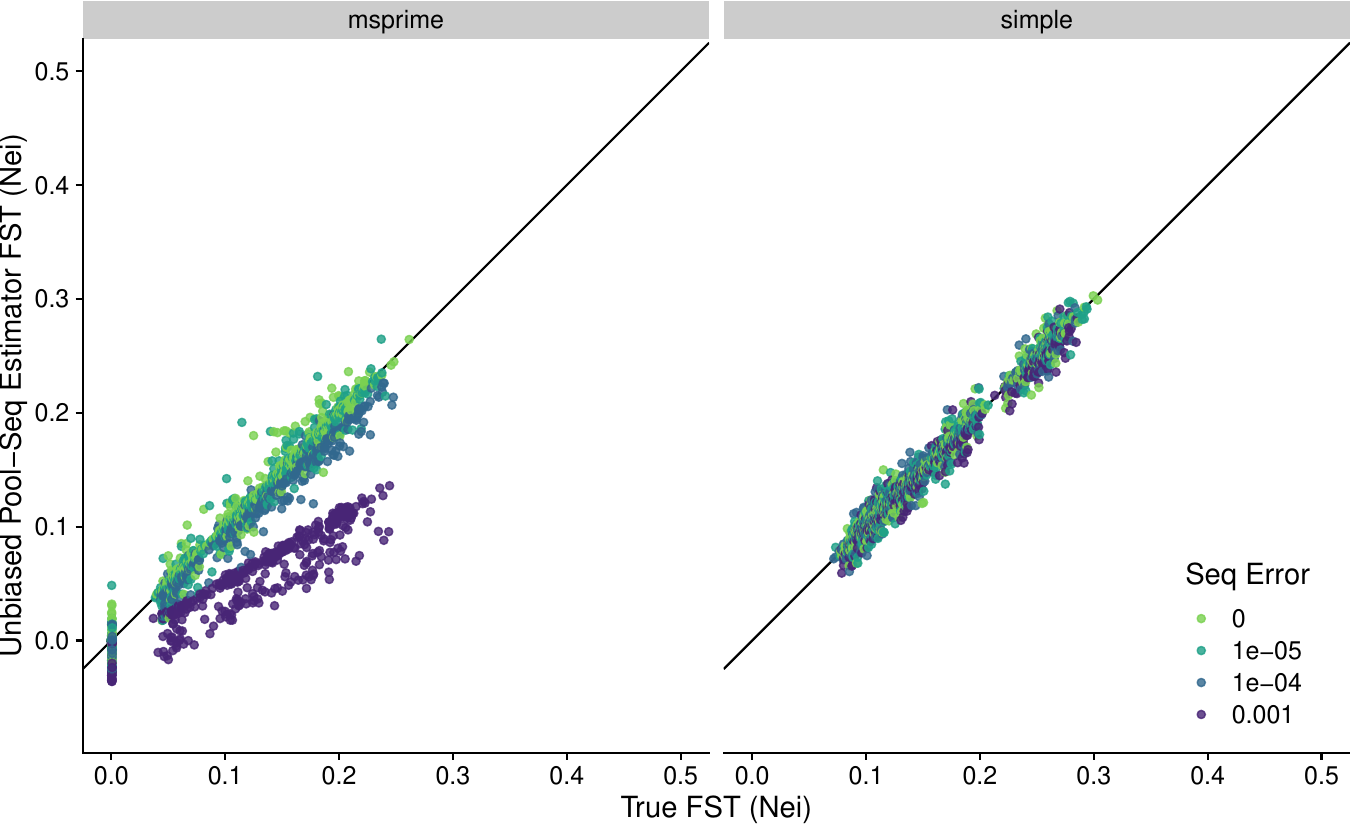}
    \vspace*{-1em}
    \caption{
        Unbiased Pool-Seq Estimator for FST (Nei) vs true FST (Nei).
    }
\label{fig:UnbiasedNei}
\end{figure*}

\begin{figure*}[p]
    \centering
    \includegraphics[width=.69\linewidth]{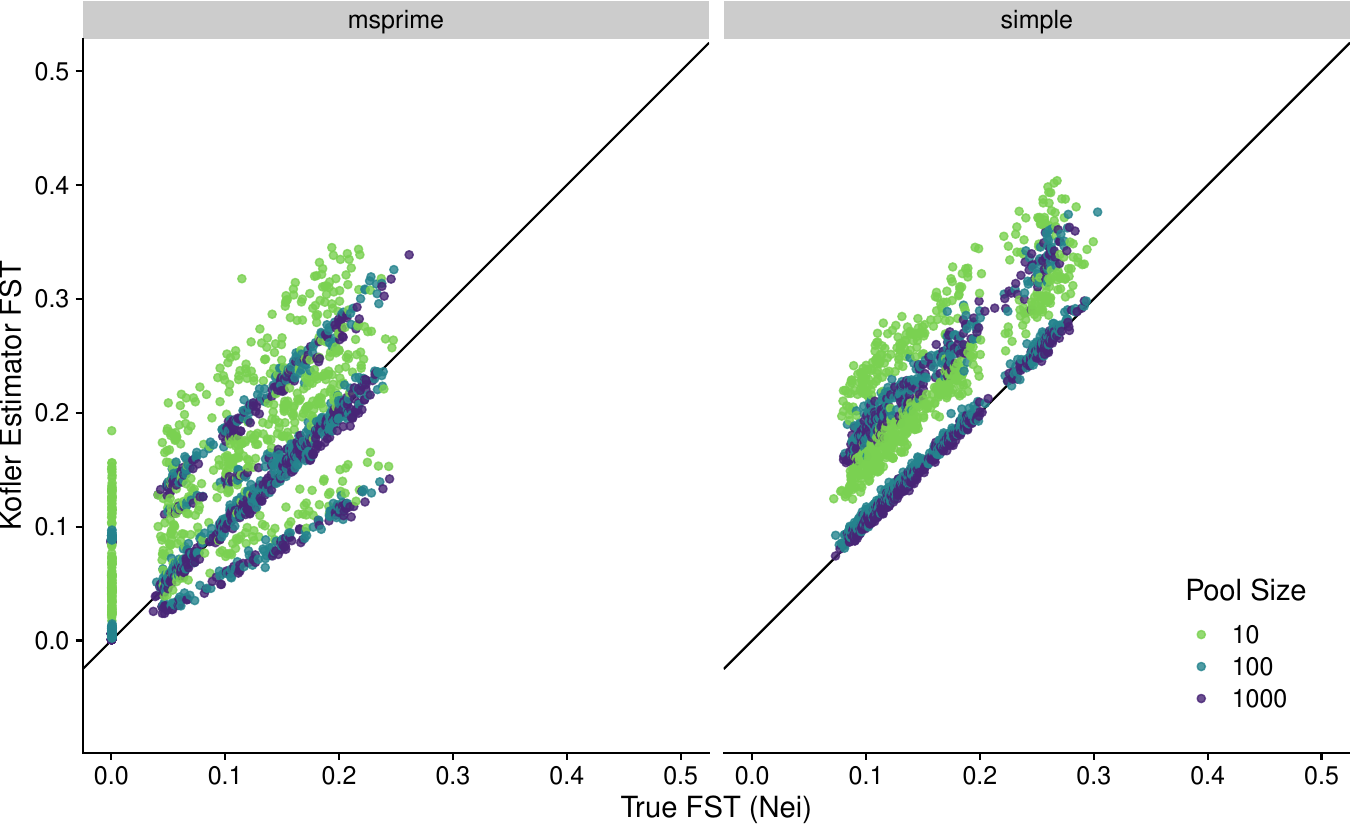}
    \includegraphics[width=.69\linewidth]{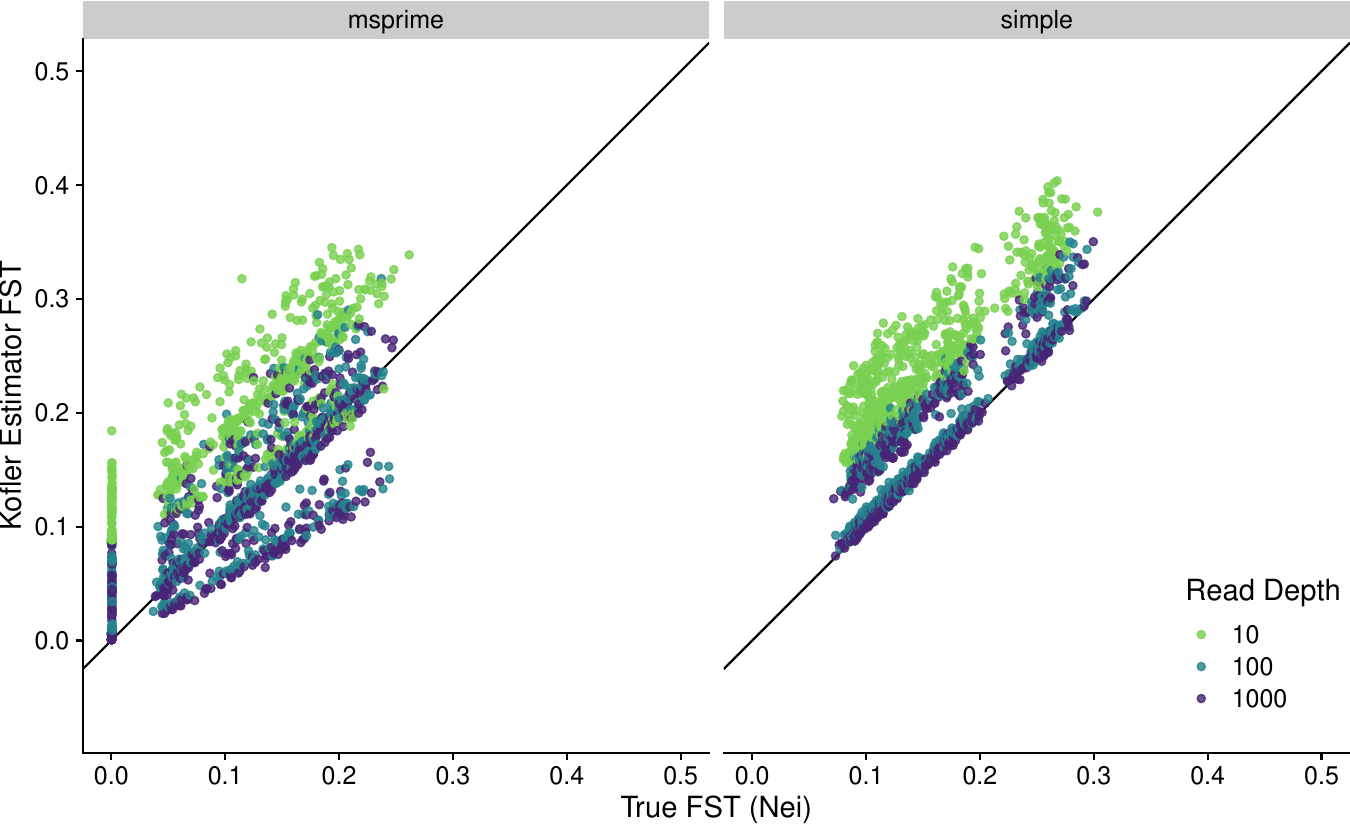}
    \includegraphics[width=.69\linewidth]{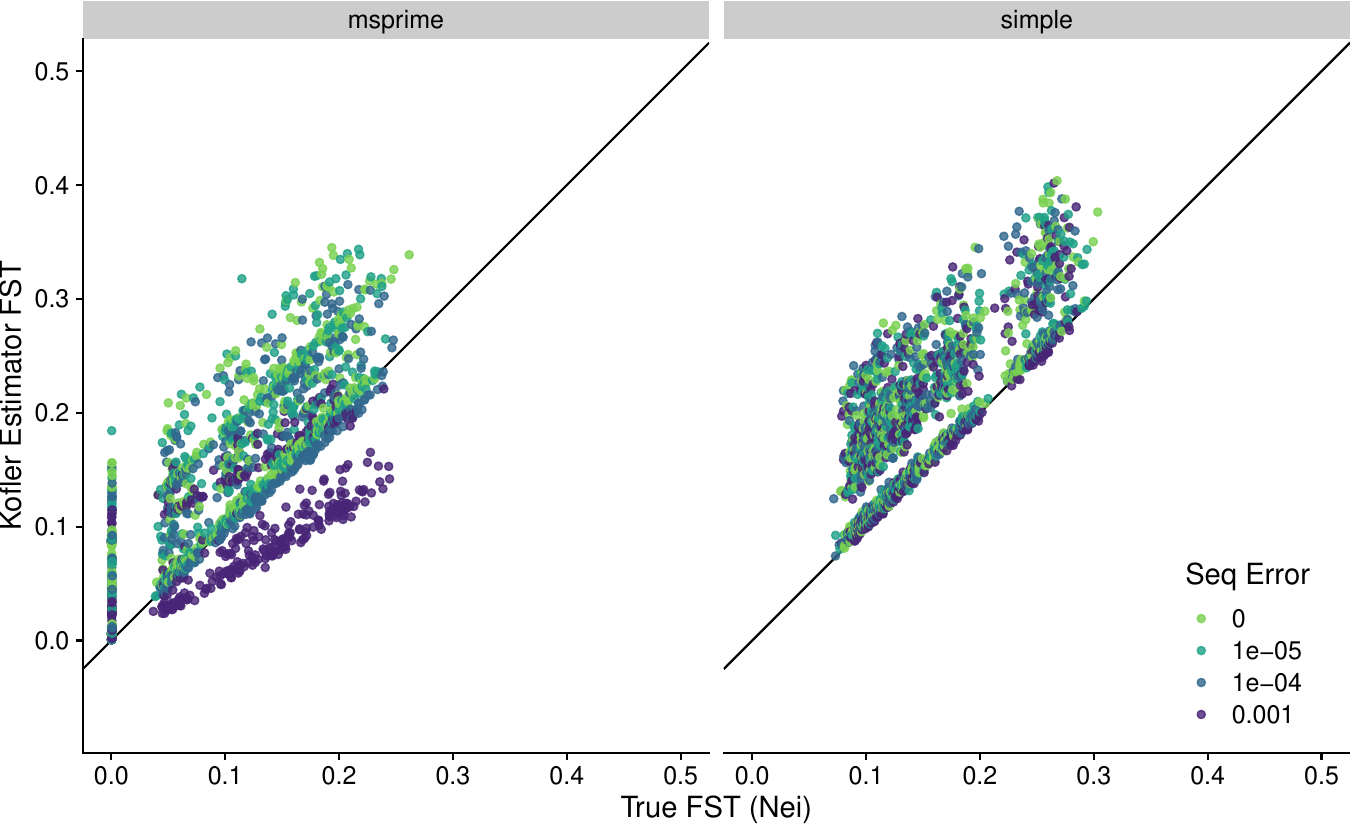}
    \vspace*{-1em}
    \caption{
        Kofler Estimator for FST vs true FST (Nei).
    }
\label{fig:Kofler}
\end{figure*}

\begin{figure*}[p]
    \centering
    \includegraphics[width=.69\linewidth]{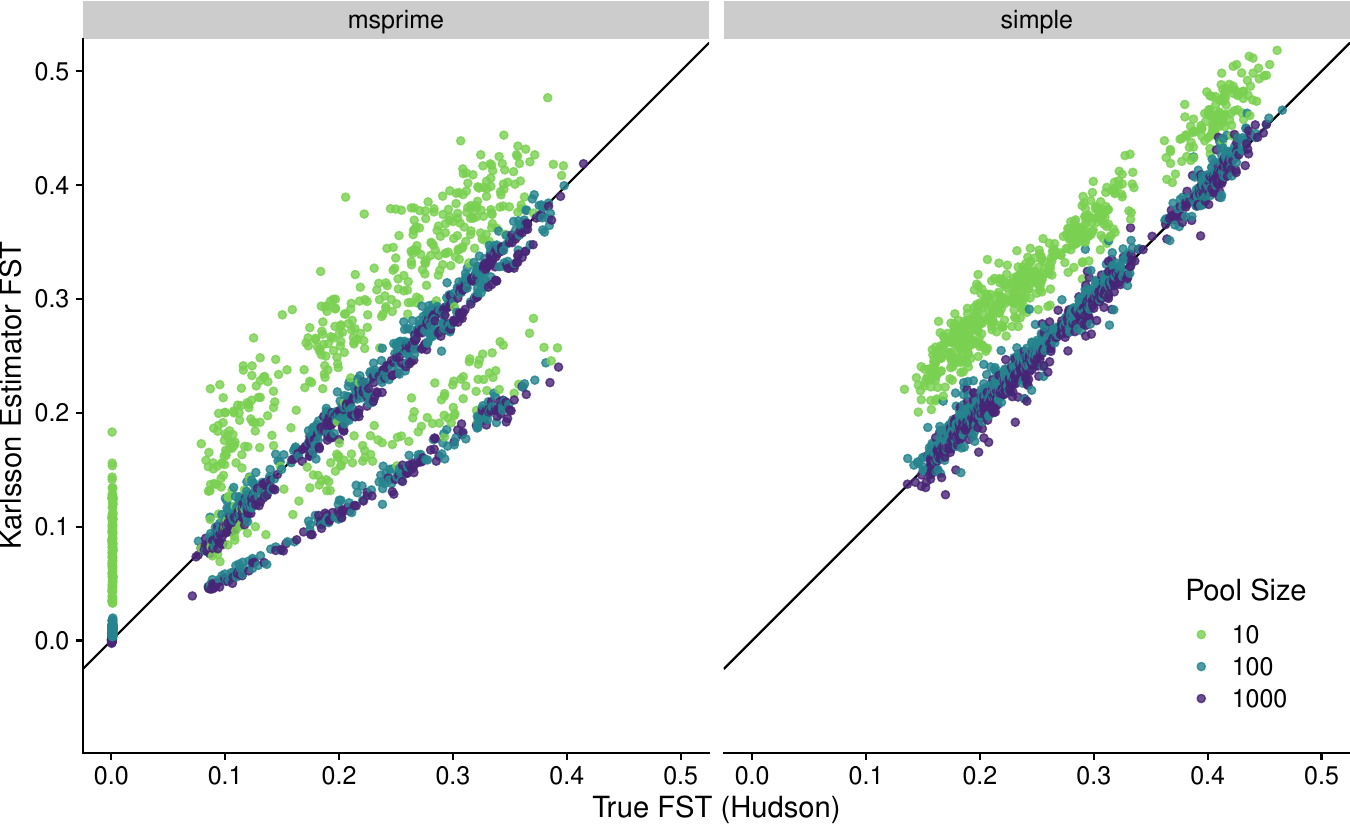}
    \includegraphics[width=.69\linewidth]{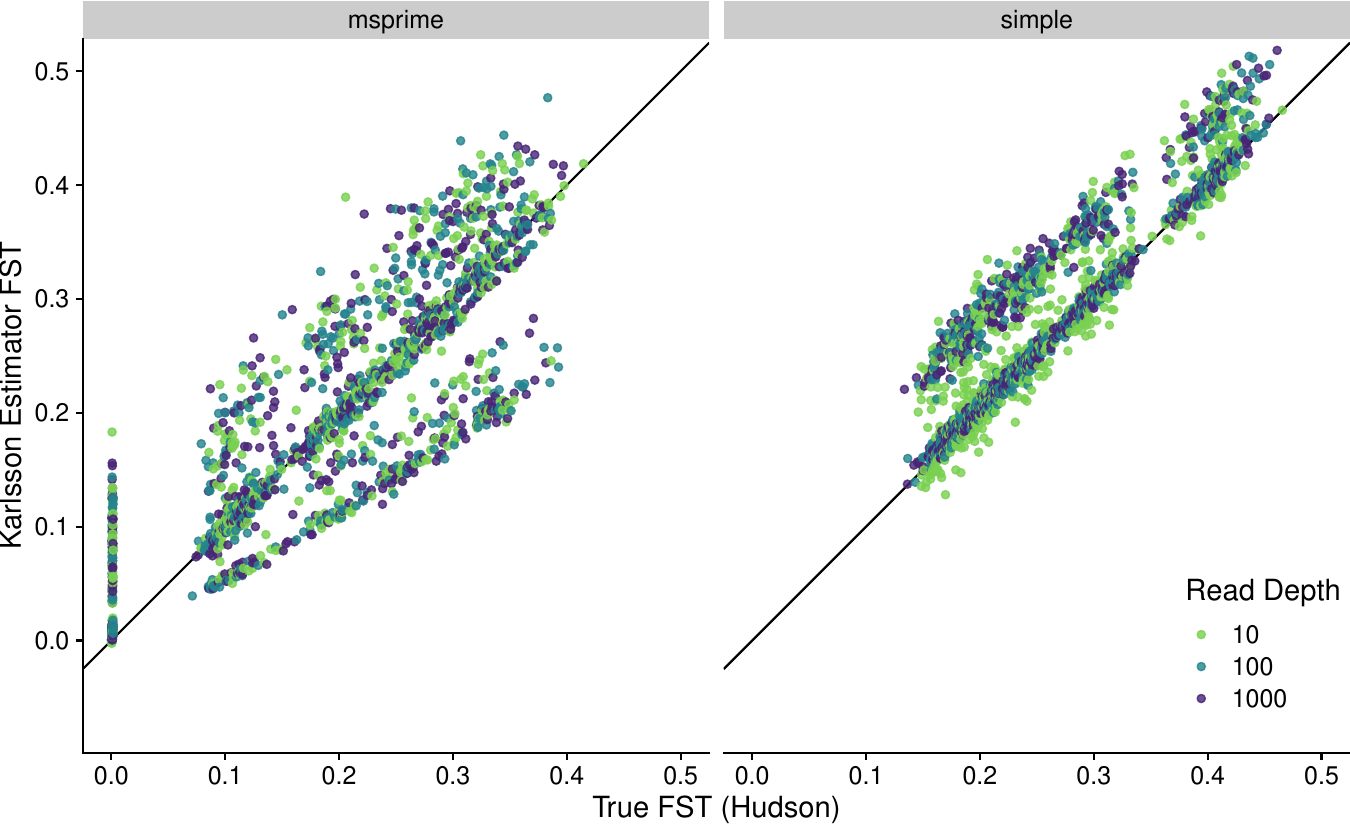}
    \includegraphics[width=.69\linewidth]{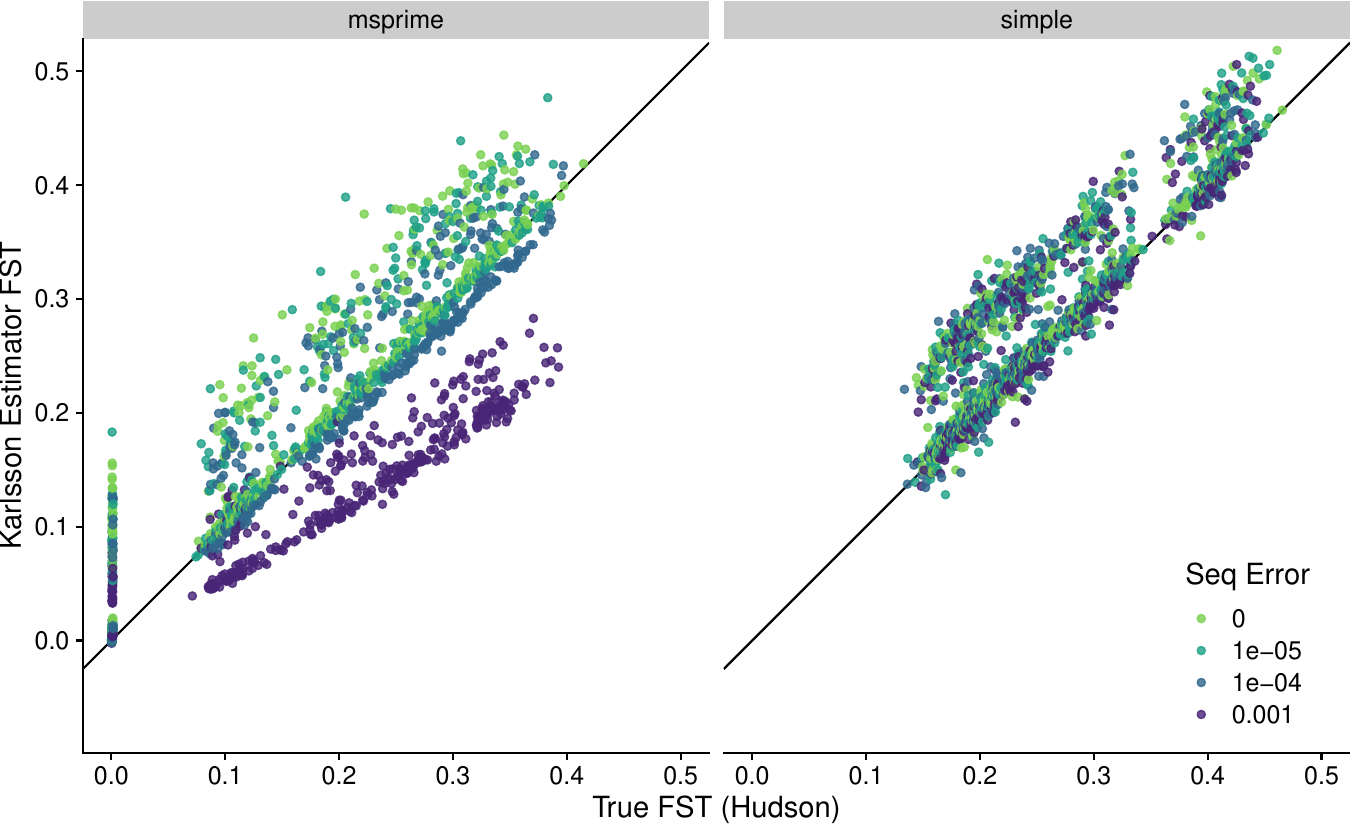}
    \vspace*{-1em}
    \caption{
        Karlsson Estimator for FST vs true FST (Hudson).
    }
\label{fig:Karlsson}
\end{figure*}

We show the results of the simulations in Figures~\ref{fig:UnbiasedHudson}-\ref{fig:Karlsson}. Each Figure shows the comparison of the estimated values versus the true values, for \texttt{msprime} on the left, and for the "simple" simulation on the right. Each plot is repeated three times, colorized by Pool Size, Read Depth, and Sequencing Error, respectively, in order to show the effects of each of these parameters on the estimation.

We find that the estimators we derived in Section~\ref{supp:sec:FST:sub:EstimatorFST} are approximately unbiased across all regimes except for the \texttt{msprime} simulations with the highest sequencing error rate, see \figref{fig:UnbiasedHudson} and \figref{fig:UnbiasedNei}. In particular, it can be observed that for larger pool sizes, the variance of the estimators reduces, as expected, as well as, to some degree, for higher read depths.

In contrast, in \figref{fig:Kofler} we find that the Kofler estimator of $\neifst$ is highly biased in a way that depends on pool size and average read depth.  This makes sense because as discussed above, the Kofler estimate of $\pi_\text{between}$ includes a Bessel correction when it should not.  That Bessel correction contains terms that depend on both the pool size and the the read depth, explaining the pool size and sample size-dependent bias.

Similarly, in \figref{fig:Karlsson} we find that the Karlsson estimator of $\hudsonfst$ is highly biased for small pool sizes, but otherwise performs well (again except for the case of high sequencing error).  Again, this makes sense as the Karlsson estimator implicitly assumes an infinite pool size, and so will be biased for small pool sizes.

For the \texttt{msprime} simulations with the highest sequencing error rate, we see that all methods severely underestimate \fst.  It makes sense that the methods should be biased that way, as none of the estimators account for sequencing error. In the \texttt{msprime} simulations, most sites are not segregating in the population.  Sites that are not segregating do not contribute to the true value of \fst{} --- they act as a $0$ added to each of the $\pi$'s. Calling the rate of sequencing error~$\epsilon$, all of the above estimators add something $\approx \epsilon$ over the number of sites to each of the $\widehat{\pi}$'s in expectation.  As a result, for $\neiestimator := 1 - \sfrac{\widehat{\pi}_\text{within}}{\widehat{\pi}_\text{total}}$, for example, we add something on the order of $\epsilon$ to our estimates of $\widehat{\pi}_\text{within}$ and $\widehat{\pi}_\text{total}$.  This inflates $\sfrac{\widehat{\pi}_\text{within}}{\widehat{\pi}_\text{total}}$ causing us to underestimate $\neifst$.  

A similar phenomenon occurs for $\hudsonestimator$, as well as the other estimators.  To see that invariant sites add $\approx\epsilon$ over the total number of sites to any of the $\widehat{\pi}$'s, consider an asymptotic regime where both the read depth and the pool size are large.  In this regime, if there were no sequencer error, the empirical frequency of each allele across reads is a very accurate estimate of the the true population allele frequency.  In particular, for invariant sites, all of the reads will show the ``0'' allele across both populations.  If we add sequencing error, then approximately $\epsilon$ proportion of the reads will show the ``1'' allele in each population.  Our estimators as introduced above ignore sequencing error, and so in this case, it would look like both populations have the ``1'' allele at frequency $\epsilon$.  Naively plugging these frequencies into $\pi_\text{within}$, $\pi_\text{between}$, or $\pi_\text{total}$ we obtain $\epsilon(1-\epsilon)\approx \epsilon$.  In practice, at invariant sites, the different estimators would include various Bessel corrections, but the intuition holds, and these sites contribute approximately $\epsilon$ to the overall estimates of the different $\pi$'s.  As a result, if $\epsilon$ is comparable to or larger than any of the $\pi$'s, then any of the  estimators will be significantly biased.  Indeed, in the \texttt{msprime} simulations, we would expect $\pi_\text{within}$ to be approximately $\theta$, which is $0.001$, matching the highest sequencing error simulations, and explaining the apparent bias.  In contrast, in the ``simple'' simulations, $\pi_\text{within}$ is approximately $0.03$ which is an order of magnitude larger than the highest sequencing error simulated and hence the bias is negligible.



\subsection{Unbiased estimates of heterozygosity in the presence of sequencing error}
\label{supp:sec:FST:sub:HeterozygositySequencingError}

In the following, we present derivations of the above estimators of \fst{} that also correct for sequencing errors. This is currently not implemented in our software, and in no other software that we are aware of. Our rationale is that many applications are based on short-read Illumina sequencing, which has an error of ~1/1000. We however present it here for completeness, in the hope that this is useful for others, especially if other types of next generation sequencing with higher error rates are used.

\todo{check whether this or something similar is implemented in poolfstat! it might be important to mention this here then!}

\todo{Jeff, do you think it would be worth showing the above bias plots here again, with the seq error correction? it might mean that we have to implement the equations though... not sure if needed now. JPS: I think that sequencing error rates are high enough and pi tends to be low enough that they could actually mess up Fst estimates in practice.  People would probably appreciate a corrected estimate of Fst, but it's also probably not super crucial.  I guess it depends on how hard it is to implement. LC: it's kind of a lot to implement... instead of just keeping counts or frequencies at each position, we'd have to keep track of each base with its respective quality score throughout the whole process... a lot of code refactoring! the average one could be done more easily though, I think.}


\subsubsection*{Overview}
\label{supp:sec:FST:sub:HeterozygositySequencingError:sub:Overview}

We will derive estimators for \piw, \pib, and \pit in the presence of sequencing error.  In the simulations above, and by a hand-wavy theoretical argument, we saw that if sequencing error is on the order of \piw, \pib, or \pit, then our estimates of $F_\text{ST}$ can be severely biased.  We consider two cases: one where we have access to read-level estimates of sequencing error (e.g., phred scores), and one where we just have access to an estimate of the average sequencing error rate.  Since the second is a special case of the first (i.e., we can just assume all reads have the same sequencing error), it will follow straightforwardly but will simplify some equations.

One interesting and important observation to note here is that when we did not account for sequencing error, our estimates of \piw, \pib, and \pit were all zero for monomorphic (non-segregating) sites.  This allowed us to sum the estimates of \piw, \pib, and \pit across only segregating sites and then divide by the total number of sites (both segregating and non-segregating).  We will see that once we account for sequencing error, monomorphic sites actually contribute a small amount to \piw, \pib, and \pit, as there will now be an \emph{additive} bias correction in addition to a multiplicative one.  As such, when computing \piw, \pib, or \pit in a window, we will need to sum across segregating sites and then add a term that is the number of non-segregating sites times the estimate of \piw, \pib, or \pit for a non-segregating sites, and then finally divide that whole expression by the total number of sites.

We will start with our general model, derive our estimators from there, and then show that they are unbiased.  


\subsubsection*{Model}
\label{supp:sec:FST:sub:HeterozygositySequencingError:sub:Model}

We will follow the notation from above where applicable, but briefly reintroduce some terms for ease of reading.  For generality, we will assume that there are $K$ possible alleles; this will allow us to switch between biallelic model and models on $\left\{A, C, G, T\right\}$.  We let $f_\tau$ be the frequency of allele $\tau$ \emph{in the population}.  Then we let $N_\tau$ be the (unobserved) number of individuals in the pool of $n$ individuals with allele $\tau$.  Finally, let there be $c$ reads at a site, and let $R_j$ be the allele of the individual sequenced on the $j^\text{th}$ read, and let $\hat{R}_j$ be the actual output of the sequencer for the $j^\text{th}$ read.  That is, if $R_j = \tau$, then the $j^\text{th}$ read comes from an individual that has the $\tau$ allele, but then once we sequence that fragment, there is a small chance that an error is introduced, so it is very likely that $\hat{R}_j$ --- the allele output by the sequencer --- is $\tau$ if $R_j$ is $\tau$, but there is some chance that the sequencer reads a different allele and then $\hat{R}_j = \tau' \ne \tau$.  In the most general case, we will assume that $\hat{R}_j$ matches $R_j$ with probability $1 - \epsilon_j$ and randomly chooses an allele other than $R_j$ with probability $\epsilon_j$ and chooses among the options uniformly at random.  For example, if we are looking at  $\left\{A, C, G, T\right\}$ and the sequencing error rate of read $j$ is $\epsilon_j$, then if $R_j = A$, we have that $\hat{R}_j = A$ with probability $1 - \epsilon_j$ and $\hat{R}_j = C$ with probability $\epsilon_j/3$, and $\hat{R}_j = G$ with probability $\epsilon_j/3$, and $\hat{R}_j = T$ with probability $\epsilon_j/3$.

Throughout we will use the following results for $j\ne j'$:
\begin{align}
\mathbb{E}\left[\mathbb{I}\left\{\hat{R}_j = \tau\right\}\right] &= \frac{\epsilon_j}{K-1} + \left(1 - \frac{K}{K-1} \epsilon_j\right)f_\tau \label{eq:first_moment}\\
\mathbb{E}\left[\mathbb{I}\left\{\hat{R}_j = \tau\right\}\mathbb{I}\left\{\hat{R}_{j'} = \tau\right\}\right] &= \frac{\epsilon_j\epsilon_{j'}}{(K-1)^2} \nonumber \\&\hspace{0.4cm}+ \frac{1}{n}\left((1-\epsilon_j)(1-\epsilon_{j'}) + \frac{n-1}{K-1}(\epsilon_j + \epsilon_{j'} - 2\epsilon_j\epsilon_{j'}) + \frac{1-2n}{(K-1)^2}\epsilon_j\epsilon_{j'}\right) f_\tau \label{eq:second_moment} \\&\hspace{0.4cm}+ \frac{n-1}{n}\left((1-\epsilon_j)(1-\epsilon_{j'}) - \frac{1}{K-1}(\epsilon_j + \epsilon_{j'} - 2\epsilon_j\epsilon_{j'}) + \frac{1}{(K-1)^2}\epsilon_j\epsilon_{j'})\right) f_\tau^2 \nonumber
\end{align}

Before going any further, these look a bit complex, so let's do a quick sanity check.  If $\epsilon_j = 0$, then we get
\begin{align*}
\mathbb{E}_{\epsilon=0}\left[\mathbb{I}\left\{\hat{R}_j = \tau\right\}\right] &= f_\tau\\
\mathbb{E}_{\epsilon=0}\left[\mathbb{I}\left\{\hat{R}_j = \tau\right\}\mathbb{I}\left\{\hat{R}_{j'} = \tau\right\}\right] &= \frac{1}{n}f_\tau + \frac{n-1}{n} f_\tau^2,
\end{align*}
which it turns out matches what we would get from looking at reads in the model with no error in the equations as described in the sections above.  Above, we considered the empirical frequency $\widehat{f}_\tau := \frac{1}{c}\sum_{j=1}^c \mathbb{I}\left\{\hat{R}_j = \tau\right\}$, and saw that if there is no sequencing error $\mathbb{E}\left[\widehat{f}_\tau(1-\widehat{f}_\tau)\right] = \frac{n-1}{n}\frac{c-1}{c}f_\tau(1-f_\tau)$.  We leave it as a straightforward exercise to see that the above (when all $\epsilon_j$ are zero) implies the same expectation for $\widehat{f}_\tau(1-\widehat{f}_\tau)$.

To see equation~\eqref{eq:first_moment}, note that by the tower property (twice),
\begin{align*}
\mathbb{E}\left[\mathbb{I}\left\{\hat{R}_j = \tau\right\}\right] &= \mathbb{E}\mathbb{E}\left[\mathbb{I}\left\{\hat{R}_j = \tau\right\} | R_j\right]\\
&= \mathbb{E}\left[(1-\epsilon_j) \mathbb{I}\left\{R_j = \tau \right\}\right] + \mathbb{E}\left[\frac{\epsilon_j}{K-1}\mathbb{I}\left\{R_j \ne \tau \right\}\right]\\
&= (1-\epsilon_j)\mathbb{E}\mathbb{E}\left[\mathbb{I}\left\{R_j = \tau \right\} | N_\tau \right] + \frac{\epsilon_j}{K-1}\mathbb{E}\mathbb{E}\left[\mathbb{I}\left\{R_j \ne \tau \right\} | N_\tau \right]\\
&= (1-\epsilon_j) \mathbb{E}\left[\frac{N_\tau}{n}\right] + \frac{\epsilon_j}{K-1}\mathbb{E}\left[\frac{n-N_\tau}{n}\right]\\
&= (1-\epsilon_j) f_\tau + \frac{\epsilon_j}{K-1}(1- f_\tau)\\
&= \frac{\epsilon_j}{K-1} + \left(1 - \frac{K}{K-1}\epsilon_j\right)f_\tau
\end{align*}

Equation~\eqref{eq:second_moment} follows the same reasoning but with substantially more algebra, so we omit it here for simplicity.

One last useful fact is that if we sum equations~\eqref{eq:first_moment}~and~\eqref{eq:second_moment} over $\tau$, we obtain something slightly nicer, since $\sum_\tau f_\tau =1$.  With this summation and some slight algebraic manipulation we see:
\begin{align}
\sum_{\tau}\mathbb{E}\left[\mathbb{I}\left\{\hat{R}_j = \tau\right\}\right] &= 1 \label{eq:first_moment_sum}\\
\sum_{\tau}\mathbb{E}\left[\mathbb{I}\left\{\hat{R}_j = \tau\right\}\mathbb{I}\left\{\hat{R}_{j'} = \tau\right\}\right] &= \frac{1}{n}\left((1-\epsilon_j)(1-\epsilon_{j'}) + \frac{n-1}{K-1}(\epsilon_j + \epsilon_{j'} - 2\epsilon_j\epsilon_{j'}) + \frac{(K-2)n + 1}{(K-1)^2}\epsilon_j\epsilon_{j'}\right)  \label{eq:second_moment_sum} \\&\hspace{0.4cm}+ \frac{n-1}{n}\left((1-\epsilon_j)(1-\epsilon_{j'}) - \frac{1}{K-1}(\epsilon_j + \epsilon_{j'} - 2\epsilon_j\epsilon_{j'}) + \frac{1}{(K-1)^2}\epsilon_j\epsilon_{j'})\right)  \nonumber \\ &\hspace{1cm}\times \sum_\tau f_\tau^2 \nonumber
\end{align}

With these preliminaries out of the way, we can get to our estimators.


\subsubsection*{Unbiased estimator for \piw}
\label{supp:sec:FST:sub:HeterozygositySequencingError:sub:UnbiasedPIW}

Recall that \piw is defined as $\frac{1}{2}\left(\pi_{(1)} + \pi_{(2)}\right)$, where $\pi_{(p)}$ is the heterozygosity in population $p$.  Since this is linear, we just need an unbiased estimator for the heterozygosity within one population, and then we can average these estimators across the populations to obtain an estimator of \piw.

Our estimator is a bit unwieldy, so first we define a number of additional variables to clean up the notation.  Note that all of these only depend upon observable quantities (i.e., the sample size, read depth, and error rates) so they can be readily computed from the observed data.
\begin{align*}
\overline{\epsilon} &:= \frac{1}{c} \sum_{j=1}^c \epsilon_j\\
\overline{\epsilon^2} &:= \frac{1}{c} \sum_{j=1}^c \epsilon_j^2\\
\epsilon^* &:= (\overline{\epsilon})^2 - \frac{\overline{\epsilon^2}}{c}\\
w_1 &:= \frac{1}{c} + \frac{1}{n}\left(\frac{c-1}{c} + 2\left(\frac{c-1}{c}\right)\left(\frac{n-K}{K-1}\right)\overline{\epsilon} + \left(1 - 2\left(\frac{n-1}{K-1}\right) + \frac{(K-2)n + 1}{(K-1)^2}\right)\epsilon^*\right)\\
w_2 &:= \frac{n-1}{n}\left(\frac{c-1}{c} - 2\left(\frac{c-1}{c}\right)\left(\frac{K}{K-1}\right)\overline{\epsilon} + \left(1 + \frac{2}{K-1} + \frac{1}{(K-1)^2}\right)\epsilon^*\right)
\end{align*}

We are now ready to define our unbiased estimator of $\pi$ within a given population.  Let $\hat{f}_\tau = \sum_{j=1}^c \mathbb{I}\left\{\hat{R}_j = \tau\right\}$ be the empirical frequency of allele $\tau$ in the observed reads.
\[
\widehat{\pi} := \frac{w_1 + w_2 - \sum_\tau \hat{f}_\tau^2}{w_2}.
\]

It is a bit of work to show, but it turns out that $\mathbb{E}[\widehat{\pi}] = \pi$. Hence, this estimator is an unbiased estimator for the heterozygosity within a population.  The overall proof is to show that
\begin{align}
\sum_\tau\mathbb{E}\left[\hat{f}_\tau^2\right] &= w_1 + w_2 \sum_\tau f_\tau^2, \label{eq:lemma}
\end{align}
which we will show below, but for now we assume this being true.  Then, we can use this formula to see
\begin{align*}
\mathbb{E}\left[\widehat{\pi}\right] = \frac{w_1 + w_2 - \sum_\tau \mathbb{E}\left[\hat{f}_\tau^2\right]}{w_2} = 1 -\sum_\tau f_\tau^2 =: \pi
\end{align*}
as required.

Now to see equation~\eqref{eq:lemma}, we write
\begin{align*}
\sum_\tau\mathbb{E}\left[\hat{f}_\tau^2\right] &= \frac{1}{c^2}\sum_{j=1}^c \sum_{j'=1}^c \sum_\tau \mathbb{E}\left[\mathbb{I}\left\{\hat{R}_j=\tau\right\}\mathbb{I}\left\{\hat{R}_{j'}=\tau\right\}\right]\\
&= \frac{1}{c^2}\left\{\left(\sum_{j=1}^c \mathbb{E}\left[\mathbb{I}\left\{\hat{R}_j=\tau\right\}\right]\right) + \left( \sum_{j\ne j'}\mathbb{E}\left[\mathbb{I}\left\{\hat{R}_j=\tau\right\}\mathbb{I}\left\{\hat{R}_{j'}=\tau\right\}\right] \right) \right\}\\
&= w_1 + w_2\sum_\tau f_\tau^2
\end{align*}
where the final line follows from using Equations~\eqref{eq:first_moment_sum}~and~\eqref{eq:second_moment_sum} and then doing quite a bit of algebra.


\subsubsection*{Unbiased estimator for \pib}
\label{supp:sec:FST:sub:HeterozygositySequencingError:sub:UnbiasedPIB}

Fortunately, the estimator for \pib is much simpler.  From equation~\eqref{eq:first_moment}, we see
\begin{align*}
\mathbb{E}\left[\hat{f}_\tau\right] &= \frac{1}{c}\sum_{j=1}^c \mathbb{E}\left[\mathbb{I}\left\{\hat{R}_j = \tau\right\}\right]\\
&= \frac{1}{c}\sum_{j=1}^c \frac{\epsilon_j}{K-1} + \left(1 - \frac{K}{K-1}\epsilon_j\right)f_\tau\\
&= \frac{\overline{\epsilon}}{K-1} + \left(1 - \frac{K}{K-1}\overline{\epsilon}\right)f_\tau
\end{align*}
Rearranging, this immediately implies that
\[
\frac{\hat{f}_\tau - \frac{\overline{\epsilon}}{K-1}}{1 - \frac{K}{K-1}\overline{\epsilon}}
\]
is unbiased for $f_\tau$.  So far we have suppressed the dependence of all of these things on the population (i.e., we've just been looking at a single sample), but for \pib we must consider two samples.  We will use a $(p)$ superscript to denote the population.  Our estimator for \pib is then
\[
\widehat{\pi}_\text{between} := 1 - \sum_\tau \left(\frac{\hat{f}^{(1)}_\tau - \frac{\overline{\epsilon}^{(1)}}{K-1}}{1 - \frac{K}{K-1}\overline{\epsilon}^{(1)}}\right)\left(\frac{\hat{f}^{(2)}_\tau - \frac{\overline{\epsilon}^{(2)}}{K-1}}{1 - \frac{K}{K-1}\overline{\epsilon}^{(2)}}\right)
\]
That this is unbiased follows immediately from the unbiasedness of our estimators of $f_\tau$ within each population and the independence of the two samples. 


\subsubsection*{Unbiased estimator for \pit}
\label{supp:sec:FST:sub:HeterozygositySequencingError:sub:UnbiasedPIT}

Recall that $\mathpit = \frac{1}{2}\left(\mathpiw + \mathpib\right)$ so with our unbiased estimators of \piw and \pib, we immediately by the linearity of expectation obtain that:
\[
\widehat{\pi}_\text{total} := \frac{1}{2}\left(\widehat{\pi}_\text{within} + \widehat{\pi}_\text{between}\right)
\]
is unbiased for \pit.


\subsubsection*{Constant error rates}
\label{supp:sec:FST:sub:HeterozygositySequencingError:sub:ConstErr}

If we assume that all reads have the same error rates, say $\epsilon$, then $\overline{\epsilon} = \epsilon$, $\overline{\epsilon^2} = \epsilon^2$, and $\epsilon^* = \frac{c-1}{c}\epsilon^2$.  Otherwise, the equations remain unchanged.


\section{PoPoolation Equations Document}
\label{supp:sec:PoPoolation2Equations}

The PoPoolations equation document also presents some simplifications and related equations that to the best of our knowledge are not implemented in their software.
We hence do not go through them in detail here, but still want to mention them, in case they might be useful for others.

\begin{itemize}
  \item They present simplified versions of $\theta_\pi$, $\theta_w$, and Tajima's D, which assume that allele frequency distribution in the reads is about the same as in the real population, and hence arrives at a simpler computation at the cost of some error. These are also useful for individual sequencing.
  \item As mentioned above in \secref{supp:sec:TajimaD:sub:PoolSequencingCorrection}, the document presents an approach to computing Tajima's D based on its variance, and extends this to windows, but (to the best of our knowledge) does not implement this, and instead implement their approach based on \citeay{Achaz2008}.
  \item They present an approach for computing \fst{} for $J$ pool-sequenced populations (instead of just two as presented above), extend this approach to large regions as well as single SNPs, and introduce weights that take the number of sequenced individuals in each population into account. More work is needed to compare this approach to their implementation and to our estimators.
\end{itemize}

These alternative approaches however need further assessment and comparison to the other approaches presented here.

\popoolissue{Lastly, we wanted to note that to us it seems that some equations in the PoPoolation equations document are not actually implemented in the code, and that the code contains compuations that are not in the document. We hence think that those were never intended to go hand in hand, and that hence the equations document is also not part of the official publication, but merely found in the code repository. Is that assessment correct, or are there other reasons for the divergence between the two?}

\bibliographystyle{natbib}

\bibliography{02-equations}




\begingroup
\let\center\flushleft
\let\endcenter\endflushleft
\maketitle
\endgroup
\vspace*{-3.5em}

\textsuperscript{a}{Department of Plant Biology, Carnegie Institution for Science, Stanford, USA.\\}
\textsuperscript{b}{Section for GeoGenetics, Globe Institute, University of Copenhagen, Denmark.\\}
\textsuperscript{c}{Department of Genetics, Stanford University, Stanford, USA.\\}
\textsuperscript{d}{Department of Biology, Stanford University, Stanford, USA.\\}
\textsuperscript{e}{Department of Global Ecology, Carnegie Institution for Science, Stanford, USA.\\}
\textsuperscript{f}{Department of Integrative Biology, University of California Berkeley, Berkeley, USA.\\}
\textsuperscript{g}{Howard Hughes Medical Institute, University of California Berkeley, Berkeley, USA.\\}

\textsuperscript{*}{To whom correspondence should be addressed: \\ \href{mailto:lczech@carnegiescience.edu}{\color{black}{lczech@carnegiescience.edu}},
and \href{mailto:moisesexpositoalonso@gmail.com}{\color{black}{moisesexpositoalonso@gmail.com}}}



\vspace*{1em}

In this supplement, we compare \toolname{grenedalf} to existing software for computing population genetics statistics for Pool-seq data, and elaborate on some of the computational improvements that \toolname{grenedalf} offers.
We evaluate the runtime and memory requirements for some representative and frequently performed tasks
by comparing \toolname{grenedalf} to libraries and tools that have similar scope and functionality.
Note that \toolname{grenedalf} internally uses our software library \toolname{genesis} \cite{Czech2020}, meaning that the benchmarks shown here are applicable to both.


\tableofcontents
\pagebreak


\section{Existing Tools}
\label{supp:sec:SoftwareComparison:sub:Tools}

We compare \toolname{grenedalf} to the following existing tools:

\begin{itemize}
    \item \toolname{PoPoolation} \cite{Kofler2011a} computes the diversity statistics $\theta_\pi$, Watterson's $\theta$, and Tajima’s D, is written in Perl, and takes \formatname{pileup} files as input. Input is traversed in windows of a given interval size, and is streamed, keeping the memory usage low.
    
    \item \toolname{PoPoolation2} \cite{Kofler2011b} computes population differentiation with \fst{}, is also written in Perl, and takes their ad-hoc \formatname{sync} file as input. It also works in interval windows and streams through the data. Note that despite the naming of the tool implying it to be an update to \toolname{PoPoolation}, it is in fact a distinct stand-alone tool.
    
    \item \toolname{poolfstat} \cite{Hivert2018,Gautier2022-zd} computes \fst{} (as well as some other $f$- and $D$-statistics, which we have not yet implemented in \toolname{grenedalf}, and hence ignore for the time being), and is written in C++ and R. Its input format is also \formatname{sync}. Due to using R datatypes, it reads all input into memory at once. It computes \fst{} in windows between pairs of samples, and also has a function to compute \fst{} between multiple samples at once, but only for the whole genome.
    
    \item \toolname{npstat} \cite{Ferretti2013} computes several diversity metrics, and is written in C. It takes \formatname{mpileup} as input, and computes statistics in windows of a fixed number of bases; it can however only compute data from one chromosome at a time.
    
    \item \toolname{samtools} \cite{Li2009b} is a toolkit for working with \formatname{sam}/\formatname{bam} and \formatname{(m)pileup} files. It does not compute any population genetic statics that we are interested in here, but we use it here as a baseline for benchmarking tasks such as file conversions.
\end{itemize}

In \toolname{grenedalf}, we implemented all the file formats that the above tools offer, as well as the considerably more complex \formatname{sam} (and its gzipped form \formatname{sam.gz}), \formatname{bam}, and \formatname{cram} files, which are what typical read mapping tools produce, and for which we internally use \toolname{htslib} \cite{Bonfield2021-wb}. In many cases, this hence eliminates the need for \textit{a priori} file conversions. Furthermore, we offer to read from \formatname{vcf} files \cite{Danecek2011-qk} by utilizing their \formatname{AD} format field (allelic depth, which is relevant when genotypes are heterozygous, polyploid, or multiple genotypes are present in a sample such as in in Pool-seq). Lastly, we can read from a variety of allele frequency table formats, similar to what is produced by \toolname{HAF-pipe} \cite{Tilk2019-jr,Kessner2013-ph} or other allele frequency estimators. To the best of our knowledge however, these file formats are not standardized for Pool-Seq data, and not implemented in any other tool; we hence do not include benchmarks based on these formats in our comparison to other tools here.

Remark on terminology: In order to avoid confusion, we have decided to use the term "read depth" throughout to mean the number of reads that are present at a particular position in the genome. This is opposed to the ambiguous term "coverage", which is instead used by PoPoolation and npstat. However, this might lead to confusion about whether this is meant to indicate "coverage depth" (that is, "read depth") or "coverage breadth". Ideally, we would even want to distinguish the average depth (e.g., 30X "coverage depth"), from the specific "read depth" at a particular locus. As we only have the latter use case here, we have hence decided to simply use "read depth" throughout.


\section{File Formats}
\label{supp:sec:SoftwareComparison:sub:FileFormats}

We here give a brief introduction of the file formats that \toolname{grenedalf} can read (at the time of writing this document), for the reader's convenience.

\begin{itemize}
    \item \formatname{sam}/\formatname{bam}/\formatname{cram} \cite{Li2009b}: Widely-used standard for storing and representing sequence alignment data, i.\,e., reads aligned to a reference genome, with \formatname{sam} being the text-based format and \formatname{bam} being its binary equivalent, and \formatname{cram} being a further compressed variant. Data is stored per-read, meaning that for the computations that we do here, we first need to construct an internal per-position representation based on reads overlapping each position.
    \item \formatname{(m)pileup} \cite{Li2009b}: Representation of the bases at each genomic position, where each line ``piles up'' the bases found across all reads that span that position. A multi-pileup (indicated by the \formatname{m} in the name) contains multiple samples, piled up on the same line in the file. This format hence already contains the data in the orientation that we need it here, but comes at the expense of extra disk space and time needed for the conversion from \formatname{sam}/\formatname{bam}. Thus, it usually does not make sense to convert to pileup when working with \toolname{grenedalf}.
    \item \formatname{sync} \cite{Kofler2011b}: A simple tab-delimited text format, where each line represents a specific genomic position and includes allele counts for multiple populations, for each of the four bases. As it only contains simple counts, it is very efficient to parse, but unfortunately (in its original form) does not support further data annotation, or even sample names. Our implementation in \toolname{grenedalf} supports some (semi-established) \textit{ad-hoc} extensions such as sample names and masked positions, which help with bookkeeping during analyses.
    \item \formatname{vcf} \cite{Danecek2011-qk}: Widely used file format that represents genetic variants per position, such as single nucleotide polymorphisms (SNPs) and insertions/deletions (indels), including their alleles, quality scores, and additional annotations. In pool sequencing, the AD (Allelic Depth) field can be utilized to represent the read depth or allele counts for each allele in a pool of sequenced individuals. We however do not recommend using \formatname{vcf} data coming from variant callers (e.\,g., GATK) directly applied to pool sequencing data, as this likely introduces bias in frequencies \cite{Czech2022}.
    \item Frequency tables: This is not a specific format, but simply any generic table format with positions along the genome in rows, and one or more columns for samples, containing allele frequencies or counts for the given position. Such tables are for example produced by \toolname{HAF-pipe} \cite{Tilk2019-jr,Kessner2013-ph}, and we can read any such format, as long as it contains a header line in a format that we can parse.
\end{itemize}


When using \toolname{grenedalf} to run some simple analyses, we recommend directly running on \formatname{bam} files that come out of the alignment step, such as when using \toolname{grenepipe} \cite{Czech2022-vf}. This is the easiest way, which avoids file conversions. However, when using \toolname{grenedalf} for some more involved analyses, where files have to be read multiple times, we instead recommend converting the \formatname{bam} files to \formatname{sync} files first. They are substantially faster to process, hence leading to overall speed improvements.


\pagebreak

\section{Benchmarks and Comparisons}
\label{supp:sec:SoftwareComparison:sub:Benchmarks}



We benchmarked the runtime and memory requirements of \toolname{grenedalf} in comparison to existing tools for several typical tasks, as well evaluated the resulting data in terms of consistency and compatibility with other tools. The results are shown in the figures below.

Unless noted otherwise, the tests were run on a laptop computer with an 8-core AMD Ryzen 7 Pro processor (with hyper-threading, for a total of 16 hardware threads), and \SI{32}{\giga\byte} of main memory, running Ubuntu 20.04.5 LTS.
We note that \toolname{genesis} and \toolname{grenedalf} offer multi-threaded processing.
They can, for example, read/parse multiple files simultaneously on distinct compute cores and parallelize computations of statistics.
We are not aware that any of the competing libraries and tools evaluated here offer this feature.
Hence, for fairness, we here also ran \toolname{grenedalf} on a single core only where possible for the tool comparisons. Note though that further speedups compared to the ones shown here can be achieved when leveraging the multi-threading, as shown below as well.

Through the comprehensive tests and figures below, we showcase that \toolname{grenedalf} consistently and significantly outperforms {\em all} other tools, is more memory-efficient (or at least not significantly worse), and never uses a prohibitively large amount of memory even for large dataset sizes. As we emphasized, the command line interface for statistics, file manipulations, and utility routines it is considerably more user-friendly than all other software, with comprehensive options for input file formats, filtering, subsetting, windowing approaches, and more. 

In particular, the most salient points of comparison of each of the existing tools are as follows:

\begin{itemize}
    \item \toolname{PoPoolation}/\toolname{PoPoolation2}: Both are generally slow (see, e.\,g., \figref{supp:fig:grenenet-diversity} and \figref{supp:fig:grenenet-fst-positions}), and do not offer a whole-genome mode; they would need to read the whole input into memory (\figref{supp:fig:random-fst-snps-genome}). Both tools only work with a single input file format (\formatname{pileup}, and \formatname{sync}, respectively), both of which need prior conversion and, in case of \formatname{pileup}, take up large amounts of disk space.
    \item \toolname{Poolfstat}: Reads the whole input into memory, which is prohibitive for large datasets (\figref{supp:fig:grenenet-fst-positions}). It does not seem to report per-SNP values with their respective positions, making it hard to track where each \fst{} value comes from, and does seem to be able to compute pairwise \fst{} in windows along the genome.
    \item \toolname{npstat}: Also needs the large \formatname{pileup} format as input, but only processes the first chromosome of the input (\figref{supp:fig:grenenet-diversity}), and produces invalid results afterwards. It pre-computes the pool-sequencing correction denominators for \emph{all} read depths up to the given maximum, instead of only computing the values based on what is actually observed in the data. This makes it prohibitively slow when working with high depth data. It however computes several other population genetic statistics as well that we have not yet implemented in \toolname{grenedalf}.
\end{itemize}

\begin{figure}
    \includegraphics[width=\textwidth]{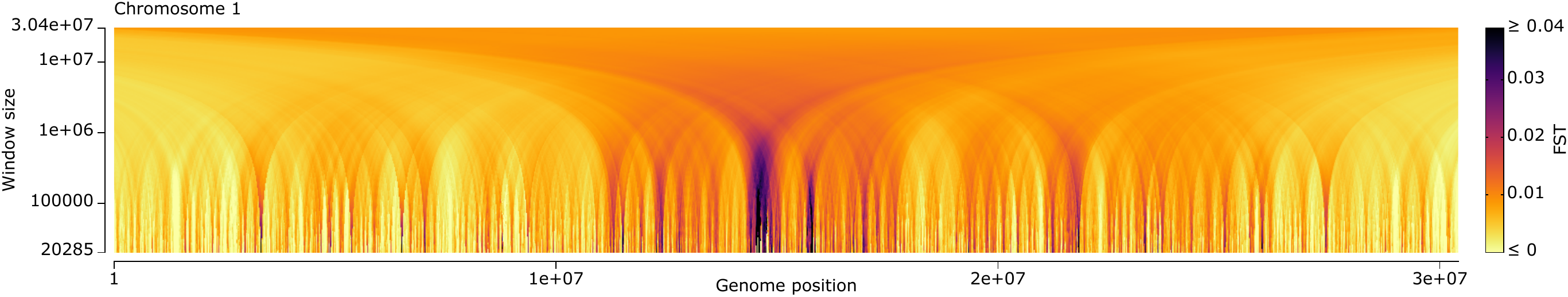}
    \centering
\caption{
    \textbf{Cathedral plot of \fst{} between two pool sequencing samples of \textit{A. thaliana}.}
    \\
    The faster speed of \toolname{grenedalf} allows us to create plots that are computationally more involved. Here, we created what we call a ``cathedral plot'' of \fst{} on chromosome 1 between two pool sequencing samples of \textit{A. thaliana}, which reveals the differentiation structure between two populations across different scales, with darker regions having a higher \fst{} between the two populations.
    \\
    The x-axis corresponds to chromosome 1 of \textit{A. thaliana}, which is $\approx$30mio bases long. The y-axis position (rows of pixels) determines the window size to be used for that row. From top to bottom, windows get smaller, with an exponentially decaying size. The top row corresponds to windows of the full length of the chromosome, and towards the bottom, windows get smaller until their size matches the resolution of the image. Here, we plotted the image with a width of 1500 pixels, so that each pixel in the bottom row represents $\approx$20k positions in the genome. For a given image width, this is the highest resolution that can be displayed; at this point, each window is exactly as wide as the pixel it corresponds to, and windows between pixels in the row are not overlapping with each other any more.
    Windows are centered around their pixel, meaning that towards the left and right of the plot (and in particular towards the top of the image, with larger window sizes), the actual number of positions within a window is truncated. 
    \\
    In other words, for each pixel, the plot involves accumulating values in a window centered on that pixel, with the width of the window determined by the row in the image. Each pixel hence corresponds to a distinct window width and position, and shows the \fst{} value in that window. Our implementation computes intermediate per-SNP values once; for \fst{} for example, we compute $\pi_\text{within}$, $\pi_\text{between}$, and $\pi_\text{total}$ for each SNP, as explained in our equations document. Per window, these values are then accumulated and the resulting \fst{} is computed.
    \\
    The plot hence ``zooms in'' from larger to smaller windows, with increasing resolution along the genome towards the bottom. This can help to visualize the scale of a statistic across different window sizes, showing its broad and fine structure, and allows picking appropriate window sizes for further analyses.
    Note that we chose exponential decay for the window size, as it offers a good balance between revealing both broad structures and fine details. This results in the y-axis being log-scaled. Other functions for the window are however also possible.
    Of course, similar types of plots can also be useful for other per-window statistics, such as diversity.
}
\label{supp:fig:cathedral-plot}
\end{figure}
\begin{figure}
    \includegraphics[width=0.7\textwidth]{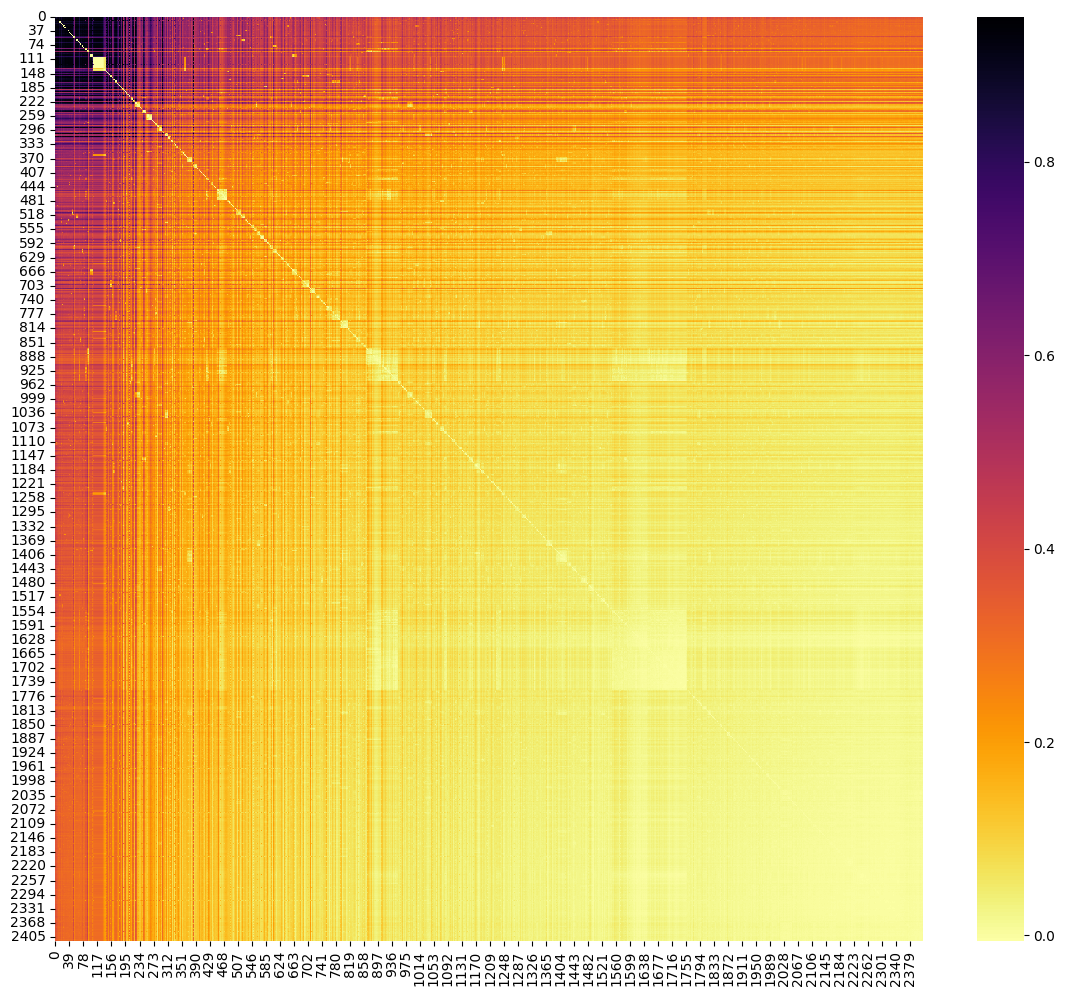}
    \centering
\caption{
    \textbf{Pairwise whole-genome \fst{} matrix between \SI{2415} pool sequencing samples.}
    \\
    Here, we computed a pairwise \fst{} matrix between \SI{2415} whole-genome Pool-Sequencing samples of \textit{A. thaliana} from our GrENE-net project, a large-scale Evolve\&Resequence project (data not published yet).
    The input data here consists of \SI{136}{\giga\byte} of frequency tables estimated with \toolname{HAF-pipe} \cite{Tilk2019-jr}, which for ease of processing were converted into \SI{12}{\giga\byte} of compressed \formatname{sync.gz} files. The files cover all five chromosomes of \textit{A. thaliana}, and contain a total of $2{,}948{,}475$ variant positions.
    \\
    For all $2{,}914{,}905$ pairs of samples ($2415 \cdot 2414 \div 2$), we then computed whole-genome \fst{}, across all variant positions, for a total of ${\approx}8.5$ trillion single-SNP \fst{} computations. We here show the resulting values as a symmetrical heat map; the rows and columns are sorted using average linkage clustering, to make it visually more appealing. We do not further analyse this data here, but simply show it as a use case of the efficiency of \toolname{grenedalf}.
    \\
    In total, the matrix took 16h 20min to compute, and used about \SI{12}{\giga\byte} of memory (mostly due to data buffering to increase efficiency, but this can be reduced if needed). We ran this on a computer cluster, as this is where the data was stored, using 8 threads on the compute node. We further used this test case for measuring thread scaling performance, see \figref{supp:fig:scaling-fst-matrix}. 
    \\
    We also tried to get \toolname{PoPoolation2} to run on this dataset (in single SNP mode though, as it would otherwise need a prohibitive amount of memory, due to its caching of all data in a window), but it failed to produce any output even after several hours of waiting. From extrapolation on subsets, we however estimate that it would take about 10 years of computation for it to produce this matrix. If it were to be run with a window size covering the whole genome, in order to obtain a single \fst{} value per pair of samples as shown here, it would require about \SI{400}{\tera\byte} of memory.
}
\label{supp:fig:pairwise-fst-matrix}
\end{figure}
\begin{figure}
    \begin{subfigure}[b]{0.49\textwidth}
    \includegraphics[width=\textwidth]{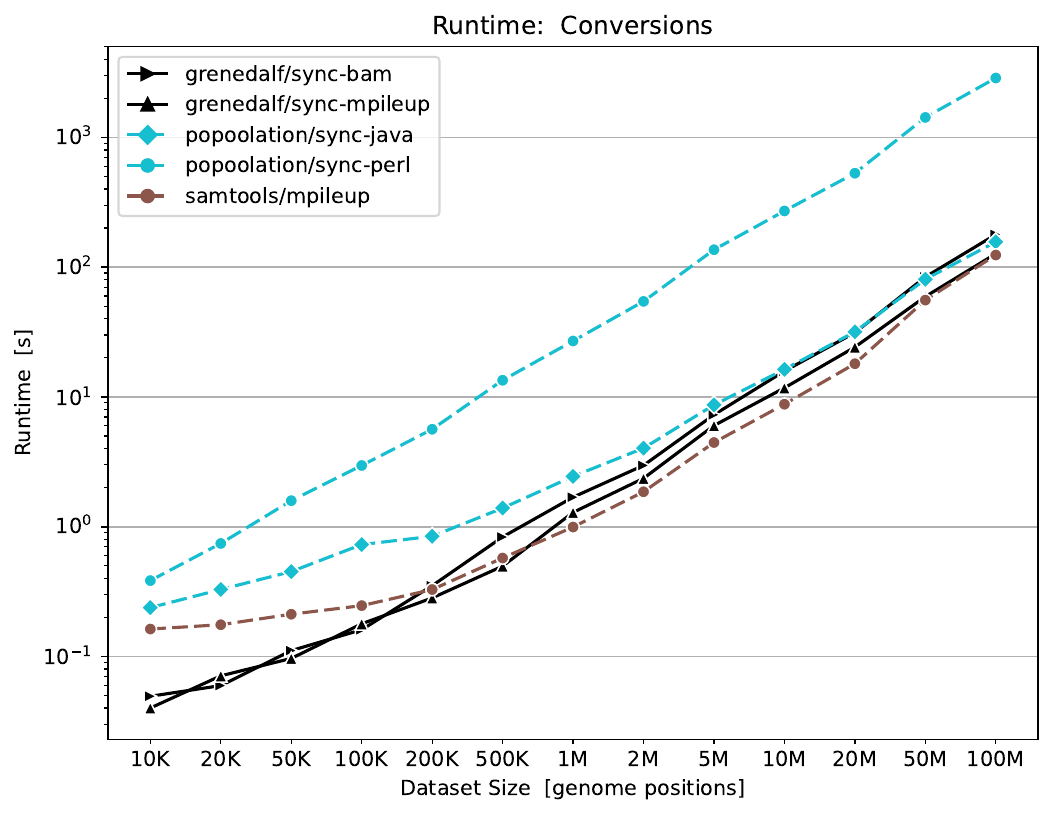}
    \centering
    \end{subfigure}
    \begin{subfigure}[b]{0.49\textwidth}
    \includegraphics[width=\textwidth]{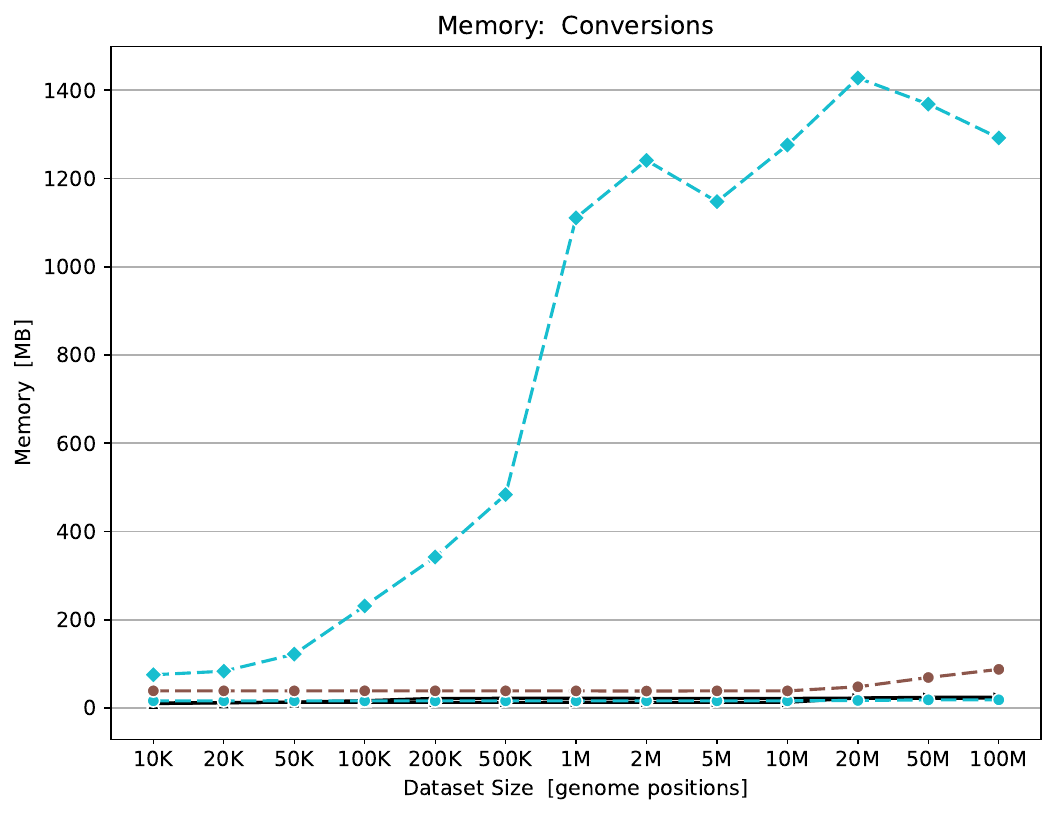}
    \centering
    \end{subfigure}
\caption{
    \textbf{Runtimes (log-log scale) and memory usage (linear–log scale) for converting input \formatname{bam} files to the intermediate file formats required by existing tools, for dataset sizes of \SI{10}{K} to \SI{100}{M} genomic positions, on real-world data.}
    \\
    Here, we show the computational requirements for the two-step process from \formatname{bam} via \formatname{(m)pileup} to \formatname{sync}, which is typically needed when using \toolname{PoPoolation2} (for \fst{}) or \toolname{poolfstat}; \toolname{PoPoolation} (for diversity estimates) only requires the former, i.\,e., \formatname{pileup} files. The processes typically uses \toolname{samtools} for the first, and one of two variants of the \toolname{PoPoolation} sync converter for the second step (written in Java, and Perl, respectively). We also show our implementations in \toolname{grenedalf}, which can produce \formatname{sync} files directly from \formatname{sam}/\formatname{bam} files (without the intermediate \formatname{pileup} step), and as well from \formatname{(m)pileup} files. In summary, the plot shows all tools that produce a \formatname{sync} file from either \formatname{bam} or \formatname{pileup} input.
    \\
    The input files consist of subsets of real Pool-Seq samples from our GrENE-net experiment, with the largest \formatname{bam} file (containing \SI{100}{M} genome positions) being about \SI{800}{\mega\byte} in size, its corresponding \formatname{pileup} file being \SI{4.0}{\giga\byte}, and the resulting \formatname{sync} file being \SI{2.4}{\giga\byte}. Note that file sizes of formats with per-read information such as \formatname{sam}/\formatname{bam} and \formatname{(m)pileup} depend on the number of reads, and are hence determined by genome size and read depth, while accumulated formats such as the \formatname{sync} format sum up nucleotide counts per position, meaning that their size only depends on genome size.
    \\
    These file conversions add overhead to the analysis, in terms of user effort for running the conversions and the associated file management, as well as the additional disk space needed. For example, for a typical Pool-Seq sample of \textit{A. thaliana} of our GrENE-net dataset, an extra $\approx$\SI{8.1}{\giga\byte} are needed for the two files, given an input \formatname{bam} file of \SI{1.0}{\giga\byte}. Hence, in \toolname{grenedalf}, we also implemented to read directly from (coordinate-sorted) \formatname{sam}/\formatname{bam}/\formatname{cram} files, which eliminates the need for these file conversions. This comes at the expense of a slight increase in runtime due to the more complex file format (and its dependency on read depth), as can be seen in the subsequent figures, which is however compensated by the generally faster runtime of \toolname{grenedalf} overall, as well as offset by the saving in time needed for the file conversions in the first place. For example, the runtime overhead for converting from  \formatname{bam} to \formatname{pileup} to \formatname{sync} with existing tools is already larger than reading the \formatname{bam} file in \toolname{grenedalf} directly.
    \\
    Still, it might be advantageous to use \toolname{grenedalf} to convert from \formatname{sam}/\formatname{bam} to \formatname{sync}, when the file is going to be read many times, or in order to combine multiple samples into one file, and make subsequent reading even faster. Note also that \toolname{grenedalf} can produce and read gzipped files, bringing down the extra disk space needed for a \formatname{sync.gz} file of a typical \textit{A. thaliana} Pool-Seq sample to about \SI{430}{\mega\byte}, with almost the same performance as uncompressed \formatname{sync} files, due to decompression being executed asynchronously.
}
\label{supp:fig:grenenet-convert}
\end{figure}
\begin{figure}
    \begin{subfigure}[b]{0.49\textwidth}
    \includegraphics[width=\textwidth]{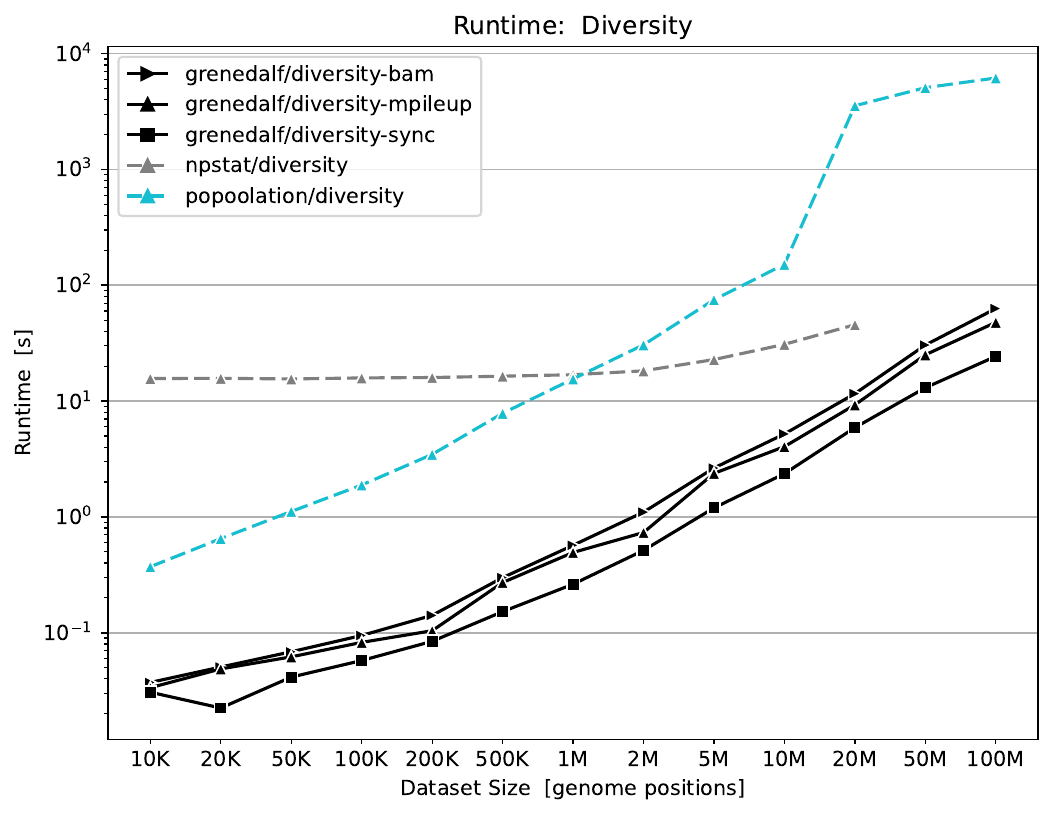}
    \centering
    \end{subfigure}
    \begin{subfigure}[b]{0.49\textwidth}
    \includegraphics[width=\textwidth]{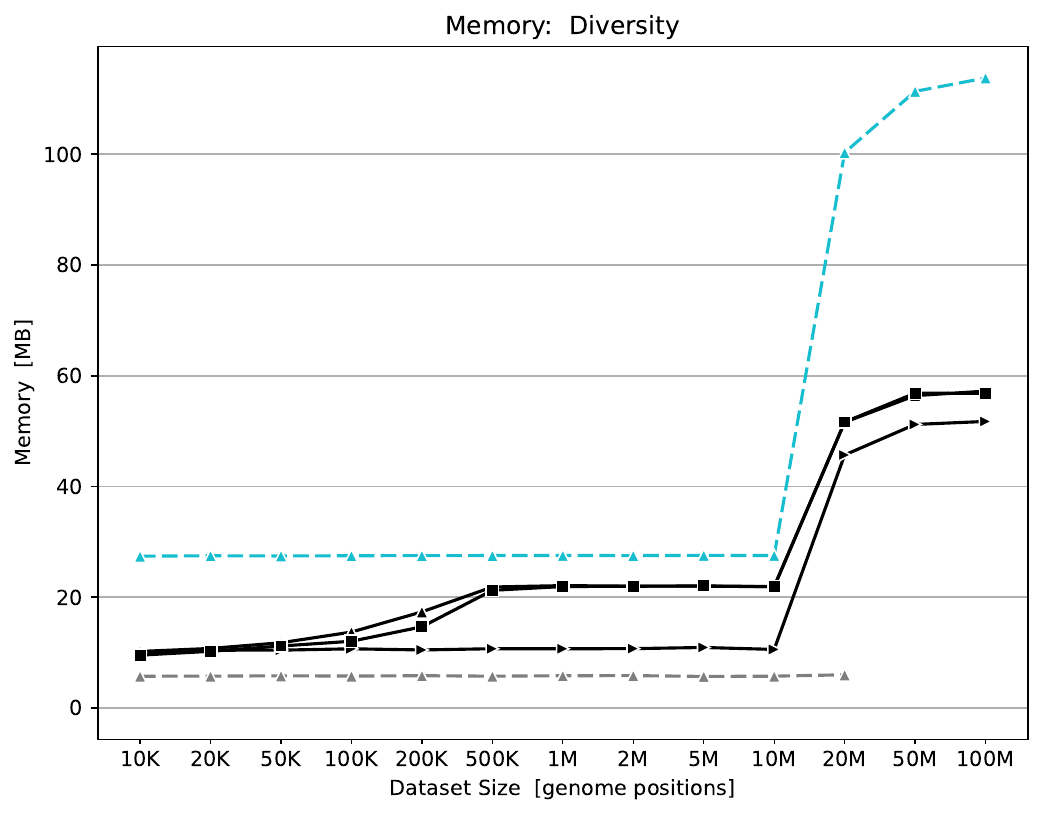}
    \centering
    \end{subfigure}
\caption{
    \textbf{Runtimes (log-log scale) and memory usage (linear–log scale) for computing diversity measures, for dataset sizes of \SI{10}{K} to \SI{100}{M} genomic positions, on real-world data.}
    \\
    The data are the same as in \figref{supp:fig:grenenet-convert}; for \toolname{grenedalf}, we also show differences in speed when using different input file formats. \toolname{PoPoolation} and \toolname{npstat} both require \formatname{pileup} files for this analysis. Unfortunately, \toolname{npstat} can only process single chromosomes at a time, which here means that we could not run tests beyond 20M base pairs, after which the second chromosome starts in the \textit{A. thaliana} genome.
    \\
    For all tools, we here computed Tajima's D, in windows of 1000 base pairs.
    The surge in runtime of \toolname{PoPoolation}, and in memory for both \toolname{PoPoolation} and \toolname{grenedalf} when processing 20M or more genome positions is due to characteristics of the dataset: The equations for computing the Pool-Seq bias correction term are dependent on the read depth of the data at a given position, requiring repeated evaluation of a binomial distribution (to be precise: the distribution is evaluated $c \times p$ times, with $c$ the read depth at the position, and $p$ the pool size of the sample).
    After 20M positions in our exemplary dataset, we first encounter higher read depth positions, hence the surge. Both tools cache the correction terms, to only compute them once per read depth level. This explains also the slight increase in memory at that point (up to $\approx$\SI{120}{\mega\byte}, still very small), as well as the flattening of the runtime for \toolname{PoPoolation}, as later on in the data, most read depth values have already been encountered, so that their correction terms are already cached.
    \\
    Note however that we here limited the maximum read depth to 1000 (see also below), in order to keep runtimes manageable for testing. With higher read depth allowed, \toolname{PoPoolation} would slow down even more, if there are positions in the data with extremely high read depth.
    On the other hand, the runtime of \toolname{grenedalf} does not suffer as much from this, as our computation of the correction term and the involved binomial distribution is highly optimized (and replaces an expensive summation of \toolname{PoPoolation} by a fast closed-form expression, as detailled in our equations document). For instance, we use fast lookup tables of log factorials as well as Stirling's approximation to compute the binomials.
    Here, that means, for the 20M position sample in \formatname{pileup} format, \toolname{grenedalf} is 380 times as fast as \toolname{PoPoolation}; for the 100M position sample, with the fixed maximum read depth of 1000, it is still more than two orders of magnitude as fast. When reading the faster \formatname{sync} format in \toolname{grenedalf} instead, the 20M position sample runs 600 times as fast as \toolname{PoPoolation}. This also indicates that the runtime of \toolname{grenedalf} is not dominated by the computation of the correction terms.
    \\
    Lastly, \toolname{npstat} instead pre-computes the Pool-seq correction terms. This is indicated by the near constant runtime for smaller files, as the amount of pre-computation is always the same, and only depends on the specified maximum read depth specified by the user. This is a severe downside of the tool; we had to limit the maximum read depth to 1000 here (for all tools, to keep it comparable), as otherwise \toolname{npstat} would have spend unreasonably long for just the pre-computation. This is not an issue for \toolname{PoPoolation} and \toolname{grenedalf} however, which only compute the correction terms as needed in the data.
}
\label{supp:fig:grenenet-diversity}
\end{figure}
\begin{figure}
    \begin{subfigure}[b]{0.49\textwidth}
    \includegraphics[width=\textwidth]{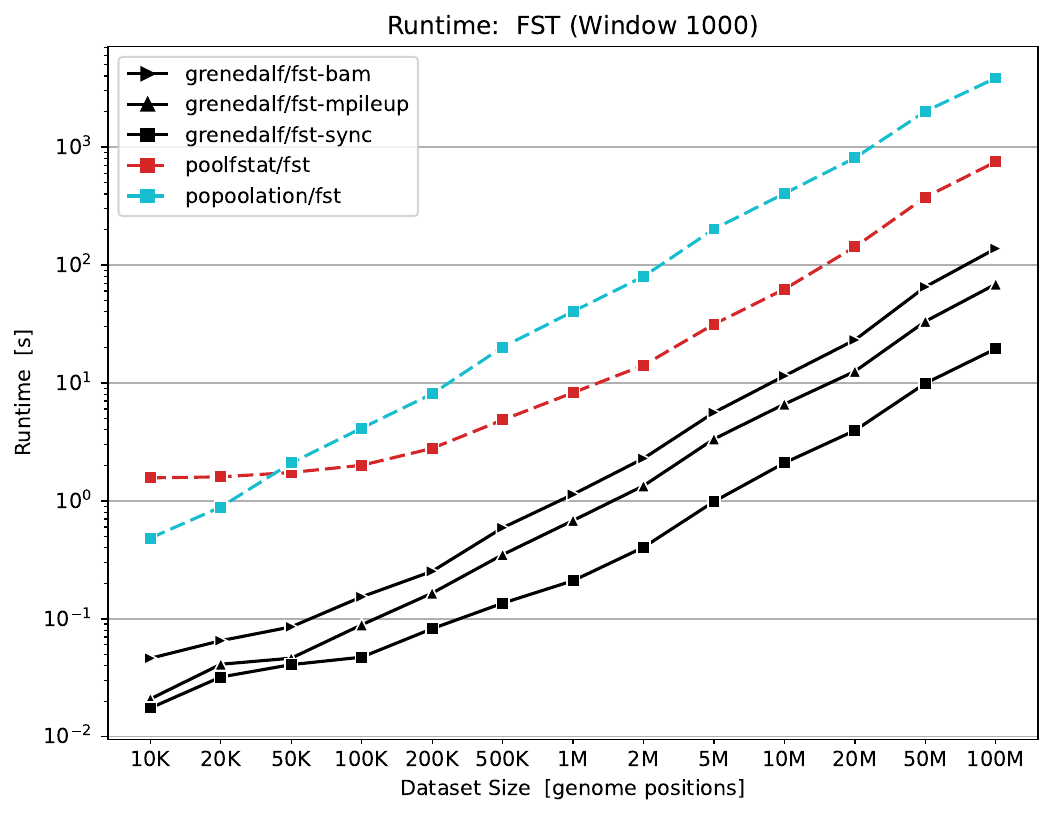}
    \centering
    \end{subfigure}
    \begin{subfigure}[b]{0.49\textwidth}
    \includegraphics[width=\textwidth]{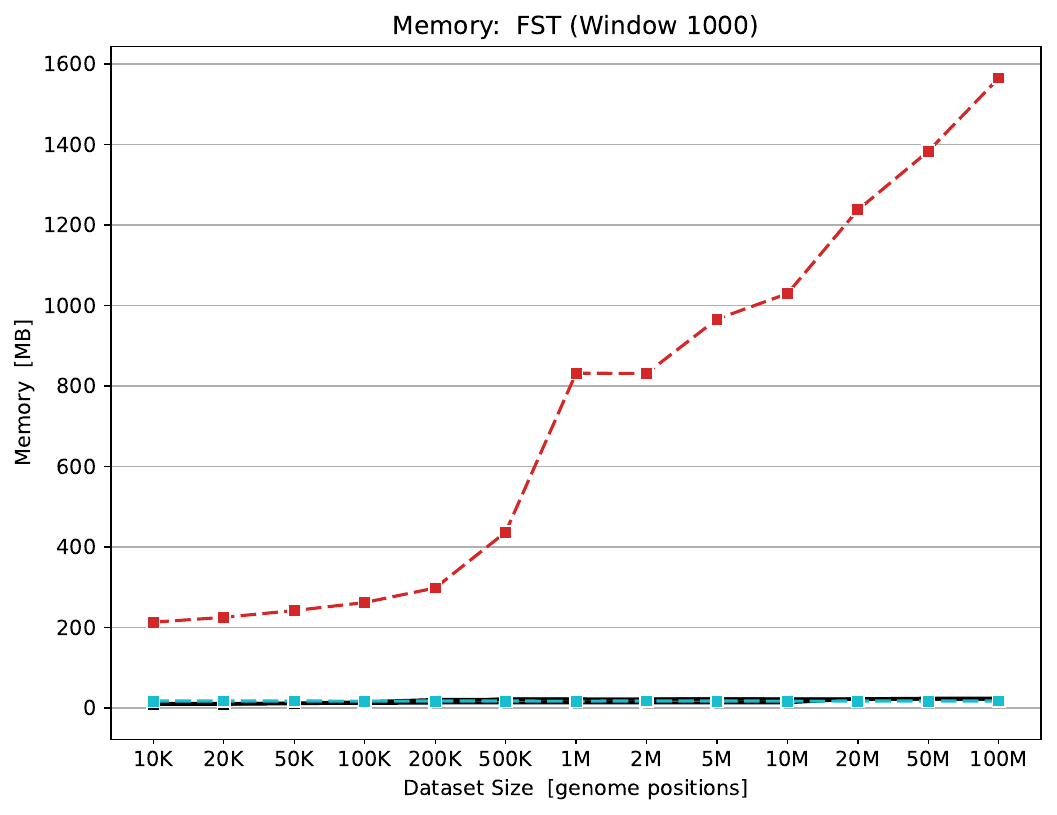}
    \centering
    \end{subfigure}
\caption{
    \textbf{Runtimes (log-log scale) and memory usage (linear–log scale) for computing \fst{}, for dataset sizes of \SI{10}{K} to \SI{100}{M} genomic positions, on real-world data.}
    \\
    We again use data from the same real world experiment as in \figref{supp:fig:grenenet-convert}; here however, we use two distinct samples, between which \fst{} is computed. For \toolname{grenedalf}, we again show the benchmark for different input file formats for comparison, with two \formatname{bam} files of two samples from GrENE-net, as well as \formatname{mpileup} and \formatname{sync} files obtained from converting and combining these two \formatname{bam} files. The \formatname{sync} file is then also used for the two tools we benchmark against here.
    \\
    The runtime is again best for \toolname{grenedalf} using \formatname{sync} input files, being about 30 to 40 times faster than \toolname{poolfstat}, and around 200 times faster than \toolname{PoPoolation2}.
    Using more involved input formats which increase in size with read depth, such as \formatname{bam} and \formatname{mpileup}, slows down \toolname{grenedalf} compared to the \formatname{sync} format, but it still maintains its advantage over the other two tools.
    \\
    The way that \toolname{poolfstat} is written in \langname{R}, it requires the input to be fully read into memory before computing \fst{}. For the largest input file here (with \SI{100}{M} genomic positions) this equates to $\approx$\SI{1.5}{\giga\byte} memory for the two samples between which we compute \fst{} here. Note that this requirement linearly scales with the genome size \emph{and} with the number of samples, making \toolname{poolfstat} not feasible for larger datasets with many samples. On the other hand, both \toolname{PoPoolation2} and \toolname{grenedalf} stream through the data, so that only data in the order of the window size is needed to be kept in memory; here, with windows of \SI{1000}{bps}, memory below \SI{25}{\mega\byte} is needed. 
    \\
    This is a particular downside of \toolname{poolfstat}, and prohibits its usage for larger number of samples. For instance (not shown here), we computed pairwise \fst{} between an increasing number of samples (ranging from 2 to 5 samples) from GrENE-net with \toolname{poolfstat}. Per sample added (each similar in size to the ones used here), the memory requirement of \toolname{poolfstat} increased linearly by about \SI{800}{\mega\byte}. For creating a plot similar to \figref{supp:fig:pairwise-fst-matrix}, the tool would hence roughly need \SI{2}{\tera\byte} of memory.
}
\label{supp:fig:grenenet-fst-positions}
\end{figure}
\begin{figure}
    \begin{subfigure}[b]{0.49\textwidth}
    \includegraphics[width=\textwidth]{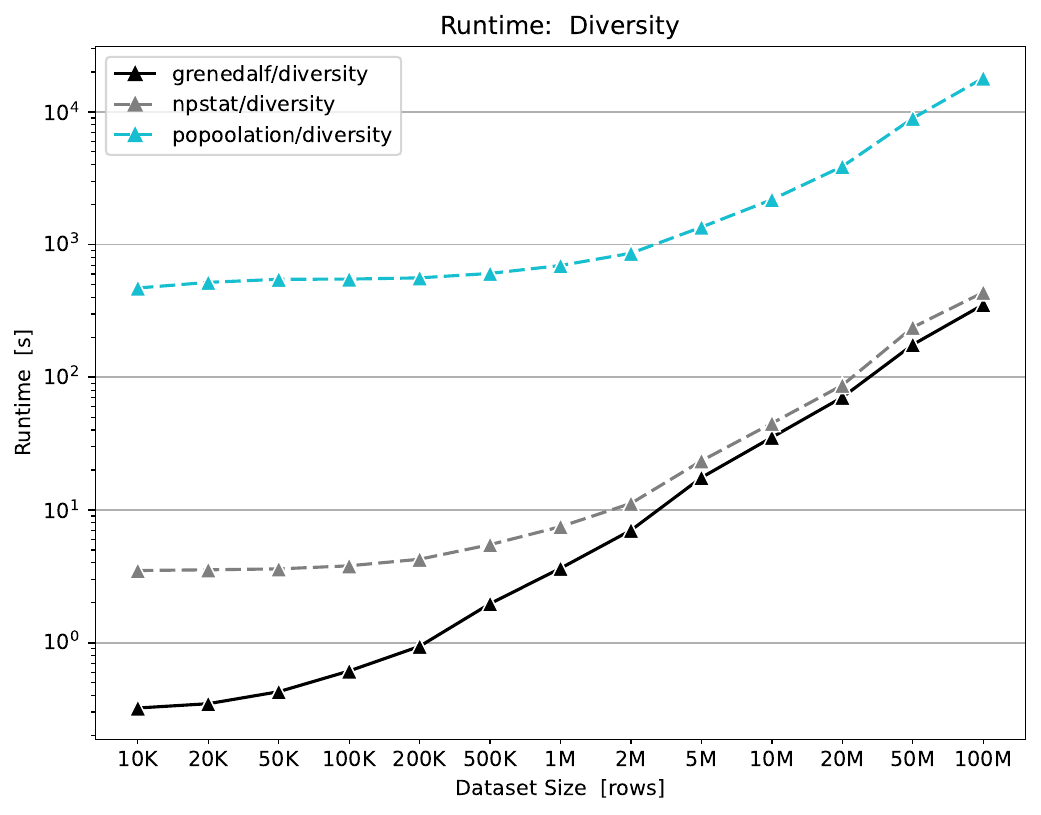}
    \centering
    \end{subfigure}
    \begin{subfigure}[b]{0.49\textwidth}
    \includegraphics[width=\textwidth]{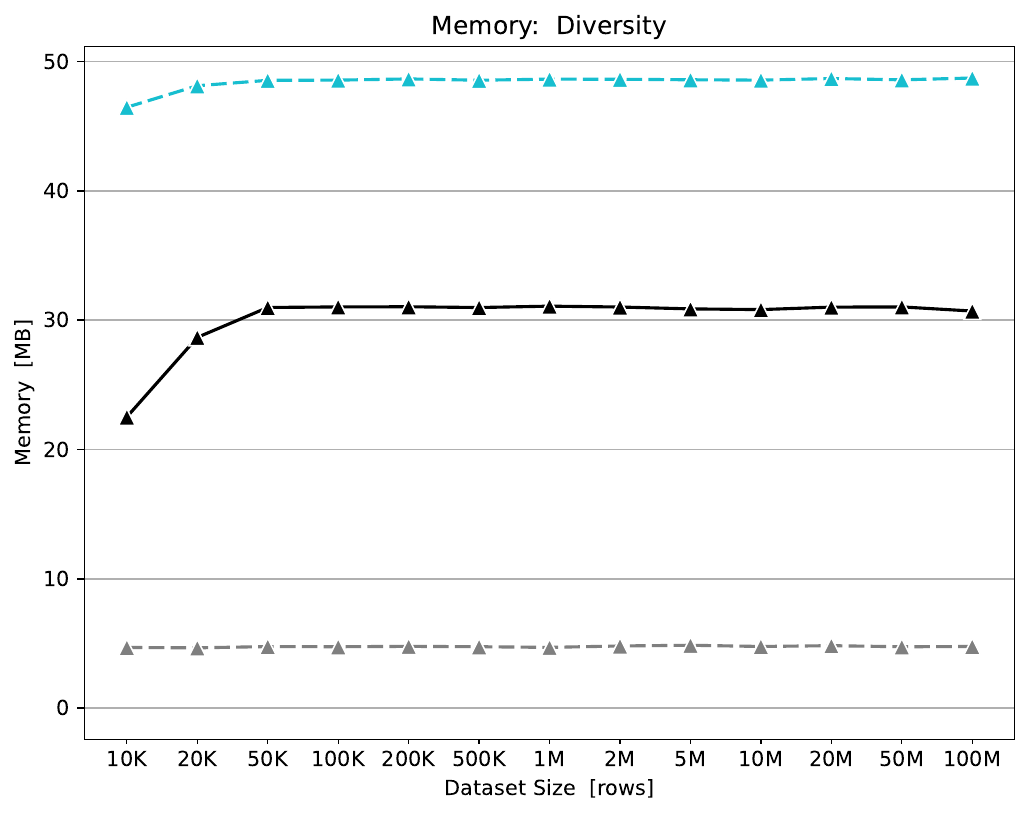}
    \centering
    \end{subfigure}
\caption{
    \textbf{Runtimes (log-log scale) and memory usage (linear–log scale) for computing diversity measures, for dataset sizes of \SI{10}{K} to \SI{100}{M} genomic positions, on randomly simulated data.}
    \\
    In order to measure the minimally expected improvements of \toolname{grenedalf} over existing tools, we simulated datasets using random read depths and allele counts to create \formatname{pileup} test files.  The read depth in these files is rather small, only up to 500, so that the computation of the Pool-seq correction terms is not dominating the runtime too much. We hence avoid data-dependent effects here, as for example seen in \figref{supp:fig:grenenet-diversity} when hitting genome positions with higher read depth.  This hence serves as a benchmark of the expected lower boundary of the speedup of grenedalf compared to other tools, showing the raw processing speed for tasks such as file reading and windowing of the data.
    \\
    We again computed Tajima's D in windows of 1000 base pairs along the genome. Due to the fixed low read depth, no surge in the runtime is observed here, and instead the overall trend in file processing speed is more evident. At larger sample sizes, \toolname{grenedalf} is about 50 times as fast as \toolname{PoPoolation}, and 1.2 times as fast as \toolname{npstat}, which is written in \langname{C}, and also comparably efficient, but lacks support for multi-threaded speedups.
}
\label{supp:fig:random-diversity}
\end{figure}
\begin{figure}
    \begin{subfigure}[b]{0.49\textwidth}
    \includegraphics[width=\textwidth]{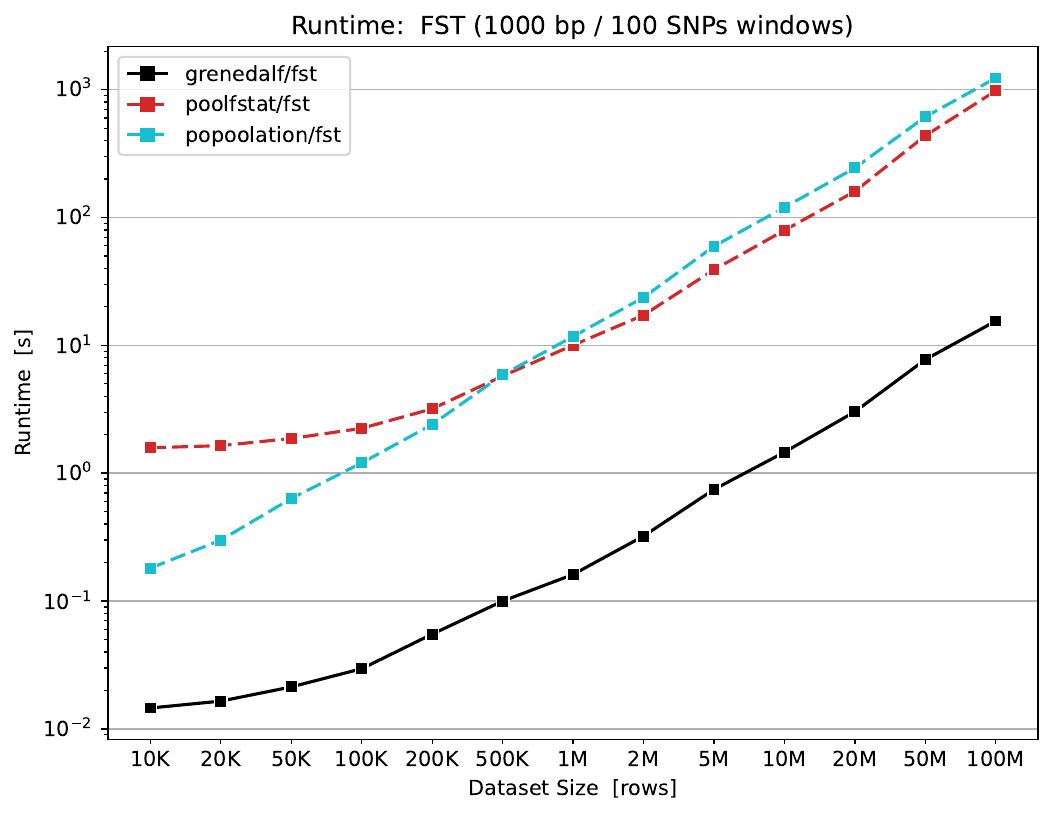}
    \centering
    \end{subfigure}
    \begin{subfigure}[b]{0.49\textwidth}
    \includegraphics[width=\textwidth]{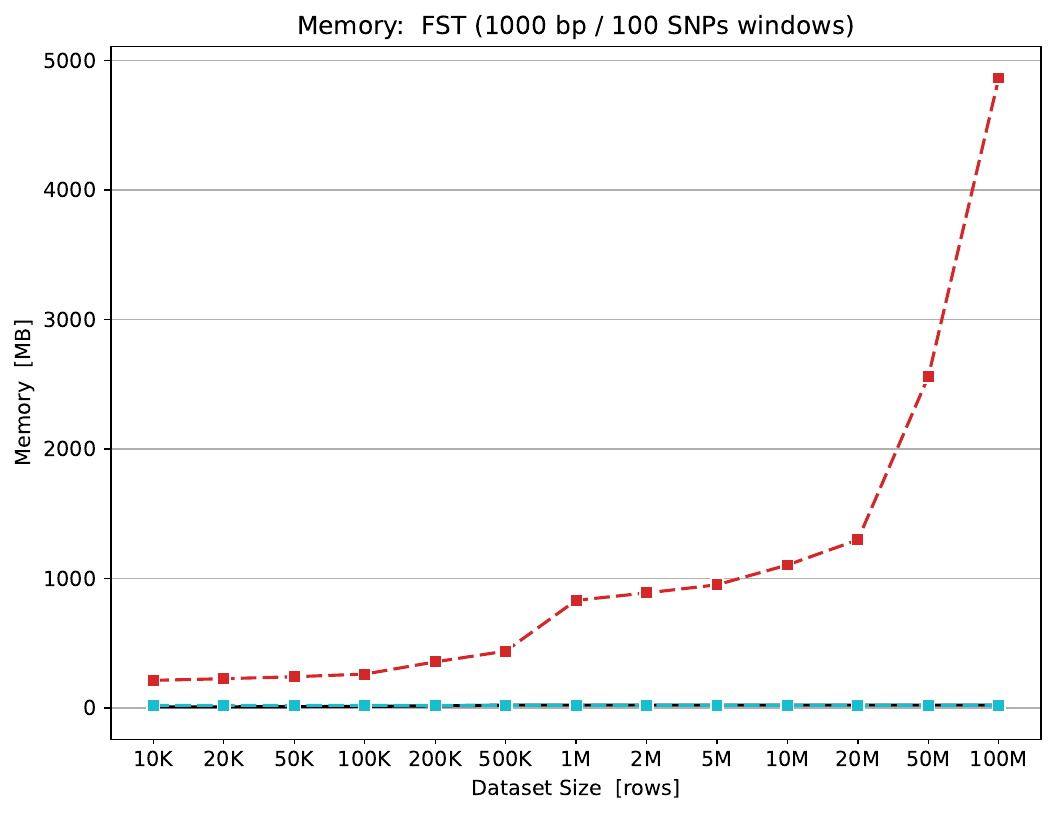}
    \centering
    \end{subfigure}
\caption{
    \textbf{Runtimes (log-log scale) and memory usage (linear–log scale) for computing \fst{}, for dataset sizes of \SI{10}{K} to \SI{100}{M} genomic positions, on randomly simulated data, in windows of 1000 base pairs and 100 SNPs.}
    \\
    For the reasons explained in \figref{supp:fig:random-diversity}, we here simulated datasets with random (small) read depths and allele counts to create \formatname{sync} test files containing two samples. We then computed \fst{} in windows along the genome. For \toolname{grenedalf} and \toolname{PoPoolation2}, we used windows of \SI{1000}{bp}. As \toolname{poolfstat} does not seem to support windows of fixed size along the genome, we instead used windows of 100 SNP length; as we simulated with 10\% variant positions, this averages out to the same total number of windows being processed. Note that \toolname{grenedalf} can also use windows with a fixed number of SNPs (as opposed to windows of a fixed length along the genome), which we omit here for simplicity.
    \\
    Again, \toolname{grenedalf} outperforms both other tools in terms of runtime by a factor of 60 and 80, respectively, for the largest input file. This is the minimally expected speed up, and should be equally as fast or even faster on real world data.
}
\label{supp:fig:random-fst-window}
\end{figure}
\begin{figure}
    \begin{subfigure}[b]{0.49\textwidth}
    \includegraphics[width=\textwidth]{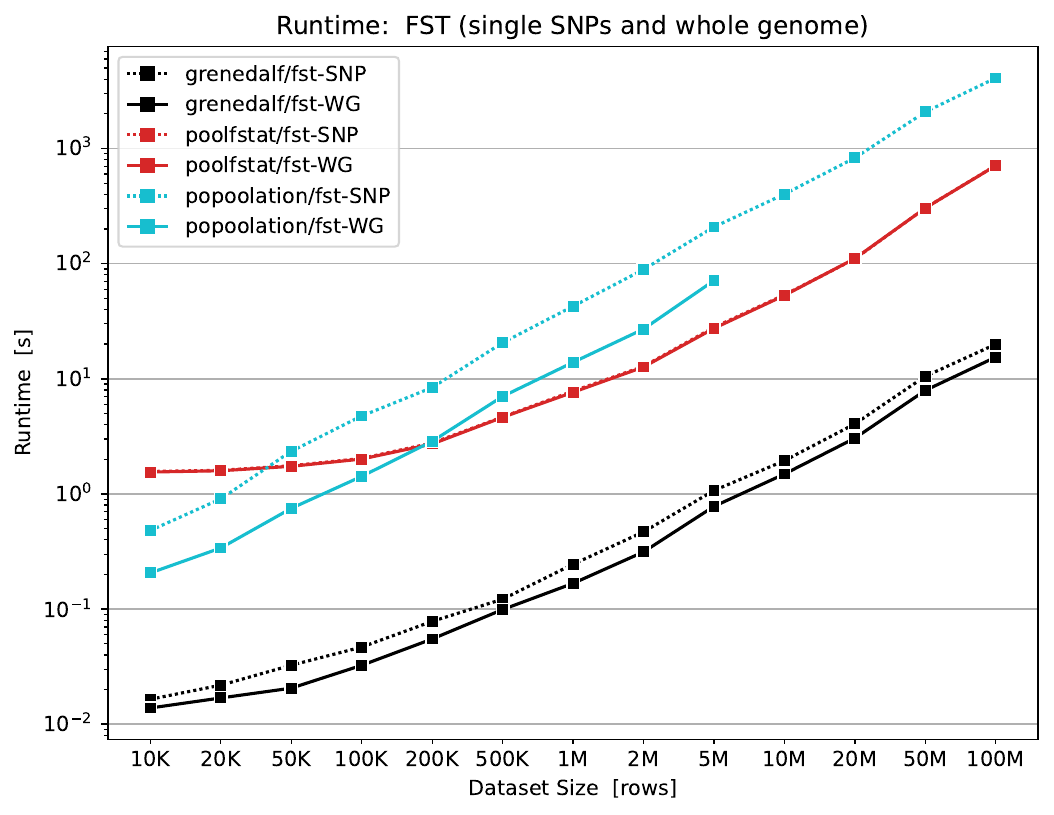}
    \centering
    \end{subfigure}
    \begin{subfigure}[b]{0.49\textwidth}
    \includegraphics[width=\textwidth]{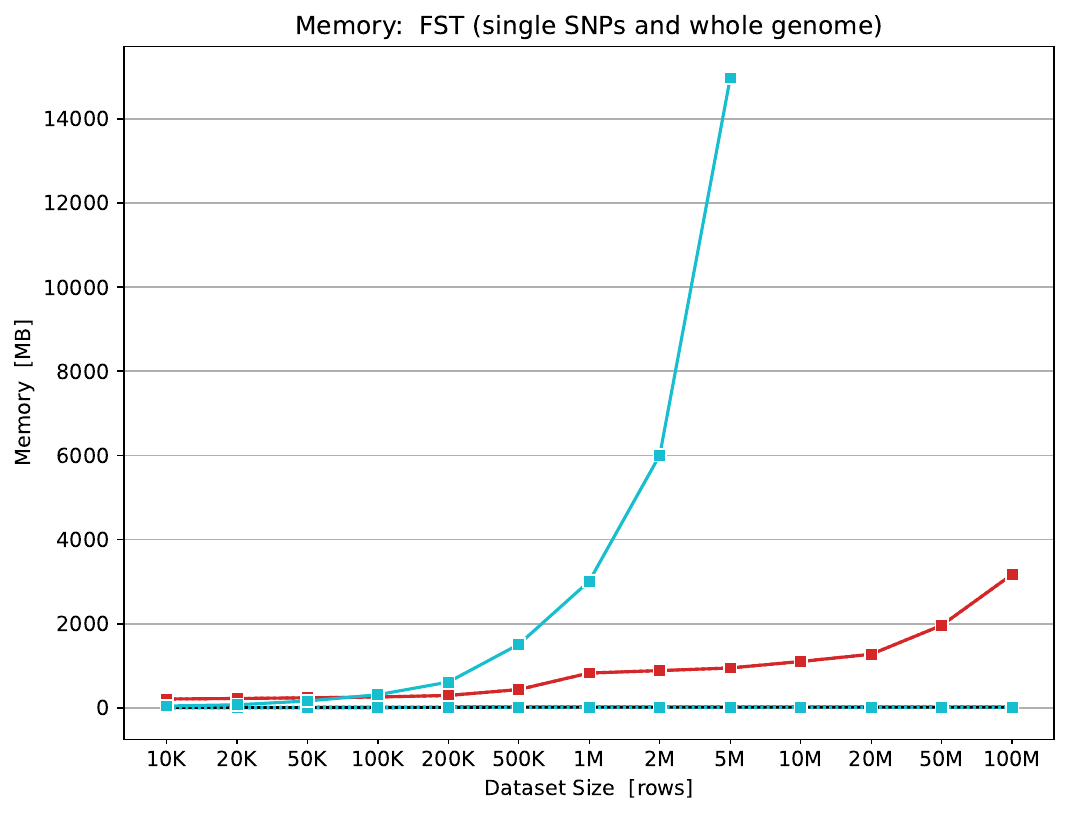}
    \centering
    \end{subfigure}
\caption{
    \textbf{Runtimes (log-log scale) and memory usage (linear–log scale) for computing \fst{}, for dataset sizes of \SI{10}{K} to \SI{100}{M} genomic positions, on randomly simulated data, on single SNPs and the whole genome.}
    \\
    This is the same data and setup as explained in \figref{supp:fig:random-fst-window}, but instead of using sliding windows, we here computed \fst{} for single SNPs, and for the whole genome, in order to show the effects of these choices.
    \\
    For \toolname{PoPoolation2}, we could not compute whole-genome \fst{} on larger datasets, as it needs to keep data in memory for the whole window, which becomes prohibitive at full genome scale.  Despite also keeping the whole genome in memory, the overall memory requirements for \toolname{poolfstat} are lower, and it hence was able to compute \fst{} for the large datasets here as well. However, this still can easily become a limitation when computing \fst{} for more samples than used in this test. The runtime and memory for \toolname{poolfstat} are virtually identical with both types of windowing.  Again, \toolname{grenedalf} outperforms both in terms of runtime. We implemented a streaming approach for whole genome data, so that even in that case, our memory requirements are not affected by the genome sizes at all (in fact, for the whole genome we need less memory than for windowed approaches, as we do not even need to keep the data for each window in memory).
}
\label{supp:fig:random-fst-snps-genome}
\end{figure}
\begin{figure}
    \begin{subfigure}[b]{0.49\textwidth}
    \includegraphics[width=\textwidth]{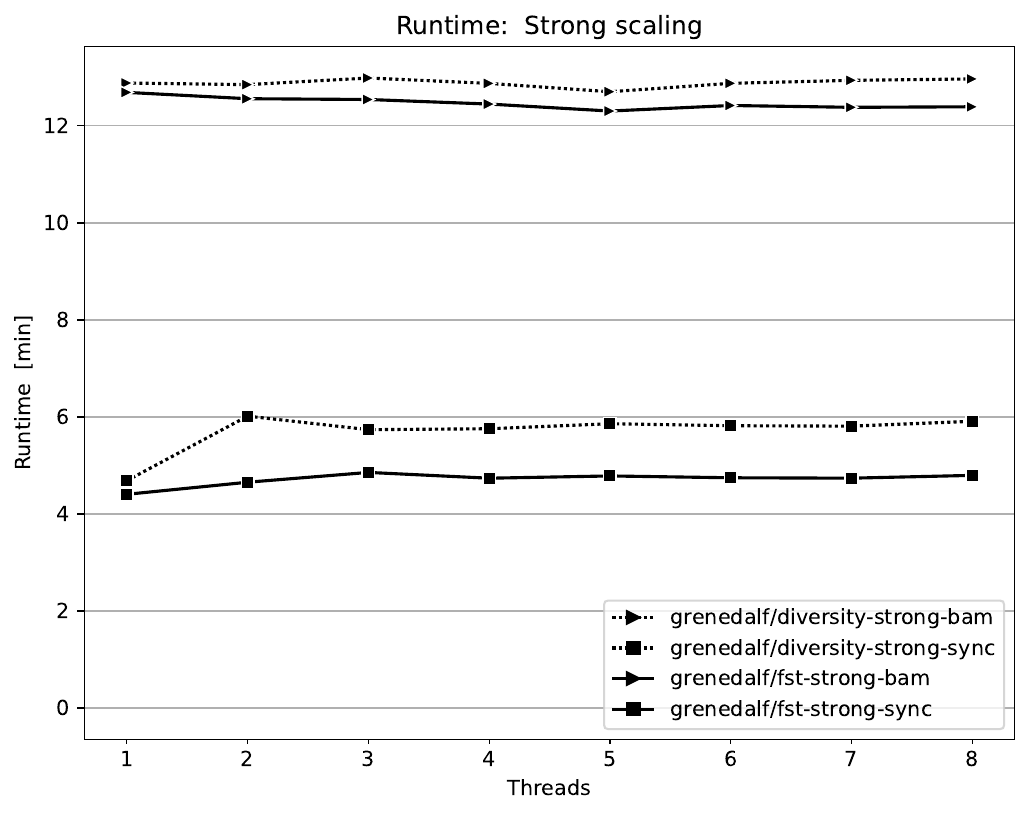}
    \centering
    \end{subfigure}
    \begin{subfigure}[b]{0.49\textwidth}
    \includegraphics[width=\textwidth]{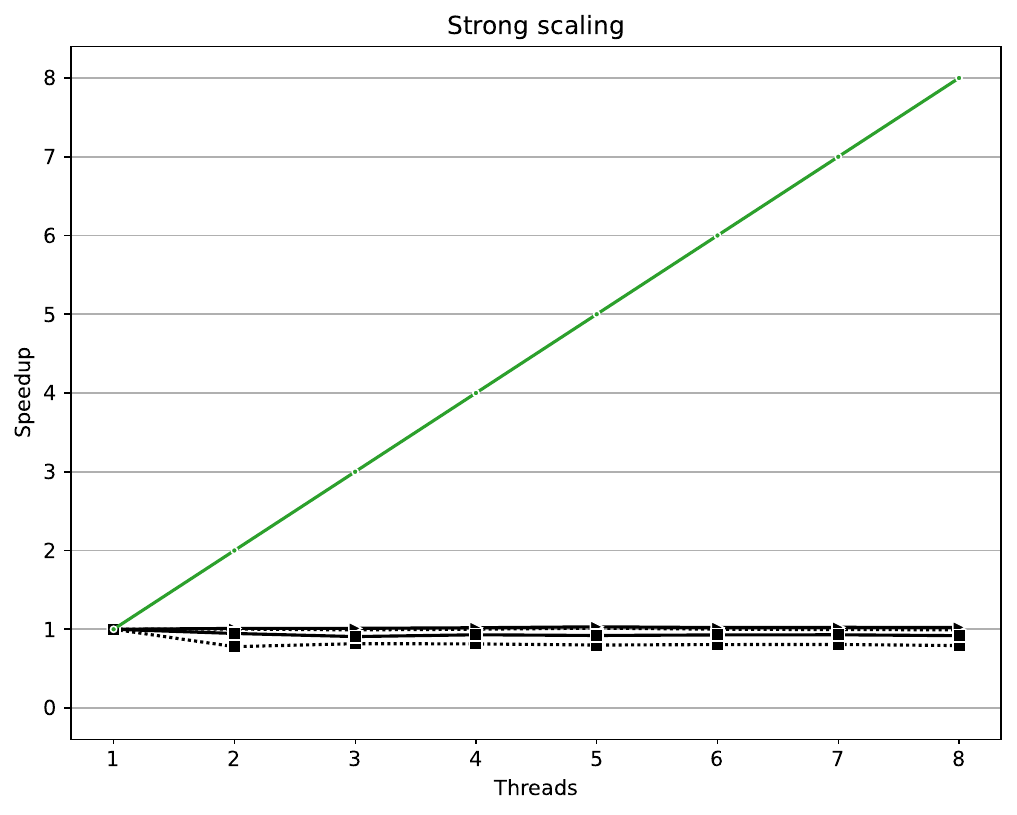}
    \centering
    \end{subfigure}
\caption{
    \textbf{Runtimes (linear scale) and strong scaling speedup for computing Tajima's D and \fst{} with multiple threads, for real-world data, in windows of 1000 base pairs.}
    \\
    We here again used real world data from our GrENE-net project, in order to test scaling efficiency of \toolname{grenedalf} when using multiple cores to speed up the computation.  In each test, we used 8 samples, in two different input file formats (\formatname{bam} and \formatname{sync}), and computed Tajima's D for each sample, as well as \fst{} between each pair of samples (28 pairs), both for the whole genome.  The left hand plot shows the total runtime of each test, with an increasing number of threads, up to the number of cores on the computer where this was tested. The right hand plot shows the strong scaling speedup based on these runtimes, i.\,e., the runtime with one thread divided by the runtime with multiple threads. The green identity line is the theoretical maximum speedup.
    \\
    As before, reading from \formatname{sync} files is significantly faster than reading from \formatname{bam} files, and accounts for a large portion of the runtime.
    This test however shows that using multiple threads on this relatively small test dataset did not yield any speedup; all runtimes are virtually identical, or even slightly worse compared to the single thread runtime.  
    \\
    This scaling inefficiency is likely due to Amdahl's law: The amount of scaling that can be achieved with multiple threads is limited by the portion of the runtime that is not parallelizable. For datasets with few samples, the parallelizable portion of the program (e.\,g., the input file reading, and the actual computation of \fst{}) is relatively small compared to the linear portion (e.\,g., the stream iteration, and the output writing).  Furthermore, with only a few samples, the thread synchronization overhead that becomes necessary when using more than one thread for the computation has a higher impact:  Our internal memory layout buffers data and computes statistics per position in the genome. With few samples, the computational work at each position is small, and hence, the synchronization between threads at every position limits the efficiency. 
    \\
    At the time being, we do not consider it worth optimizing this further though. This is because for larger datasets, where more work is done per position (and hence the parallelizable portion of the program is larger), and where runtime starts to matter more, we do see a significant speedup with multiple threads, as shown in \figref{supp:fig:scaling-fst-matrix}.
}
\label{supp:fig:scaling-strong}
\end{figure}
\begin{figure}
    \begin{subfigure}[b]{0.49\textwidth}
    \includegraphics[width=\textwidth]{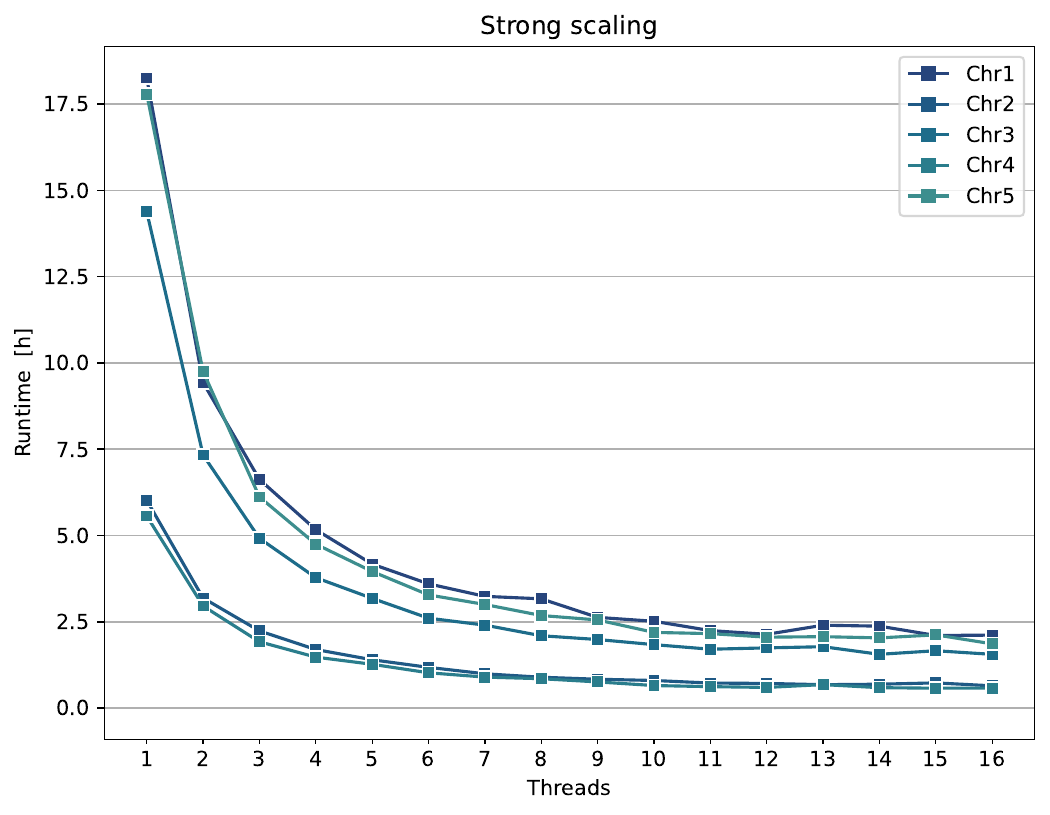}
    \centering
    \end{subfigure}
    \begin{subfigure}[b]{0.49\textwidth}
    \includegraphics[width=\textwidth]{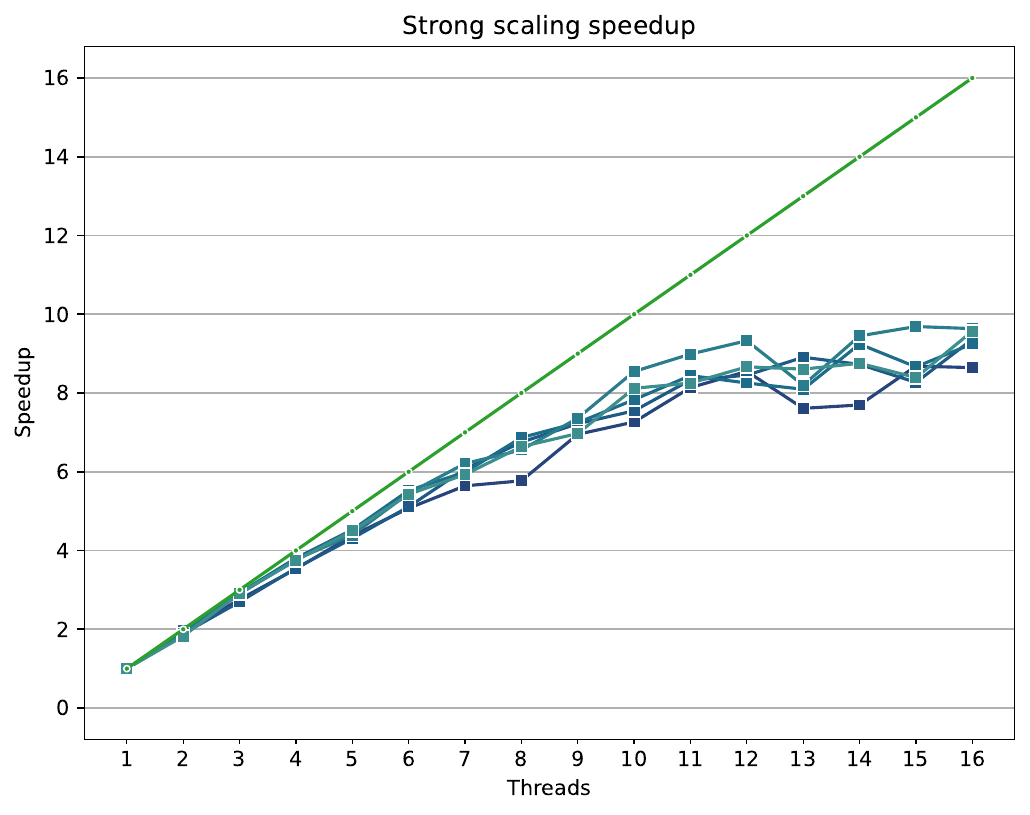}
    \centering
    \end{subfigure}
\caption{
    \textbf{Runtimes (linear scale) and strong scaling speedup for computing pairwise whole-genome \fst{} with multiple threads, for \SI{2415} samples.}
    \\
    Here, we used the dataset of \figref{supp:fig:pairwise-fst-matrix}, with $2{,}415$ samples, for a large-scale strong scaling test. We computed pairwise \fst{} between all $2{,}914{,}905$ pairs of samples, for the whole genome with $2{,}948{,}475$ variant positions.
    As this was run on a computer cluster, we were able to test with up to 16 threads.  Each cluster node was allocated with 4 extra cores on top of the cores needed for each computational thread, in order to avoid core over-subscription. To compensate for cluster usage fluctuations, the results shown here are the minimum times of three independent runs for each measurements.  We furthermore split the dataset into its 5 chromosomes, in order to get more measurement points, and shorter overall waiting time to run the tests.
    \\
    The left hand plot shows the total runtimes.  In contrast to \figref{supp:fig:scaling-strong}, we do see a significant speedup here for increasing numbers of threads. With this large dataset, the computation of all ${\approx}3$ million pairs of \fst{} constitutes a larger portion of the runtime, meaning that a larger portion of the work is parallelized. Hence, more computational work is being executed per position, also meaning that thread synchronization overhead has less influence here.
    \\
    The right hand plot again shows the strong scaling speedup, that is, the runtime with one thread divided by the runtime with multiple threads. For up to ${\approx}10$ threads, we observe almost ideal speedup (green line), levelling off after that due to Amdahl's law. This is expected, and can be interpreted to indicate that (for this dataset) using more than 10 threads has diminishing returns. We note though that this is likely fast enough for any contemporary dataset: The overall runtime across all $2{,}914{,}905$ pairs of samples and all five chromosomes using 10 threads was $\approx$\SI{8}{\hour}.
}
\label{supp:fig:scaling-fst-matrix}
\end{figure}
\begin{figure}
    \begin{subfigure}[b]{0.49\textwidth}
    \includegraphics[width=\textwidth]{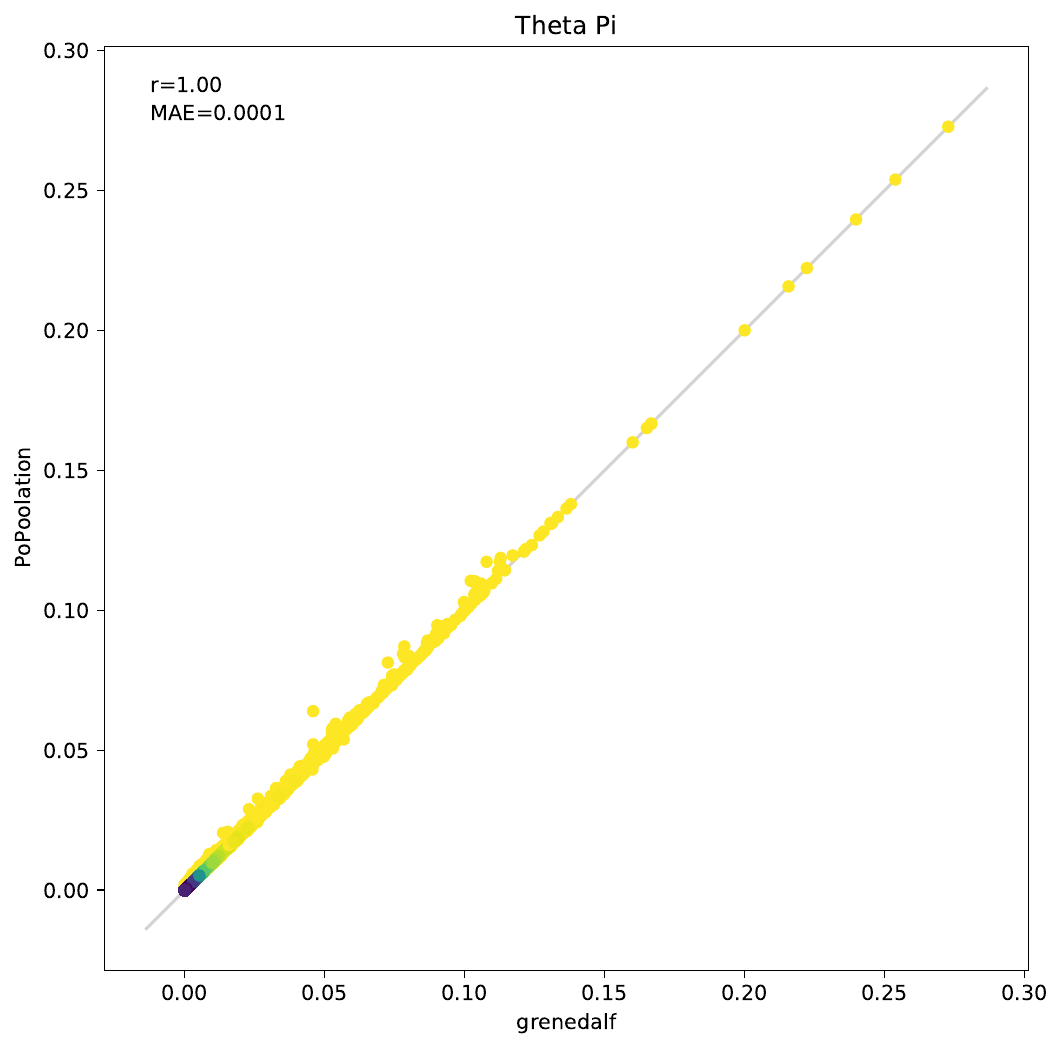}
    \centering
    \end{subfigure}
    \begin{subfigure}[b]{0.49\textwidth}
    \includegraphics[width=\textwidth]{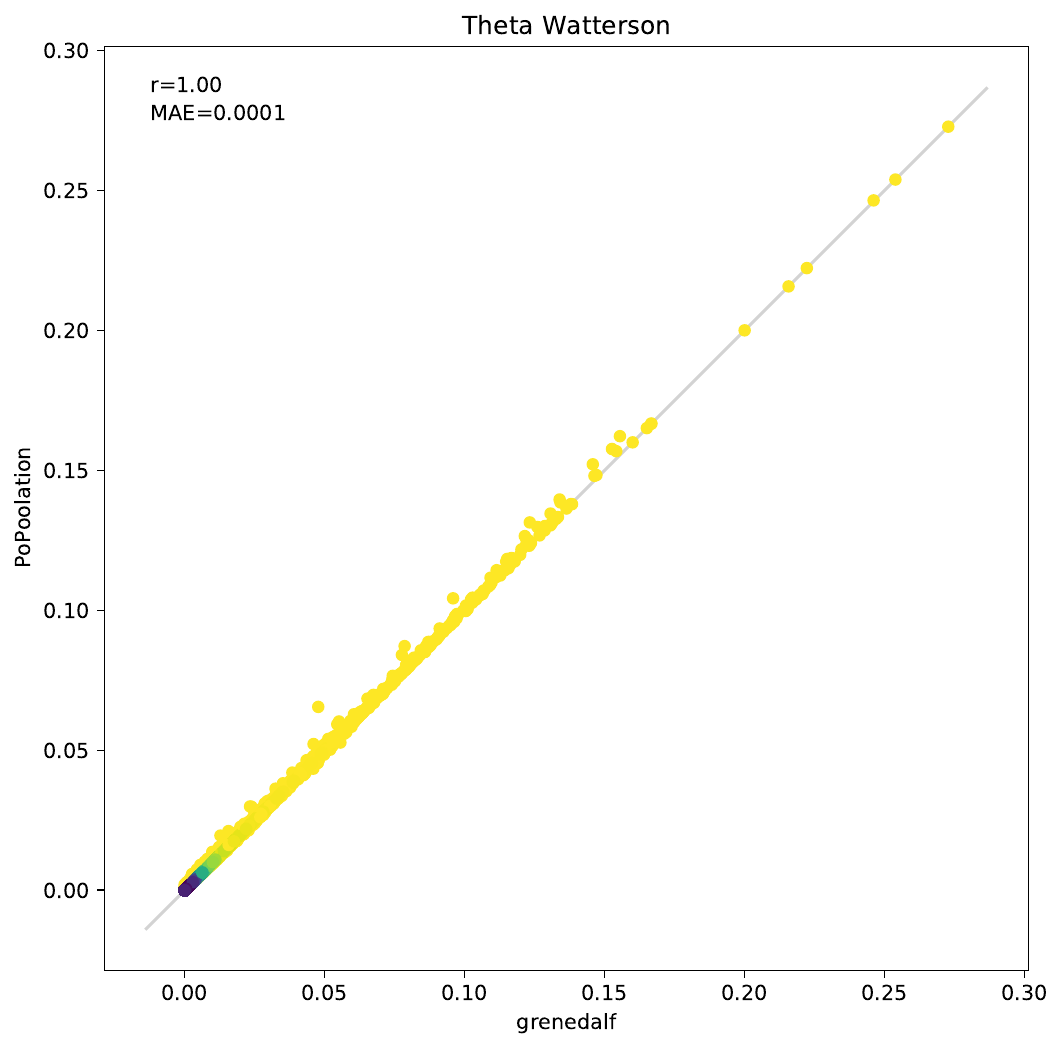}
    \centering
    \end{subfigure}
\caption{
    \textbf{Comparison of \toolname{grenedalf} and \toolname{PoPoolation} for Theta Pi and Theta Watterson on real-world data.}
    \\
    We here compare the results of the implementation in \toolname{grenedalf} to those obtained with \toolname{PoPoolation} for Theta Pi (left) and Theta Watterson (right). The input data is the same as used in \figref{supp:fig:grenenet-diversity}, 
    namely Pool-Seq samples from our GrENE-net experiment, in order to check for any discrepancies between the tools on real-world data. Such data often contains types of noise and variation that are hard to capture realistically in simulated data.  We used a window size of 1000bp, and matching parameters for minimum counts and read depths.  The plots compare the values per window obtained with both tools, with the coloring based on a Gaussian kernel density estimate, indicating that most values of the diversity estimators are rather small here.
    \\
    Our results match closely with those obtained from \toolname{PoPoolation}. During development of \toolname{grenedalf}, we however noted some initial discrepancies between the two tools that resulted from differences in the filtering of positions with a high number of deletions in the data. This was particularly prevalent in windows with a low number of variant positions and low read depths, where omitting a single SNP due to some filter setting already has a large impact on the overall estimate for the window.  We have since adapted our filtering of deletions to match more closely with that of \toolname{PoPoolation}, as seen here. The remaining slight differences between the tools are likely resulting from similar edge cases in the implementation details, or might even be of numerical nature, but are small enough to not warrant further investigation.
}
\label{supp:fig:correlation-thetas-popoolation}
\end{figure}
\begin{figure}
    \begin{subfigure}[b]{0.49\textwidth}
    \includegraphics[width=\textwidth]{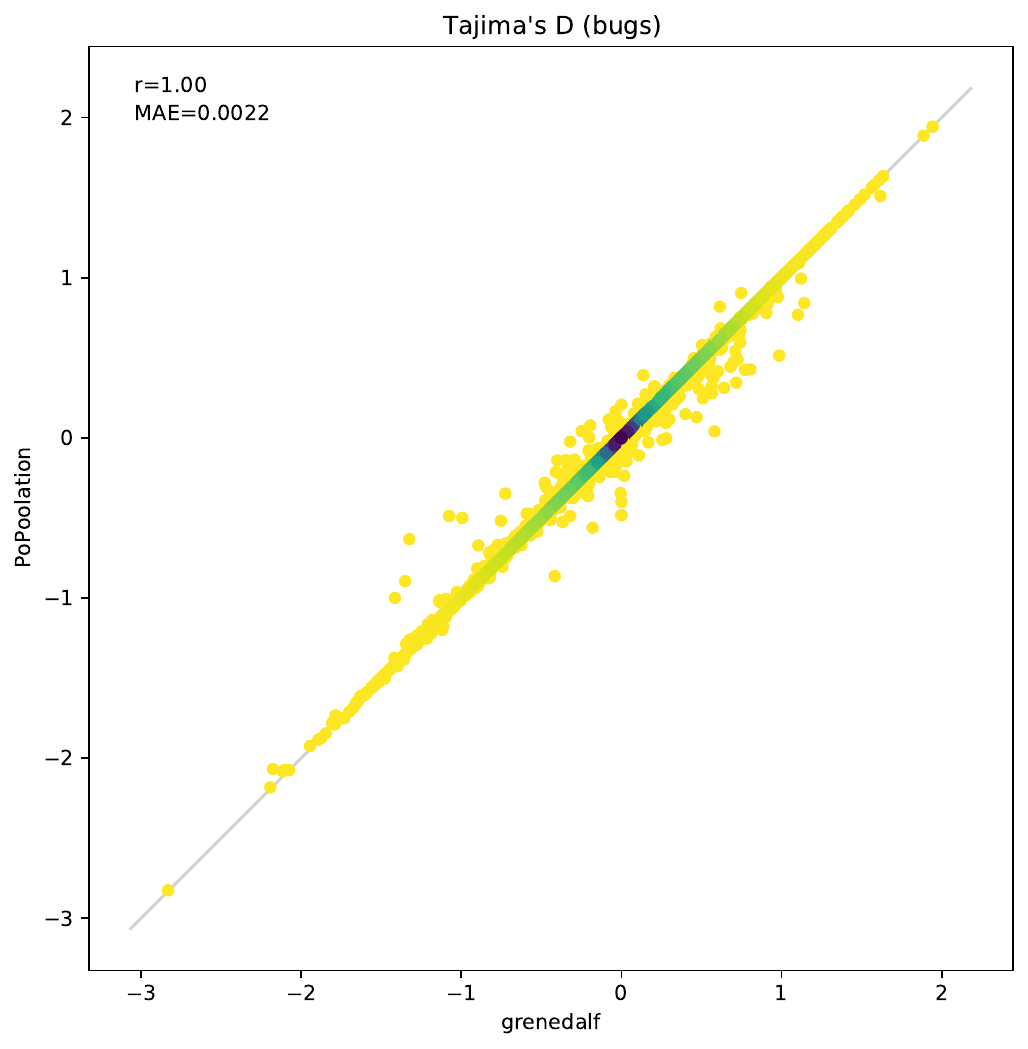}
    \centering
    \end{subfigure}
    \begin{subfigure}[b]{0.49\textwidth}
    \includegraphics[width=\textwidth]{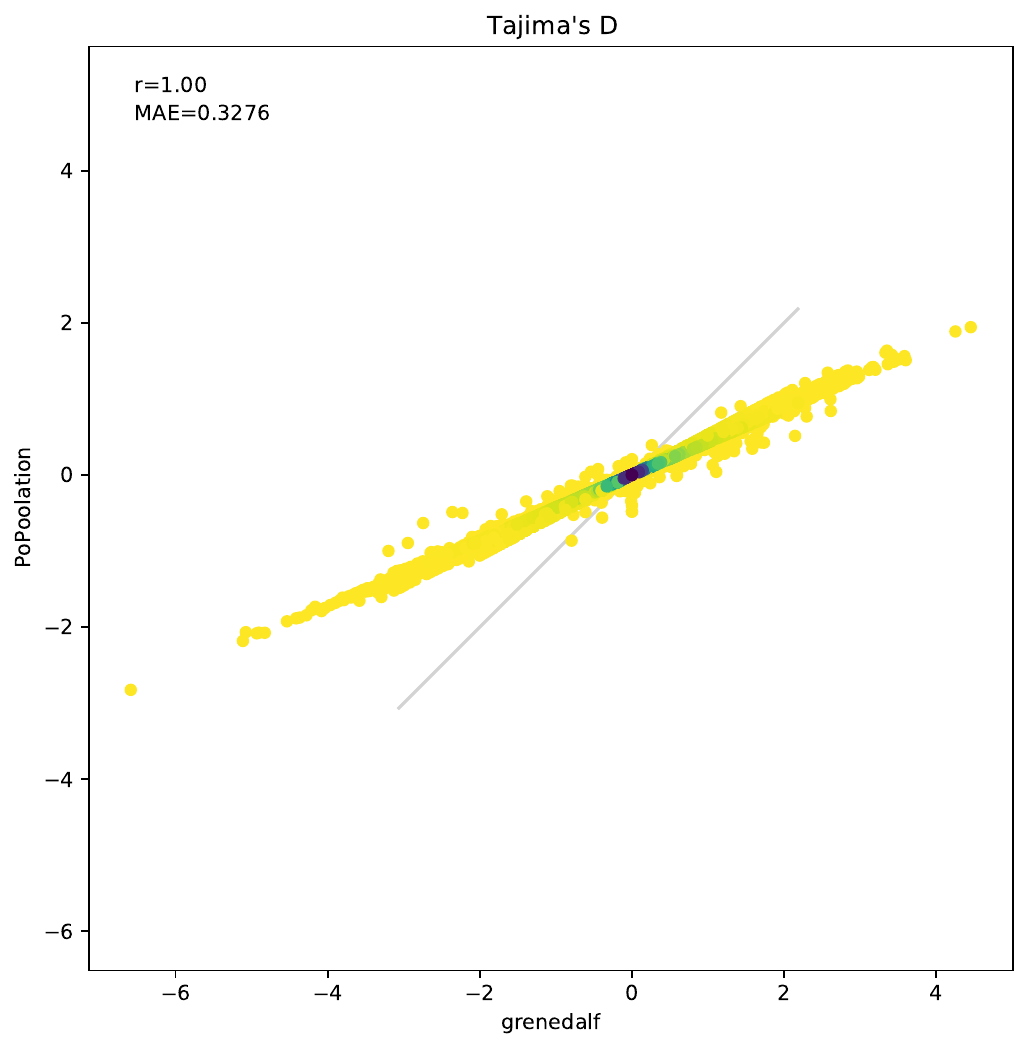}
    \centering
    \end{subfigure}
    %
\caption{
    \textbf{Comparison of \toolname{grenedalf} and \toolname{PoPoolation} for Tajima's D on real-world data.}
    \\
    The setup for this figure is equivalent to \figref{supp:fig:correlation-thetas-popoolation}, but shows the comparison for Tajima's D. As explained in our equations document, the implementation of Tajima's D in \toolname{PoPoolation} contains several bugs that affect the results; we hence show the comparison to \toolname{grenedalf} here with our intentional re-implementation of those bugs for purposes of comparison (left). Similar to \figref{supp:fig:correlation-thetas-popoolation}, the results are highly consistent between the tools, with only a few positions not fully matching. 
    \\
    We also show the effect of those bugs (right): At fixed pool sizes with relatively consistent read depth throughout the sample, the bugs act as a constant factor applied to the values. However, as detailed in our equations document, even with those bugs removed, this estimator of Tajima's D does not fully capture the complexity of the data, and is in fact not comparable with the numerical results from a ``classical'' estimator of Tajima's D based on individual data. In the Pool-seq case, other population characteristics need to be taken into account, which the estimator does not do. We hence recommend to not interpret values obtained from this estimator numerically. For details, see our equations document.
}
\label{supp:fig:correlation-tajima-d-popoolation}
\end{figure}
\begin{figure}
    \begin{subfigure}[b]{0.49\textwidth}
    \includegraphics[width=\textwidth]{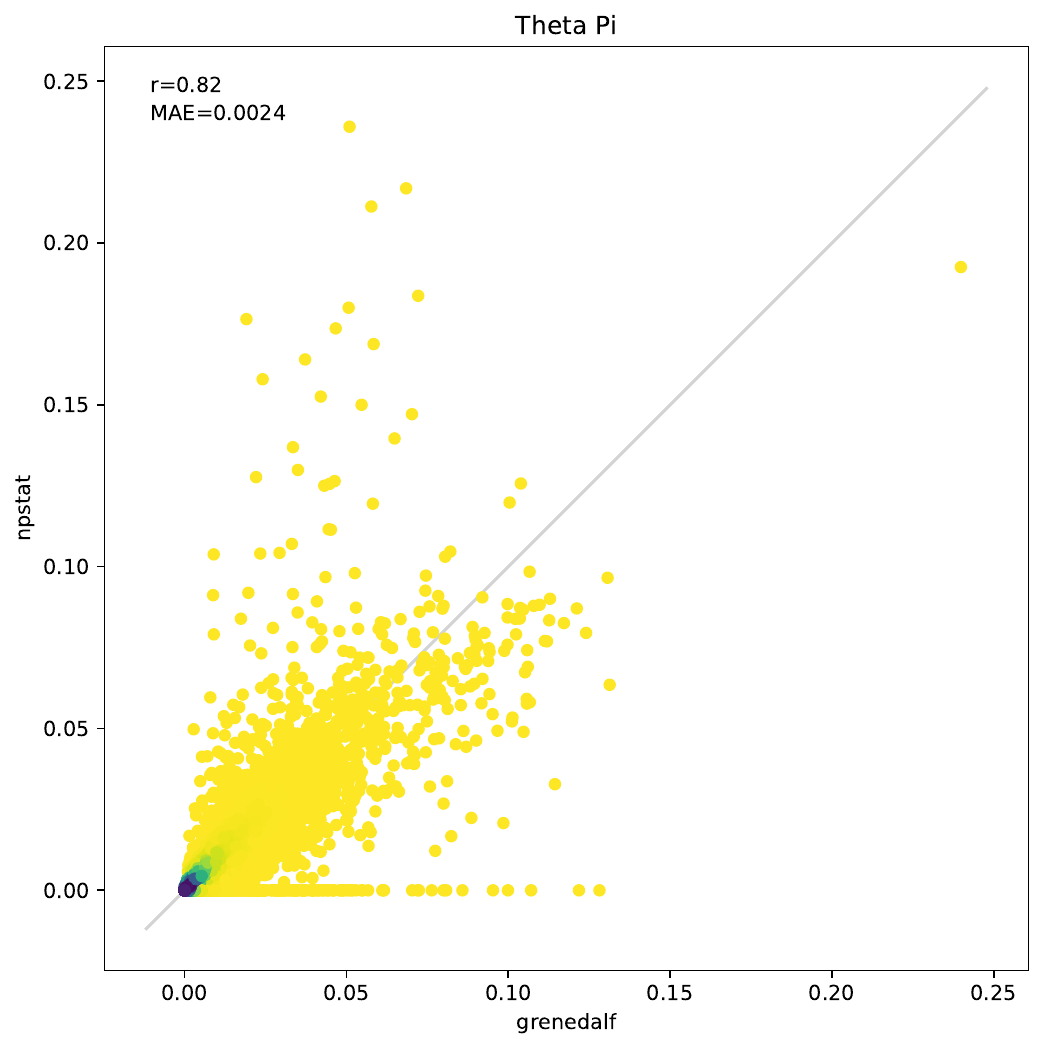}
    \centering
    \end{subfigure}
    \begin{subfigure}[b]{0.49\textwidth}
    \includegraphics[width=\textwidth]{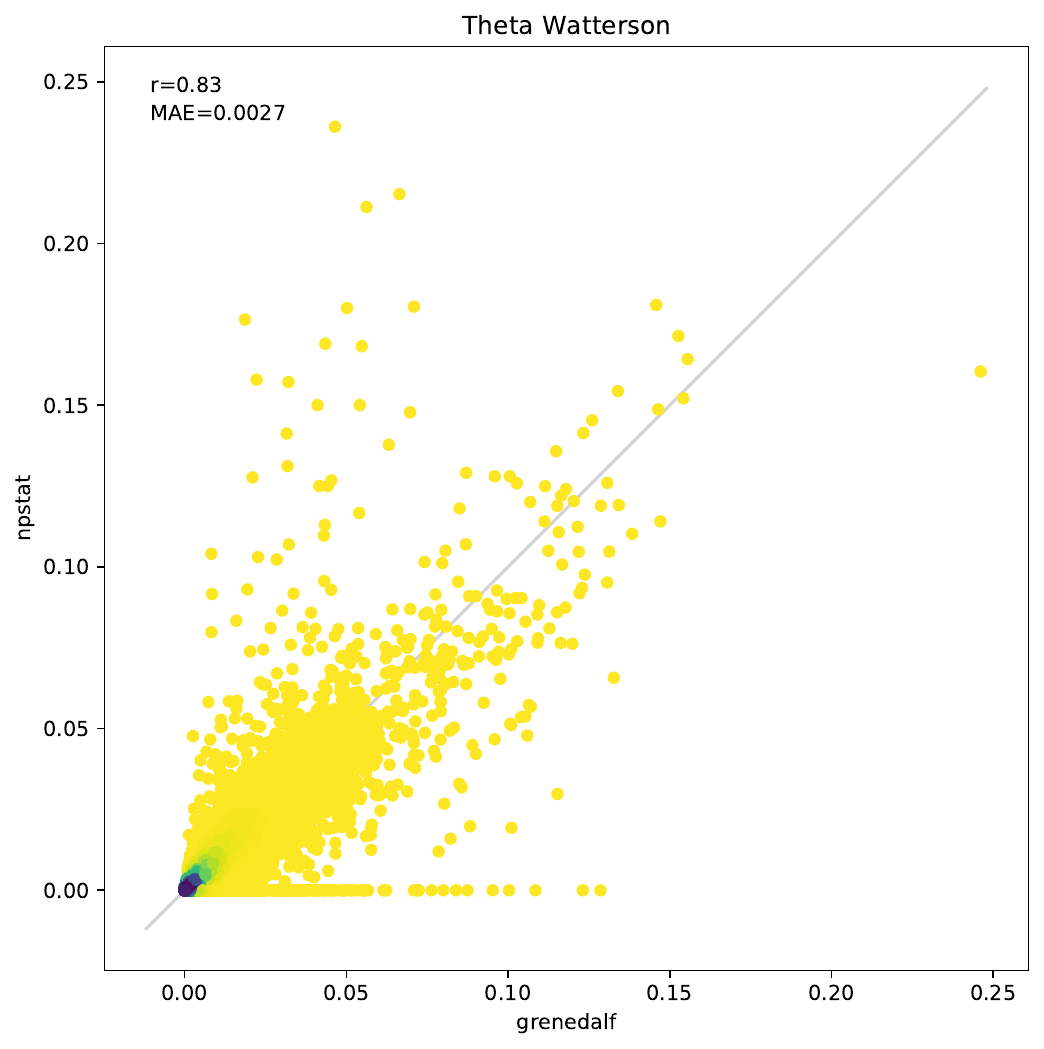}
    \centering
    \end{subfigure}
\caption{
    \textbf{Comparison of \toolname{grenedalf} and \toolname{npstat} for Theta Pi and Theta Watterson on real-world data, with a window size of 1000bp.}
    \\
    This figure is the equivalent of \figref{supp:fig:correlation-thetas-popoolation} for comparing the results of the implementation in \toolname{grenedalf} to those obtained with \toolname{npstat} for Theta Pi (left) and Theta Watterson (right).
    The difference between the tools is more pronounced here. Based on our inspection, and following the reasoning as outlined in \figref{supp:fig:correlation-thetas-popoolation}, we find that this is due to slight differences in the conditions and order of filters applied to the data, and similar implementation details that can have large effect in windows with few variants and low read depth. We however did not attempt to track down the exact origins of these differences in the source code. 
    \\
    Instead, we compared the values obtained with \toolname{grenedalf} and \toolname{npstat}.  
    We find that the differences in values obtained with both tools between the two estimators (Theta Pi and Theta Watterson) are highly consistent (correlation coefficient of 0.98). In other words, positions in the data where \toolname{grenedalf} and \toolname{npstat} give different estimates of Theta Pi also give systematically different estimates of Theta Watterson (and vice versa), with the difference between values obtained from both tools at a given position being almost identical for both statistics.
    This is also visible in the two plots here, where the pattern of scattering between the two tools looks almost identical for both estimators.
    \\
    This corroborates our finding that the differences between \toolname{grenedalf} and \toolname{npstat} are indeed artifacts of filter implementation (or related implementation details such as which positions are counted in the denominators of the estimators), and not in the actual statistics implementations.  We further show that this is the case in \figref{supp:fig:correlation-thetas-npstat-100000}.
}
\label{supp:fig:correlation-thetas-npstat-1000}
\end{figure}
\begin{figure}
    \begin{subfigure}[b]{0.49\textwidth}
    \includegraphics[width=\textwidth]{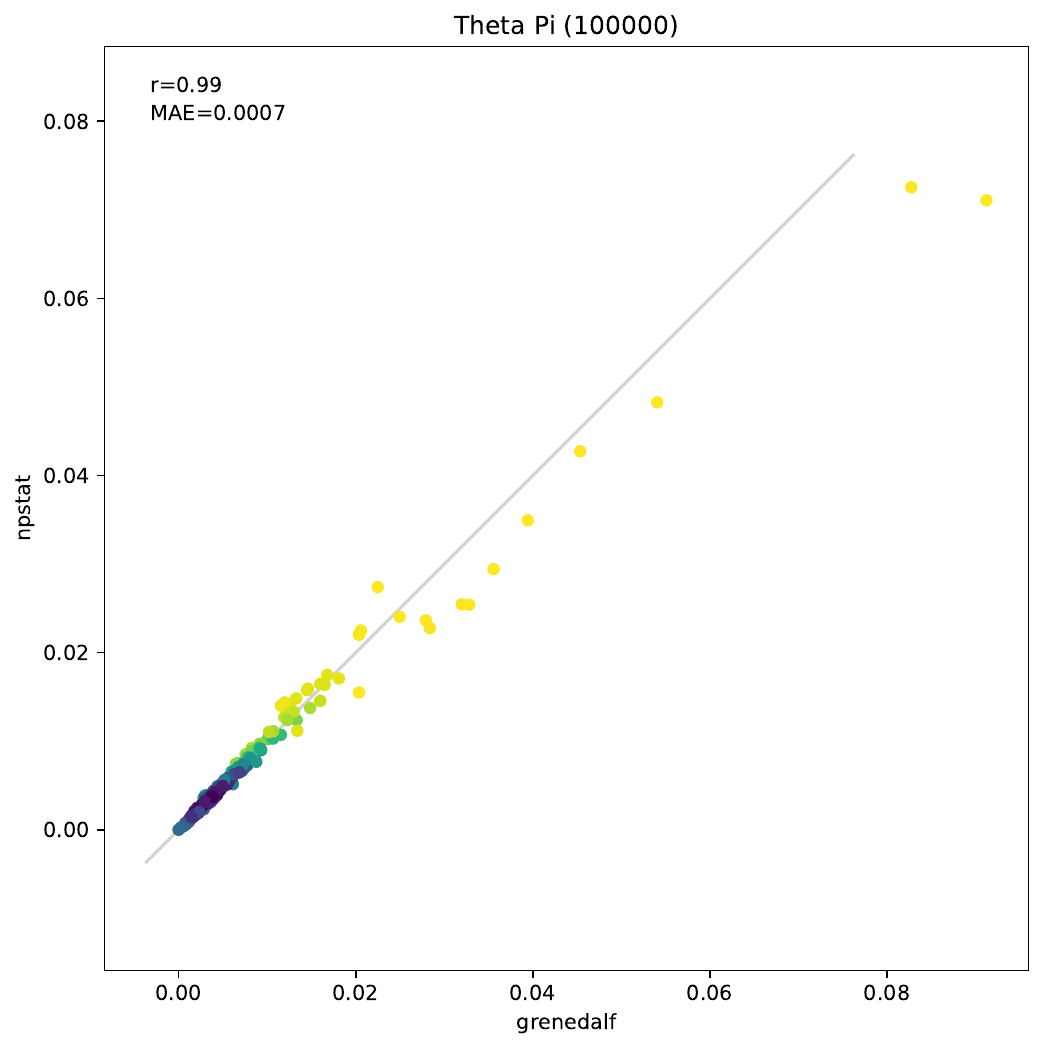}
    \centering
    \end{subfigure}
    \begin{subfigure}[b]{0.49\textwidth}
    \includegraphics[width=\textwidth]{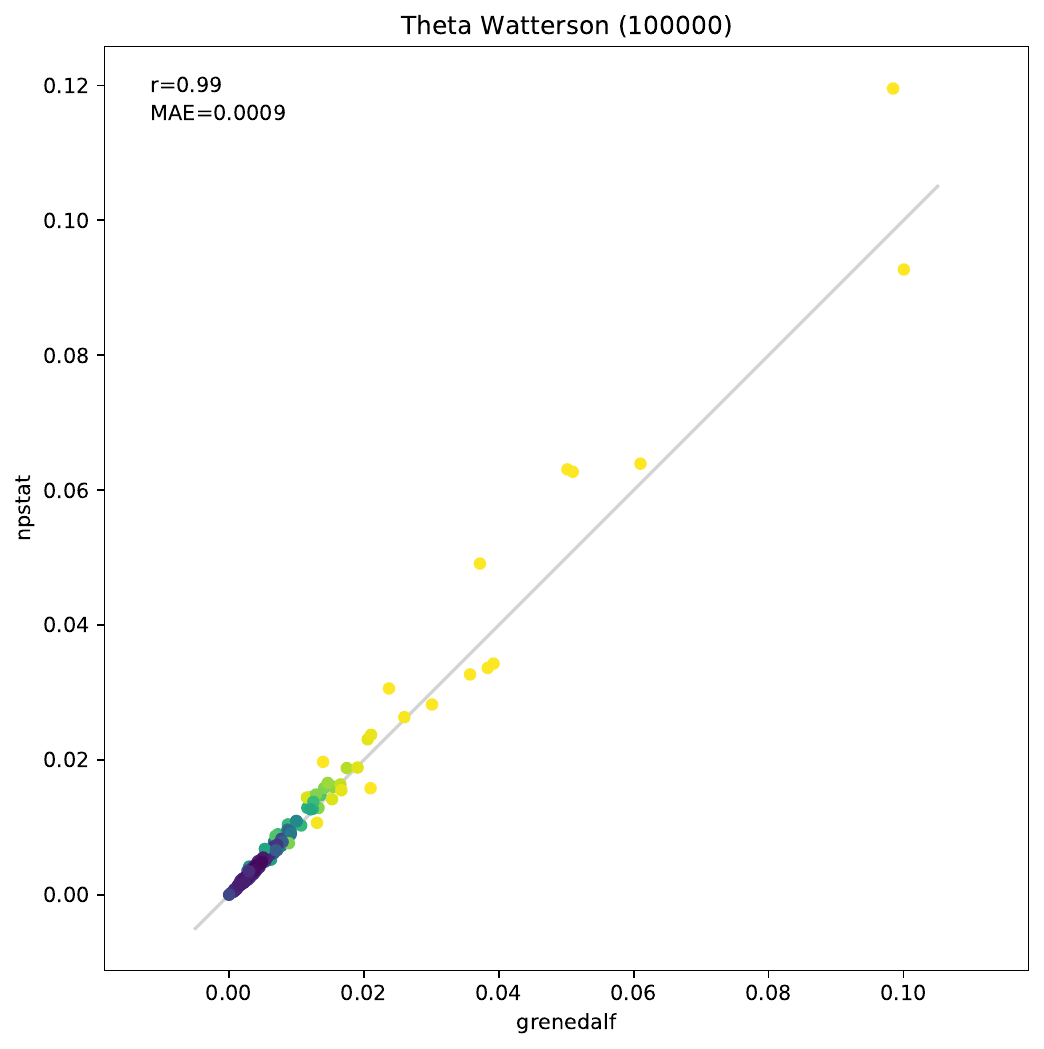}
    \centering
    \end{subfigure}
\caption{
    \textbf{Comparison of \toolname{grenedalf} and \toolname{npstat} for Theta Pi and Theta Watterson on real-world data, with a window size of 100,000bp.}
    \\
    This figure is the equivalent of \figref{supp:fig:correlation-thetas-npstat-1000} with a larger window size. As predicted there, we here indeed observe a stronger match between \toolname{grenedalf} and \toolname{npstat}. This is because the windows here contain more variants, so that variants that are treated differently between the tools based on slight implementation details do not exert large effects. 
    Thus, differences in values between the tools are likely caused by slight edge cases in the implementation details for small windows with low number of variants and low read depth.
    We hence do not further investigate this here, as these decisions are to some degree arbitrary, and diminish at larger read depths and in larger windows with more variant positions. 
    \\
    We note that this implies to not over-interpret high values of either statistic in windows with low read depth and low variation: estimates in such windows tend to have high variance, where slight implementation details can have a large effect. Larger window sizes can be used to average out the effect of these outliers.
}
\label{supp:fig:correlation-thetas-npstat-100000}
\end{figure}
\begin{figure}
    \begin{subfigure}[b]{0.49\textwidth}
    \includegraphics[width=\textwidth]{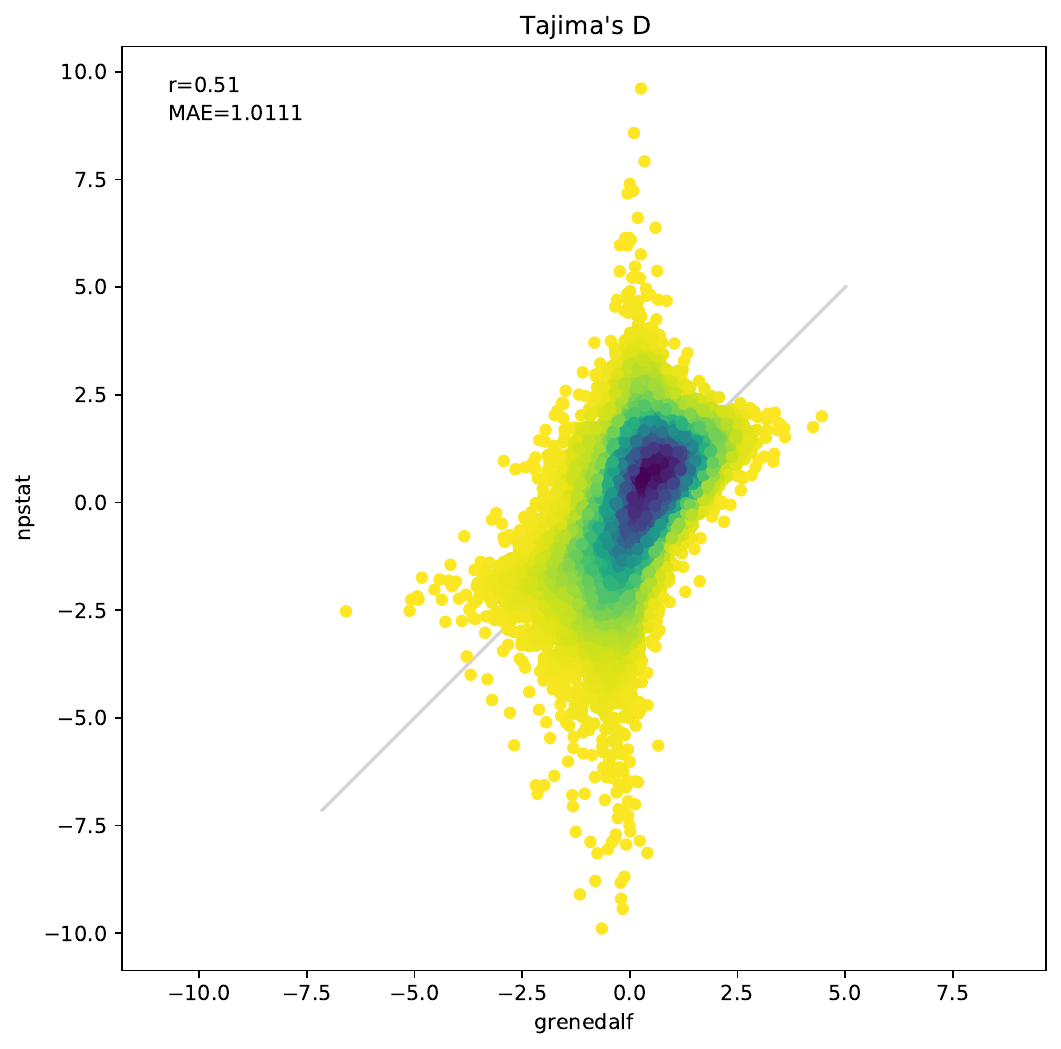}
    \centering
    \end{subfigure}
    \begin{subfigure}[b]{0.49\textwidth}
    \includegraphics[width=\textwidth]{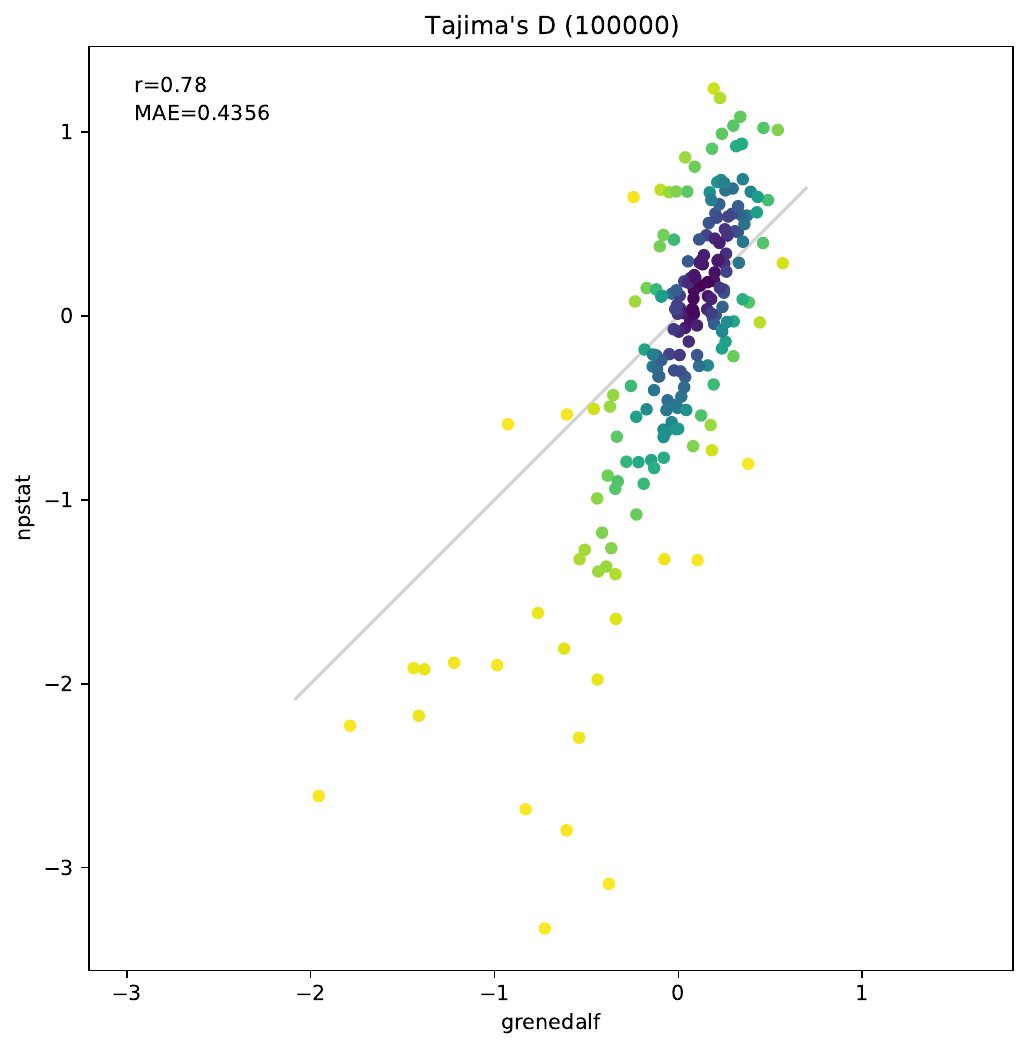}
    \centering
    \end{subfigure}
\caption{
    \textbf{Comparison of \toolname{grenedalf} and \toolname{npstat} for Tajima's D on real-world data.}
    \\
    Based on \figref{supp:fig:correlation-thetas-npstat-1000} and \figref{supp:fig:correlation-thetas-npstat-100000}, we here show the comparison between \toolname{grenedalf} and \toolname{npstat} for Tajima's D, for both window sizes of 1000bp (left) and 100,000bp (right).
    The discrepancies between the two tools is most pronounced at the small window size, where no strong correlation is present. Given the results of the previous figures, this is expected, as the discrepancies between the tools are exacerbated when computing the difference between Theta Pi and Theta Watterson in order to get the estimator for Tajima's D. At the larger window size, there is a more significant correlation between the tools, but estimates seem to be stretched by a constant factor in \toolname{npstat}.
    \\
    Given the difficulties and shortcomings of the estimator of Tajima's D for pool sequencing that we describe in our equations document, we decided to not further investigate this. Likely, the differences seen here are a combination of noise in the data in low read depth conditions, amplified by slight implementation differences, and further obscured by statistical deficits of the estimator itself. We hence urge practitioners to exercise caution when interpreting values of Tajima's D numerically.
}
\label{supp:fig:correlation-tajima-d-npstat}
\end{figure}
\begin{figure}
    \begin{subfigure}[b]{0.49\textwidth}
    \includegraphics[width=\textwidth]{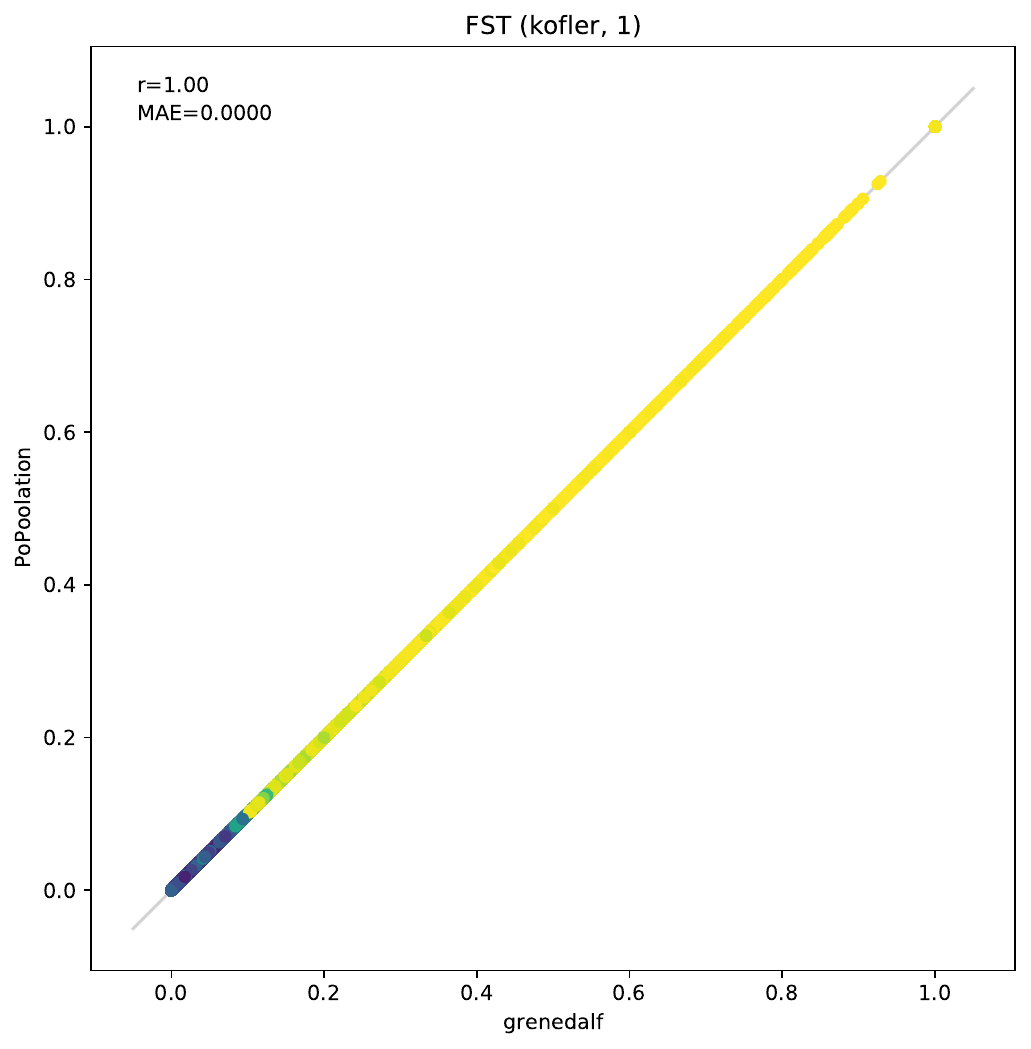}
    \centering
    \end{subfigure}
    \begin{subfigure}[b]{0.49\textwidth}
    \includegraphics[width=\textwidth]{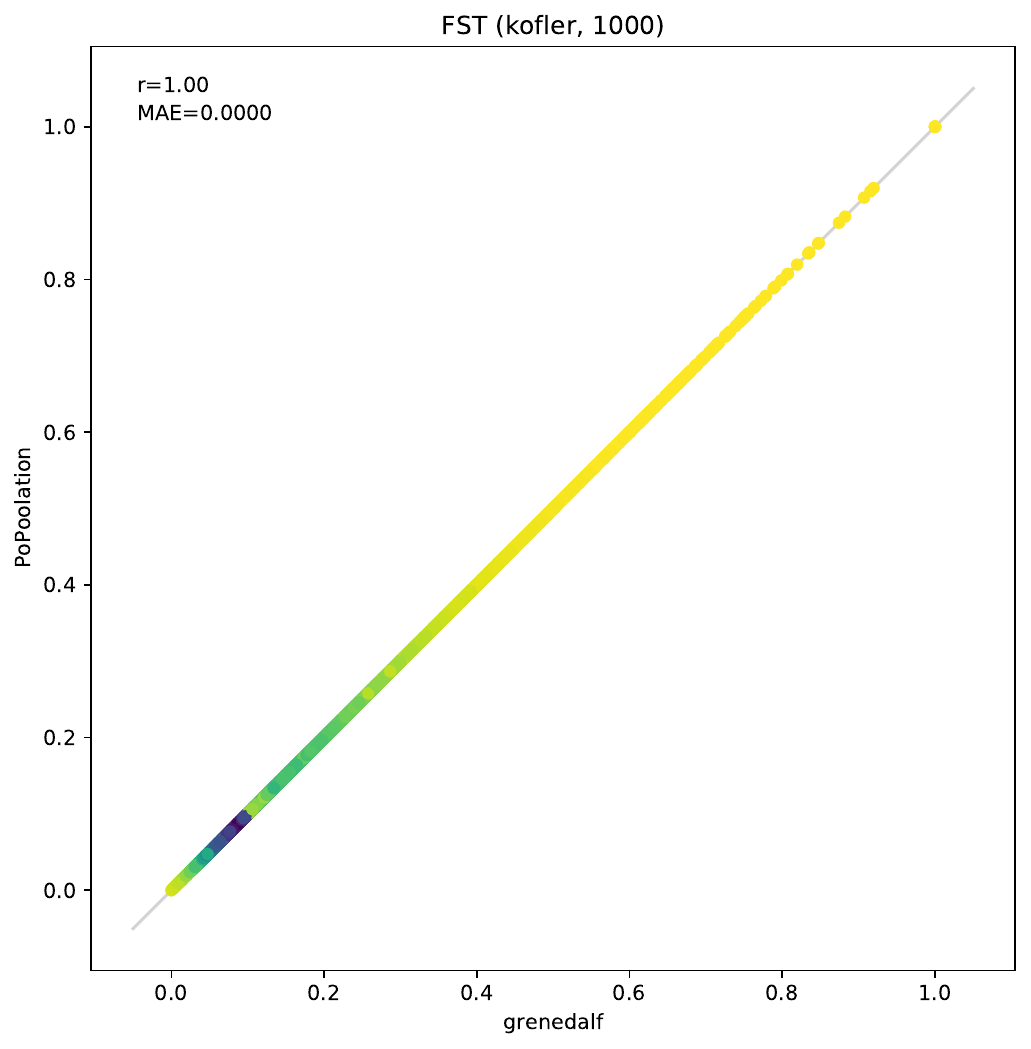}
    \centering
    \end{subfigure}
\caption{
    \textbf{Comparison of \toolname{grenedalf} and \toolname{PoPoolation2} of values for \fst{} Kofler on real-world data.}
    \\
    We here compare the results of the implementation of the ``Kofler'' estimator of \fst{} in \toolname{grenedalf} to those obtained with \toolname{PoPoolation2} for different window sizes. The input data is the same as used in \figref{supp:fig:grenenet-fst-positions}, and as explained in the plots above. We ran both tools with window sizes of 1bps and 1000bps, respectively, in order to evaluate potential single SNP differences as well as window normalization differences. Other parameters such as minimum counts and read depths were matching otherwise. The plots show the values per window obtained with each tool, with coloring being based on a Gaussian kernel density estimate. Both tools produce almost identical results. 
}
\label{supp:fig:correlation-fst-popoolation-kofler}
\end{figure}
\begin{figure}
    \begin{subfigure}[b]{0.49\textwidth}
    \includegraphics[width=\textwidth]{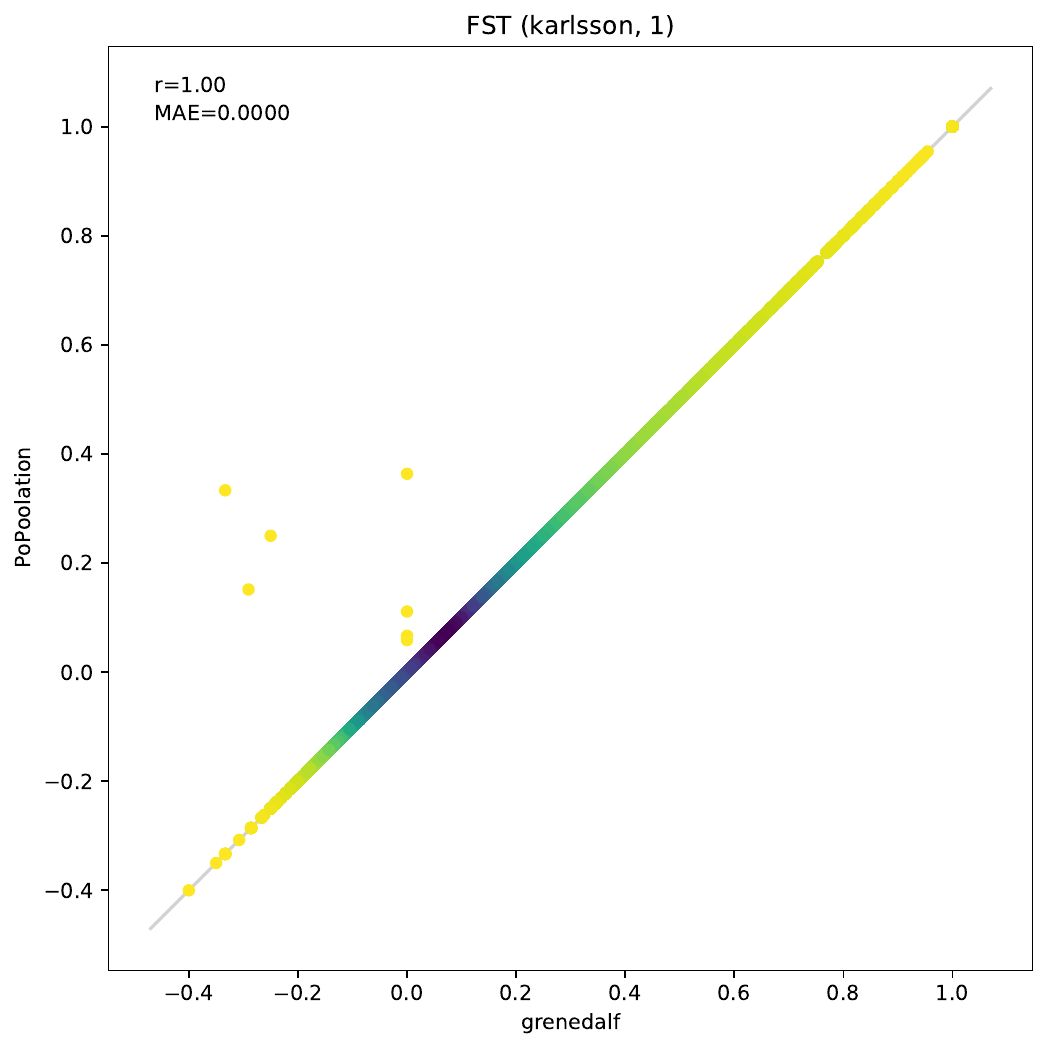}
    \centering
    \end{subfigure}
    \begin{subfigure}[b]{0.49\textwidth}
    \includegraphics[width=\textwidth]{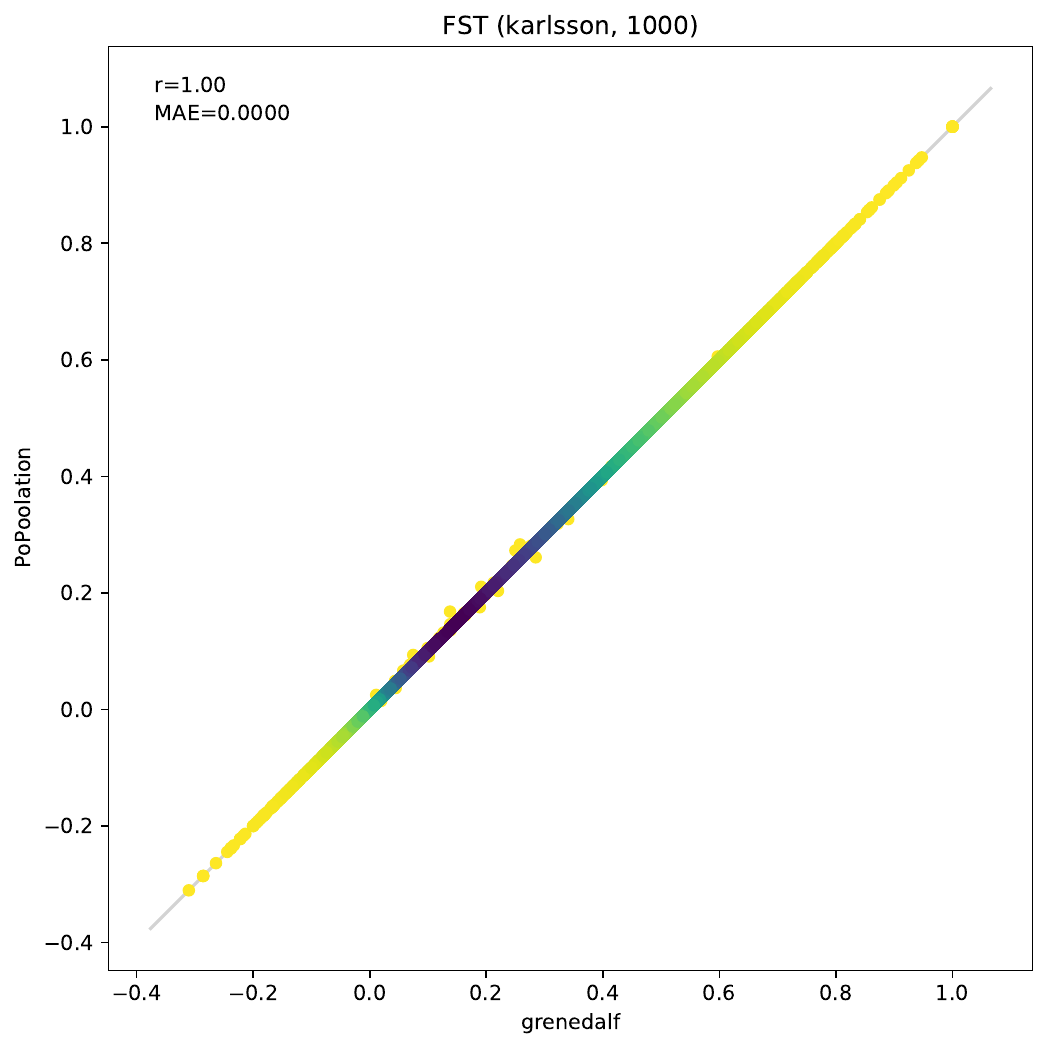}
    \centering
    \end{subfigure}
\caption{
    \textbf{Comparison of \toolname{grenedalf} and \toolname{PoPoolation2} of values for \fst{} Karlsson on real-world data.}
    \\
    This figure is the equivalent of \figref{supp:fig:correlation-fst-popoolation-kofler} for the ``Karlsson'' estimator of \fst{}. The setup, results, and conclusions are exactly the same; see there for details. Note that due to the corrections applied in the estimator, negative values of \fst{} can be observed here. This is expected; see our equation document for details.
}
\label{supp:fig:correlation-fst-popoolation-karlsson}
\end{figure}
\begin{figure}
    \begin{subfigure}[b]{0.49\textwidth}
    \includegraphics[width=\textwidth]{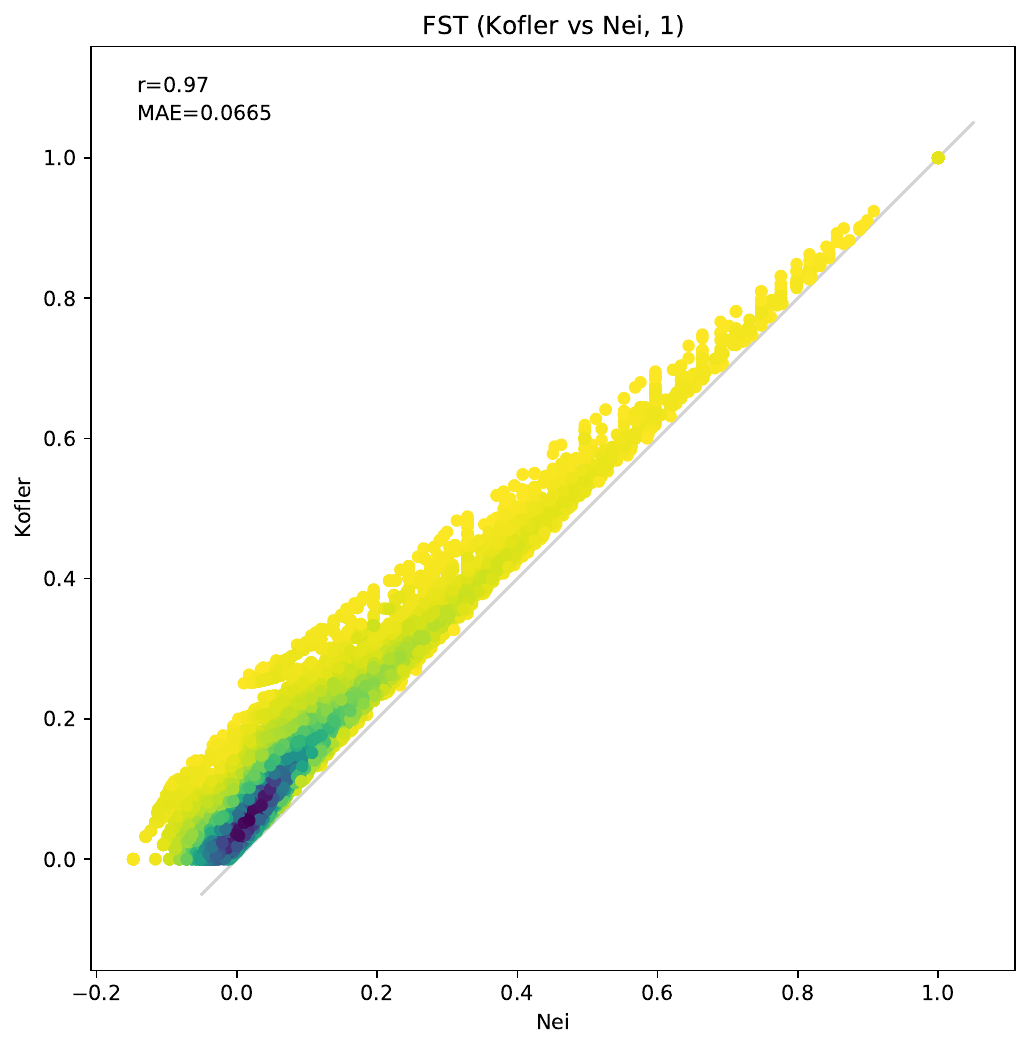}
    \centering
    \end{subfigure}
    \begin{subfigure}[b]{0.49\textwidth}
    \includegraphics[width=\textwidth]{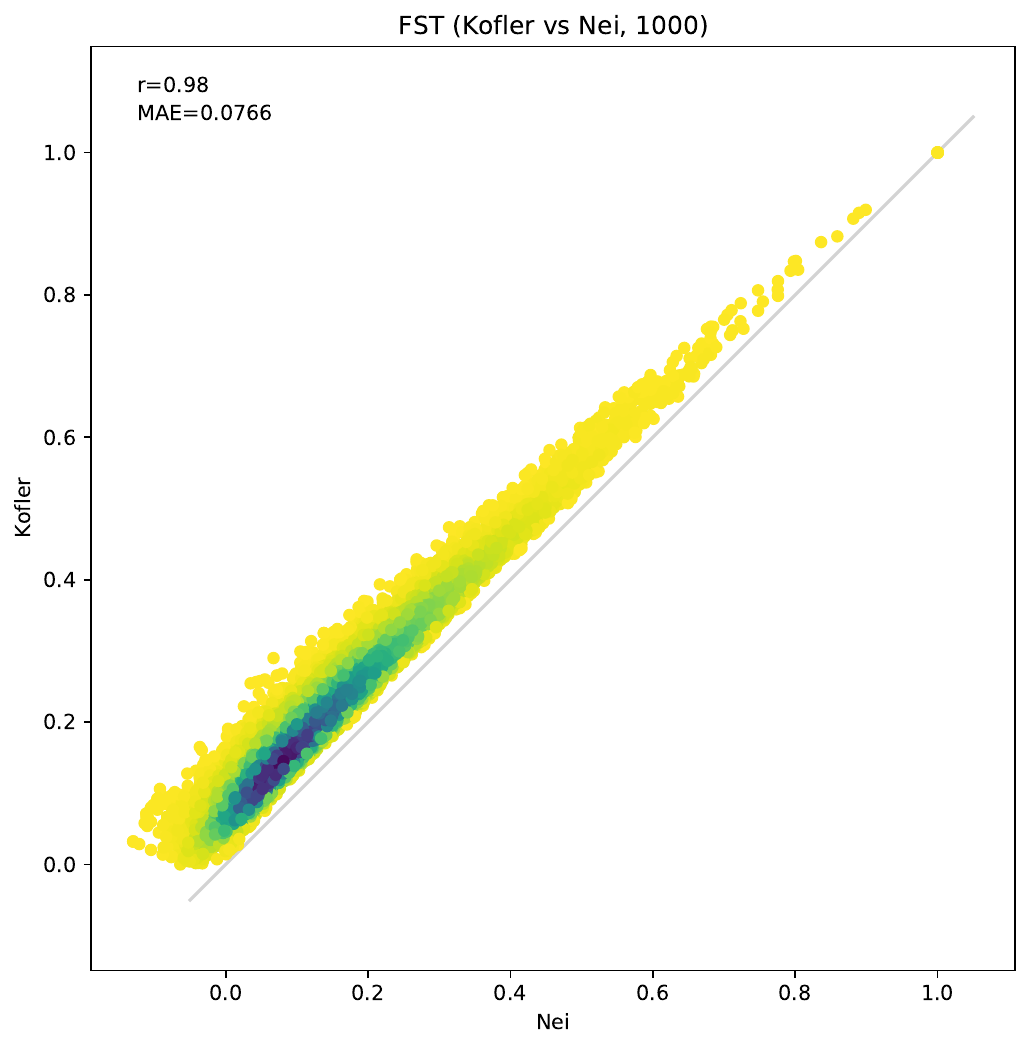}
    \centering
    \end{subfigure}
\caption{
    \textbf{Comparison of the Kofler vs Nei estimators of \fst{} on real-world data.}
    \\
    As shown in our assessment of the Pool-seq correction equations for \fst{}, the estimator as implemented in \toolname{PoPoolation2}, which we call the ``Kofler'' estimator, is biased upward for small pool sizes and small read depths.  Here, we show the effect of this bias, compared to our asymptotically unbiased estimator of Nei's definition of \fst{}, on single SNPs and in windows of 1000bps. The data again consists of two samples from GrENE-net, as introduced above. All computations where conducted with \toolname{grenedalf}. To focus on the effect of read depth, we used a pool size of $100$ here (although in reality, the data came from a smaller pool of individuals). See also \figref{supp:fig:correlation-fst-karlsson-hudson}.
    \\
    It is evident that the Kofler estimator has consistently larger values, independent of the window size. This corroborates our theoretical findings of the estimator being upwards biased, as explained in our equations document.
}
\label{supp:fig:correlation-fst-kofler-nei}
\end{figure}
\begin{figure}
    \begin{subfigure}[b]{0.49\textwidth}
    \includegraphics[width=\textwidth]{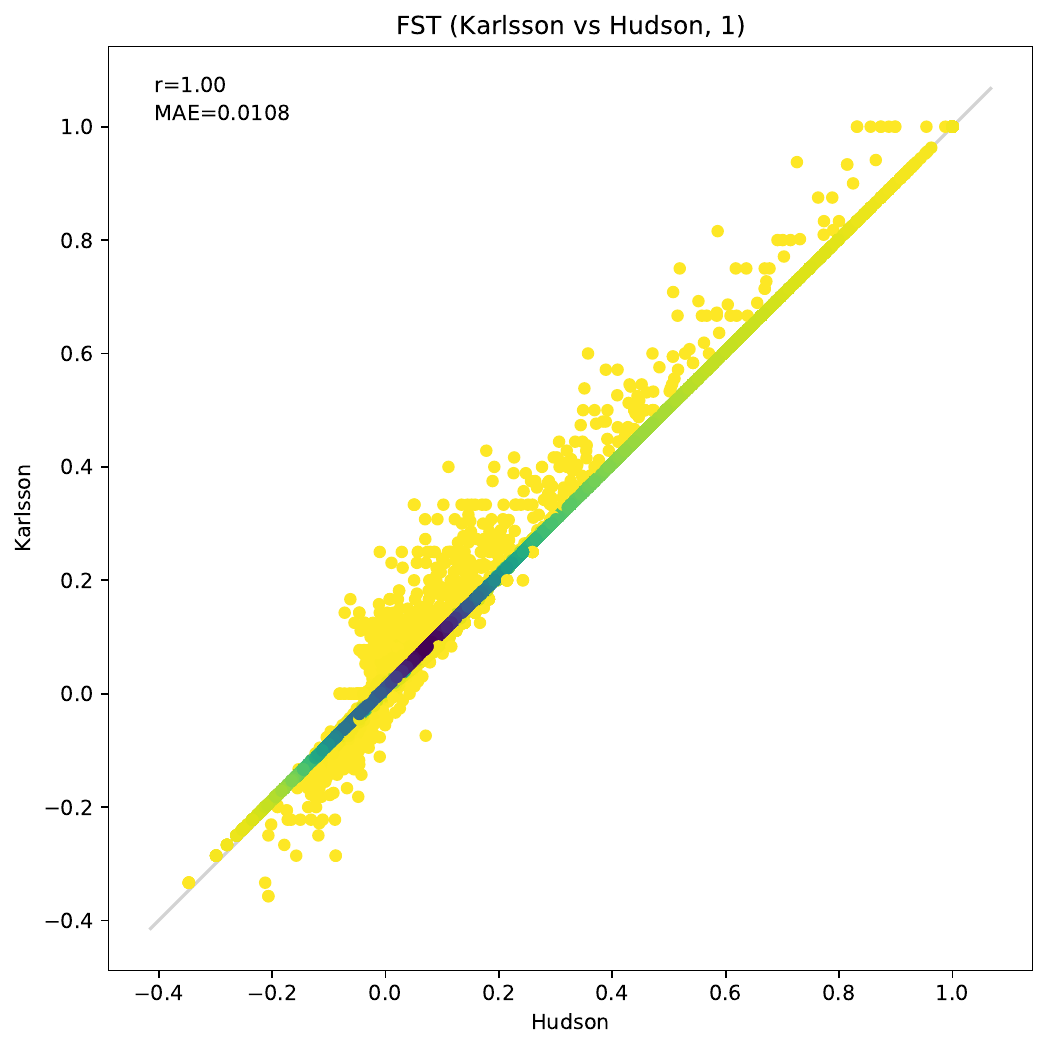}
    \centering
    \end{subfigure}
    \begin{subfigure}[b]{0.49\textwidth}
    \includegraphics[width=\textwidth]{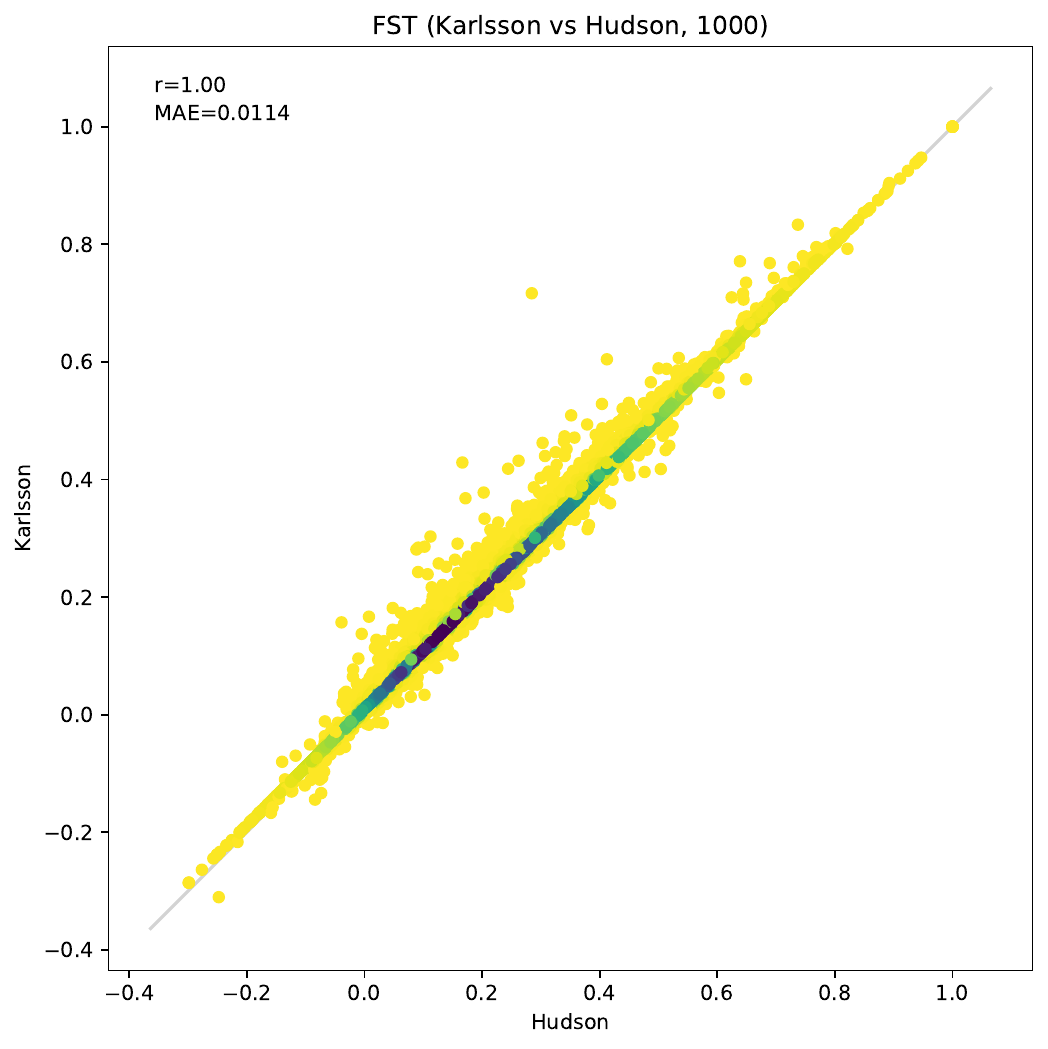}
    \centering
    \end{subfigure}
\caption{
    \textbf{Comparison of the Karlsson vs Hudson estimators of \fst{} on real-world data.}
    \\
    Similar to \figref{supp:fig:correlation-fst-kofler-nei}, here we evaluate the differences of the ``Karlsson'' estimator compared to our asymptotically unbiased estimator following Hudson's definition of \fst{}, on single SNPs and in windows of 1000bp.  We again used a pool size of $100$ here. As the Karlsson estimator is only biased for low pool sizes, it behaves virtually identical to our Hudson estimator at that pool size, as also shown in the equations assessment. The scattering seen in the plots here solely come from multiallelic positions, where the third and forth allele are taken into account by the Hudson estimator, but ignored by the Karlsson estimator (as a consequence of it being defined for biallelic SNPs). The differences completely disappear when additionally filtering the input data for biallelic SNPs or constraining the data by removing the counts of the third and fourth alleles.
}
\label{supp:fig:correlation-fst-karlsson-hudson}
\end{figure}

Below, we show several benchmarks and case studies, evaluating runtime and memory requirements in detail:

\begin{itemize}
    \item \figref{supp:fig:cathedral-plot} and \figref{supp:fig:pairwise-fst-matrix}: Exemplary plots based on computing \fst{}, showing what the improved speed of \toolname{grenedalf} can be used for.
    \item \figref{supp:fig:grenenet-convert}, \figref{supp:fig:grenenet-diversity}, \figref{supp:fig:grenenet-fst-positions}: Benchmarks based on real world data, as a showcase with realistic data and assumptions.
    \item \figref{supp:fig:random-diversity}, \figref{supp:fig:random-fst-window}, \figref{supp:fig:random-fst-snps-genome}: Benchmarks based on randomly simulated data, as a baseline test with very simple input, showing the minimally expected improvement provided by \toolname{grenedalf}.
    \item \figref{supp:fig:scaling-strong} and 
    \figref{supp:fig:scaling-fst-matrix}: Scaling efficiency of \toolname{grenedalf} when using multi-threading.
    \item \figref{supp:fig:correlation-thetas-popoolation} and \figref{supp:fig:correlation-tajima-d-popoolation}:  Comparison of the diversity estimators (Theta Pi, Theta Watterson, and Tajima's D) between \toolname{grenedalf} and \toolname{PoPoolation}, based on real world data.
    \item \figref{supp:fig:correlation-thetas-npstat-1000}, \figref{supp:fig:correlation-thetas-npstat-100000}, \figref{supp:fig:correlation-tajima-d-npstat}:  Comparison of the diversity estimators (Theta Pi, Theta Watterson, and Tajima's D) between \toolname{grenedalf} and \toolname{npstat}, based on real world data.
    \item \figref{supp:fig:correlation-fst-popoolation-kofler} and \figref{supp:fig:correlation-fst-popoolation-karlsson}:  Comparison of the Fst estimators between \toolname{grenedalf} and \toolname{PoPoolation2}, based on real world data.
    \item \figref{supp:fig:correlation-fst-kofler-nei} and \figref{supp:fig:correlation-fst-karlsson-hudson}:  Evaluation of the biases of different estimators of \fst{} on real world data, corroborating our findings on simulated data as shown in our assessment of the equations.
\end{itemize}

The scripts for running the tests and creating the plots shown here are provided at
\url{https://github.com/lczech/grenedalf-paper}.

Additionally, in order to double-check the correctness of our implementation of the equations, we also re-implemented a minimal version of all estimators in Python, based on our equations document. This independent implementation follows the document closely, and was used for simulated test under different read depth and pool size conditions. We find that results obtained with our implementation in grenedalf match this simple independent implementation, confirming that the equations are implemented as described.

\pagebreak



\clearpage

\clearpage

\clearpage

\clearpage

\bibliographystyle{pnas-new}

\bibliography{03-benchmarks}